\documentclass[12pt]{article} 
\usepackage[sectionbib]{natbib}
\usepackage{array,epsfig,fancyheadings,rotating}
\usepackage[]{hyperref}  

\usepackage{sectsty, secdot}
\sectionfont{\fontsize{12}{14pt plus.8pt minus .6pt}\selectfont}
\renewcommand{\theequation}{\thesection\arabic{equation}}
\subsectionfont{\fontsize{12}{14pt plus.8pt minus .6pt}\selectfont}

\usepackage{amsfonts,amsmath,amssymb,bm,mathtools,mathrsfs}
\usepackage{graphicx,booktabs,caption,subfigure,multirow,nameref,makecell}
\usepackage{xr,times,xcolor,natbib,setspace,comment}
\usepackage[ruled]{algorithm2e}

\newcommand{\Bcal}{\mathcal{B}}
\newcommand{\trans}{^{\mathrm{\scriptscriptstyle T} }}
\usepackage{enumerate}   

\newcommand{\change}[1]{{\leavevmode\color{black}{#1}}}

\newtheorem{theorem}{Theorem}
\newtheorem{lemma}{Lemma}

\newtheorem{proposition}{Proposition}
\newtheorem{assumption}{Assumption}

\newtheorem{definition}{Definition}

\usepackage[toc,page]{appendix}
\usepackage{titlesec}

\makeatletter
\renewcommand{\algocf@captiontext}[2]{#1\algocf@typo. \AlCapFnt{}#2} 
\def\@algocf@capt@plain{top}
\renewcommand{\algocf@makecaption}[2]{%
  \addtolength{\hsize}{\algomargin}%
  \sbox\@tempboxa{\algocf@captiontext{#1}{#2}}%
  \ifdim\wd\@tempboxa >\hsize
    \hskip .5\algomargin%
    \parbox[t]{\hsize}{\algocf@captiontext{#1}{#2}}
  \else%
    \global\@minipagefalse%
    \hbox to\hsize{\box\@tempboxa}
  \fi%
  \addtolength{\hsize}{-\algomargin}%
}
\makeatother

\def\eop{\hfill {\large $\Box$}}

\makeatletter
\newcommand*{\addFileDependency}[1]{
  \typeout{(#1)}
  \@addtofilelist{#1}
  \IfFileExists{#1}{}{\typeout{No file #1.}}
}
\makeatother

\newcommand*{\myexternaldocument}[1]{%
    \externaldocument{#1}%
    \addFileDependency{#1.tex}%
    \addFileDependency{#1.aux}%
}

\myexternaldocument{supp-09022024}

\begin{document}


\renewcommand{\baselinestretch}{2}

\markright{ \hbox{\footnotesize\rm Statistica Sinica
}\hfill\\[-13pt]
\hbox{\footnotesize\rm
}\hfill }

\markboth{\hfill{\footnotesize\rm FIRSTNAME1 LASTNAME1 AND FIRSTNAME2 LASTNAME2} \hfill}
{\hfill {\footnotesize\rm FILL IN A SHORT RUNNING TITLE} \hfill}

\renewcommand{\thefootnote}{}
$\ $\par


\fontsize{12}{14pt plus.8pt minus .6pt}\selectfont \vspace{0.8pc}
\centerline{\large\bf Bayesian Inference of Spatially Varying Correlations }
\vspace{2pt} 
\centerline{\large\bf via the Thresholded Correlation Gaussian Process}
\vspace{.4cm} 
\centerline{Moyan Li$^1$, Lexin Li$^2$ and Jian Kang$^1$} 
\vspace{.4cm} 
\centerline{\it $^1$University of Michigan, Ann Arbor and $^2$University of California, Berkeley}
 \vspace{.55cm} \fontsize{9}{11.5pt plus.8pt minus.6pt}\selectfont


\begin{quotation}
\noindent {\it Abstract:}
{ A central question in multimodal neuroimaging analysis is to understand the association between two imaging modalities and to identify brain regions where such an association is statistically significant. In this article, we propose a  Bayesian nonparametric spatially varying correlation model to make inference of such regions. We build our model based on the thresholded correlation Gaussian process (TCGP). It ensures piecewise smoothness, sparsity, and jump discontinuity of spatially varying correlations, and is well applicable even when the number of subjects is limited or the signal-to-noise ratio is low. We study the identifiability of our model, establish the large support property, and derive the posterior consistency and selection consistency. We also develop a highly efficient Gibbs sampler and its variant to compute the posterior distribution. We illustrate the method with both simulations and an analysis of functional magnetic resonance imaging data from the Human Connectome Project.  
}\\

\vspace{9pt}
\noindent {\it Key words and phrases:}
Bayesian modeling; Gaussian process; Multimodal correlation analysis; Neuroimaging analysis.
\par
\end{quotation}\par

\def\thefigure{\arabic{figure}}
\def\thetable{\arabic{table}}

\renewcommand{\theequation}{\thesection.\arabic{equation}}

\fontsize{12}{14pt plus.8pt minus .6pt}\selectfont

\section{Introduction}
\label{sec:introduction}

Multimodal neuroimaging is now prevailing in neuroscience research, where different types of brain images are collected for a common set of subjects. Common imaging modalities include  anatomic magnetic resonance imaging (MRI),  resting-state or task-based functional MRI (fMRI), diffusion tensor imaging (DTI), positron emission tomography (PET), among many others. Multimodal neuroimaging analysis aggregates such diverse but often complementary information, consolidates knowledge across different modalities, and produces improved understanding of neurological disorders. See \citet{uludaug2014general} \change{and \citet{zhu2023statistical} for some excellent reviews on multimodal neuroimaging analysis.}

A central question in multimodal neuroimaging analysis is to understand the association between two imaging modalities and to identify brain regions where such an association is statistically significant. This question is of great scientific interest. For instance, \citet{zhu2014fusing} surveyed and showed joint analysis of fMRI and DTI reveals important interplays between brain functions and structures. \citet{cavaliere2018multimodal} showed fMRI and PET together improve the characterization of patients with consciousness disorder. \citet{li2018spatially} jointly analyzed two PET modalities with different nuclear tracers, and identified brain regions where the tau protein and glucose metabolism are strongly correlated to facilitate the understanding of Alzheimer's disease pathology. \citet{harrewijn2020combining} studied resting-state and task-based fMRI, and found that functional connectivities during the rest and the dot-probe task are positively correlated, which conforms to and further extends the current studies of human cognitive behaviors.

In this article, we propose a  Bayesian nonparametric spatially varying correlation model to address the question of estimation and inference of spatial regions where two imaging modalities are significantly correlated. We build our model based on the thresholded correlation Gaussian process (TCGP), which ensures piecewise smoothness, sparsity, as well as jump discontinuity of spatially varying correlations, and works well even when the number of subjects is limited or the signal-to-noise ratio is small. We study the identifiability of our model, establish the large support property, and derive the posterior consistency and selection consistency. We derive the full conditional distributions, propose a Gibbs sampling algorithm that is highly efficient, and propose a hybrid mini-batch MCMC to further improve the computational efficiency.  We apply our proposed method to jointly analyze the resting-state and working memory task-based fMRIs from a study of the Human Connectome Project, and identify a number of scientifically meaningful brain regions that offer useful insights for cognitive neuroscience research.  

Our proposal is related to but also substantially different from the existing literature on multimodal correlation analysis and Bayesian inference. 

For multimodal correlation analysis, there are, broadly speaking, three categories of solutions. The first category is voxel-wise analysis, which estimates the correlation at each voxel separately, then conducts massive voxel-wise significance tests with false discovery control. This approach is computationally easy to implement, but it does not incorporate any spatial or scientific knowledge into statistical inference. Besides, the number of voxels, and thus the number of tests, is huge, whereas the number of subjects in most studies is  limited. As a result, voxel-wise analysis often suffers from a particularly low detection power. Although the random field theory has been suggested for multiple testing correction so to improve voxel-wise analysis \citep{worsley2004unified}, it does not fully address the low power issue, and is also not directly applicable in our problem due to the complex structure of spatially varying correlations. The second category is region-wise analysis, which first summarizes, usually by averaging, the imaging signals within each brain region defined by some pre-specified brain atlas, then carries the correlation analysis at the region level. Although region-wise analysis generally enjoys a better power than voxel-wise analysis, it is sensitive to the choice of brain atlas. More importantly, the voxels within the same region may not always share the same correlation patterns. Averaging the signals by regions may weaken or cancel out significant correlations. The third category merges voxel-wise and region-wise analysis. In particular, \citet{li2018spatially} adapted the spatially varying coefficient model, which is widely used in neuroimaging analysis but generally for a different purpose \citep[e.g.,][]{zhu2014spatially, LiZhu2017, li2021sparse}, to the problem of multimodal correlation analysis. They proposed a multi-step procedure, which first fits a spatially varying coefficient model and obtains a smoothed correlation estimate at the voxel level, then applies a graph clustering algorithm to partition the brain into regions with homogeneous correlations, and finally carries out a likelihood ratio test at the region level to identify the regions where two imaging modalities are significantly correlated. However, this procedure involves multiple tuning parameters, and the testing results may be sensitive to their choices. In addition, due to multiple steps of estimation, it is difficult to establish the theoretical guarantees for the final inference method.

For Bayesian modeling and inference, our proposal also makes a number of useful contributions. First of all, we propose a new Bayesian nonparametric prior, i.e, the thresholded correlation Gaussian process, for spatially varying correlation coefficients that are sparse and piecewise smooth over the space. It is constructed under a Bayesian hierarchical model, by thresholding a Gaussian process of the variances for another two correlated Gaussian processes. Our model targets the second-order correlations between two modalities. Relatedly, \citet{bhattacharya2011sparse} proposed a multiplicative Gamma process shrinkage prior with latent factors to model high-dimensional covariance matrices. Nevertheless, their method places the sparsity on the individual latent factors, whereas we need to impose the sparsity at the voxel level, and the sparsity on the latent factors does not lead to the sparsity on the voxels. In addition, our model hinges on the idea of thresholding a Gaussian process. A similar strategy has been adopted in prior constructions for modeling sparse regressions or spatially varying functions, i.e., either thresholding Gaussian random variables \citep{nakajima2013bayesian_a, ni2019bayesian,cai2020bayesian}, or thresholding Gaussian processes \citep{kang2018scalar, wu2024bayesian}. However, none of those priors are readily applicable for Bayesian analysis of spatially varying correlations as in our setting. Recently, there are several literature focuses on Bayesian brain connectivity analysis. For instance,
\citet{chen2016bayesian} introduced a novel Bayesian hierarchical model designed to infer brain connectivity. This model unifies voxel-level and region-level connectivity by acknowledging the distribution of voxel connectivity between regions.  \citet{lukemire2021bayesian} developed an integrative Bayesian approach for jointly modeling multiple brain networks, which provides a systematic inferential framework for network comparisons. However, our proposal addresses a completely different problem, which aims to infer spatially varying correlations between two image modalities.

Second, we contribute to posterior computations for Bayesian models with thresholding type priors. Most existing solutions resort to gradient based MCMC algorithms \citep{roberts1998optimal, girolami2011riemann}, where a smooth approximation of the thresholding function is required to get the analytically tractable first derivative \citep{cai2020bayesian, wu2024bayesian}. There have also been recent advances in developing new sampling algorithms \citep[e.g.,][]{ahn2012bayesian, chen2014stochastic, nishimura2020discontinuous}. However, these algorithms usually converge relatively slowly, and require multiple tuning parameters. By contrast, instead of using a gradient-based MCMC, we  derive the full conditional distributions, and propose a Gibbs sampler algorithm that is highly efficient. Besides, the proposed posterior computation algorithm is general, and can be applied to other Bayesian models with thresholding priors as well. 

Finally, we are among the first to study the theoretical properties of Bayesian analysis of spatially varying correlations. Particularly, we show that the proposed TCGP has a large prior support on a wide class of sparse, piecewise smooth, and spatially varying correlation functions. We establish the posterior consistency based upon the foundational work of \citet{choi2005posterior, ghosal2006posterior, tokdar2007posterior}. However, it is far from a simple extension, as it involves a two-level Bayesian hierarchical model, multiple Gaussian processes, as well as some thresholding functions. To address these challenges, we propose an equivalent model representation for the transformed data, where the spatially varying correlation coefficients become model parameters that specify the mean of the transformed data. This equivalent formulation substantially simplifies the theoretical analysis in the original model. In light of the sparsity, we further establish the selection consistency of activation regions with nonzero correlation coefficients.  

The rest of the article is organized as follows. We develop our spatially varying correlation model in Section \ref{sec:model}. We derive the theoretical properties in Section \ref{sec:theory}, and the Gibbs sampling algorithm in Section \ref{sec:estimation}. We carry out the simulations in Section \ref{sec:simulations}, and analyze the fMRI data in Section \ref{sec:realdata}. We relegate all technical proofs to the Supplementary Material.

\section{Spatially Varying Correlation Model}
\label{sec:model}

In this section, we first propose our Bayesian spatially varying correlation model and the correlation Gaussian process. We then present an equivalent model formulation.

\subsection{Nonparametric model and correlation Gaussian process}
\label{subsec:model_setup}

Suppose the observed data consist of $n$ subjects, each with two imaging modalities. Suppose these two imaging modalities are well aligned in a $d$-dimensional compact spatial space $\Bcal \subset \mathbb{R}^{d}$, which is generally true for multimodal neuroimaging. Suppose each image consists of measurements at $m$ voxel locations $\Bcal_m = \{v_1, \ldots, v_m\} \subseteq \Bcal$, and we often use $v, v' \in \Bcal$ to denote some generic voxel locations in $\Bcal$. Let $Y_{1,i}(v)$ and $Y_{2,i}(v)$ denote the two imaging measures at location $v$, for subject $i = 1, \ldots, n$. We consider the following model:
\vspace{-3mm}
\begin{align}\label{eq:svcm}
\begin{split}
Y_{k,i} (v) & = \mu_{k,i}(v)+ \varepsilon_{k,i}(v),\quad \varepsilon_{k,i}(v) \sim \mathrm{N}\big(0,\tau^2_k(v)\big), \quad \textrm{for}\;\;  k = 1, 2, 
\end{split}
\end{align}
where  $\mu_{k,i}(v)$ are the spatially varying functions that represent the expected values of $Y_{k,i}(v)$, $\varepsilon_{k,i}(v)$ are the random noises that are mutually independent over $k, i, v$, and follow a normal distribution $\textrm{N}(\cdot, \cdot)$ with mean zero and variance $\tau^2_k(v)$, $k=1,2$. 

We next propose a novel prior model for $\mu_{1,i}(v)$ and $\mu_{2,i}(v)$: 
\vspace{-3mm}
\begin{equation}\label{eq:sdgp}
\resizebox{0.9\hsize}{!}{
$
\begin{split}
& \mu_{1,i}(v) = \eta_{+,i}(v) + \eta_{-,i}(v),  \quad \mu_{2,i}(v) = \eta_{+,i}(v) - \eta_{-,i}(v), \quad \textrm{for}\;\; i = 1, \ldots, n, \\
& \eta_{+,i} \sim \mathrm{GP}(0, \kappa_{+}), \qquad \eta_{-,i} \sim \mathrm{GP}(0, \kappa_{-}), 
\end{split}
$}
\end{equation}
\change{where $\eta_{+,i}$ and $\eta_{-,i}$ are two independent Gaussian processes given $\kappa_{+}$ and $\kappa_{-}$,} and $\eta_{+,i}$ and $\eta_{-,i}$ capture the positive and negative correlations between the two modalities, respectively. We further assume that $\kappa_{+}$ and $\kappa_{-}$ are of the form,
\vspace{-3mm}
\begin{equation} \label{eq:covkern}
\kappa_{+}(v, v') = \sigma_{+}(v)\sigma_{+}(v')\kappa(v,v'), \qquad \kappa_{-}(v, v') = \sigma_{-}(v)\sigma_{-}(v')\kappa(v,v'),
\vspace{-3mm}
\end{equation}
where $\sigma^2_{+}(v)$ and $\sigma^2_{-}(v)$ are the spatially varying variance functions for $\eta_{+,i}(v)$ and $\eta_{-,i}(v)$, respectively, and $\kappa(\cdot, \cdot)$ is a stationary correlation kernel function. There are various choices for the kernel $\kappa(\cdot,\cdot)$; for instance, we employ a Mat{\'e}rn kernel in our implementation, 
\vspace{-3mm}
\begin{equation} \label{eq: matern}
\kappa(v,v') = \frac{2^{1-\gamma_1}}{\Gamma(\gamma_1)}\left(\surd{\left(2 \gamma_1\right)} \frac{\|v-v'\|}{\gamma_2}\right)^{\gamma_1} B_{\gamma_1}\left(\surd{\left(2 \gamma_1\right)} \frac{\|v-v'\|}{\gamma_2}\right), 
\end{equation}
where $\Gamma(\cdot)$ is the gamma function, $B_{\gamma_1}(\cdot)$ is the modified Bessel function of the second kind, and $\gamma_1$ and $\gamma_2$ are two positive hyperparameters that can be determined by the Bayes factor. To impose the sparsity as well as to ensure the identifiability, we require that $\sigma^2_{+}(v) \sigma^2_{-}(v) = 0$. In other words, only one of the two terms $\sigma^2_{+}(v)$ and $\sigma^2_{-}(v)$ is nonzero. 

Finally, we impose that the variance functions are of the form, 
\vspace{-3mm}
\begin{align} \label{eq:varfunc}
\begin{split}
\sigma_{+}(v) =  G_\omega\{\xi(v)\},  \quad  \sigma_{-}(v) = G_\omega\{-\xi(v)\},  \quad  \xi(v) \sim \mathrm{GP}(0, \kappa), 
\end{split}
\end{align}
where $G_\omega(x) = xI(x>\omega)$ is a thresholding function with the thresholding parameter $\omega \geq 0$ and $I(\cdot)$ the indicator function, $\xi(v)$ is a spatially varying function that determines both $\sigma_{+}(v)$ and $\sigma_{-}(v)$ through $G_\omega(x)(\cdot)$. As a prior specification, we assume $\xi(v)$ follows another Gaussian process with mean zero and correlation kernel $\kappa(\cdot, \cdot)$, and $\kappa(\cdot, \cdot)$ is the same as that in \eqref{eq:covkern}. Note that the construction in \eqref{eq:varfunc} ensures $\sigma^2_+(v)$ and $\sigma^2_{-}(v)$ are uniquely determined by $\xi(v)$, and $\sigma^2_{+}(v) \sigma^2_{-}(v) = 0$.

Following the prior specifications \eqref{eq:sdgp} to \eqref{eq:varfunc}, and integrating out $\mu_{1,i}(v)$ and $\mu_{2,i}(v)$ in \eqref{eq:svcm}, we obtain the spatially varying correlation function between $Y_{1,i}(v)$ and $Y_{2,i}(v)$ as, 
\vspace{-3mm}
\begin{equation}\label{eq:spat_cor}
\resizebox{0.9\hsize}{!}{
$\begin{split}
\rho(v) \; & =  \; \mathrm{Corr}\Big\{ Y_{1,i}(v),Y_{2,i}(v) \ \big| \ \xi(v), \tau^2_1(v), \tau^2_2(v) \Big\} \\
\; & = \; \frac{G^2_\omega\{\xi(v)\}-G^2_\omega\{-\xi(v)\}}{\left[G^2_\omega\{\xi(v)\}+G^2_\omega\{-\xi(v)\}+\tau^2_1(v)\right]^{1/2}\left[G^2_\omega\{\xi(v)\}+G^2_\omega\{-\xi(v)\}+\tau^2_2(v)\right]^{1/2}}.
\end{split}$}
\end{equation}
We say that $\rho(v)$ in \eqref{eq:spat_cor} follows a \emph{thresholded correlation Gaussian Process}, as formally defined below. 

\begin{definition}
Given any nonzero spatially varying variance functions $\tau^2_1(v)$ and $\tau^2_2(v)$, and the thresholding parameter $\omega \geq 0$, suppose $\xi(v)\sim \textrm{GP}(0,\kappa)$, then $\rho(v)$ in \eqref{eq:spat_cor} follows a thresholded correlation Gaussian process, denoted as $\rho \sim \textrm{TCGP}(\omega,\kappa,\tau^2_1,\tau^2_2)$.
\end{definition}
Under this construction, with probability one, a correlation Gaussian process is between -1 and 1, and enjoys both piecewise smoothness and sparsity. 

\change{We make a few remarks regarding our model setup. 

First, our proposed model encompasses a large class of spatially varying functions that are piecewise smooth, sparse, and jump discontinuous, the features that we commonly encounter in neuroimaging data \citep{zhu2014spatially}. Moreover, instead of specifying a voxel-wise prior, our TCGP incorporates the spatial information of the image, leading to potentially more accurate detection. We note that, our choice to model spatial correlation as a piecewise smoothness function is rooted in the observation that spatially contiguous voxels often exhibit similar correlation patterns \citep{li2018spatially}. Due to the anatomical and functional connections between different brain regions, neighboring regions often show similar patterns of structural or functional organizations. Consequently, the correlation between two modalities tends to vary smoothly within these contiguous regions, resulting in piecewise smoothness in the correlation image. In addition, the correlation image derived from brain images of two modalities may also display jump discontinuities, attributed to abrupt changes or transitions in the underlying relations between the two modalities. Jump discontinuities frequently manifest at boundaries between different brain regions or tissue types. For instance, abrupt shifts in correlation values may occur at boundaries between grey matter and white matter, or between cortical and subcortical structures. These boundaries signify distinct anatomical or functional transitions, leading to discontinuities in the correlation image.

Second, our prior model focuses on the correlation between the two image modalities, and does not directly specify the correlation between the two modalities at two different voxels, mainly because it is not our primary target of interest. Nevertheless, our prior specifications implicitly take into account the cross-voxel correlations through Gaussian process. 

Third, the introduction of $\xi(v)$ ensures that at least one of $\sigma^2_+(v)$ and $\sigma^2_-(v)$ equals 0. If $\sigma^2_+(v)=0$ and $\sigma^2_-(v)\neq 0$, then $\eta_+(v)=0$ and $\eta_-(v)\neq 0$, indicating a negative correlation between the two imaging modalities at this voxel. Conversely, if $\sigma^2_+(v)\neq 0$ and $\sigma^2_-(v)=0$, then a positive correlation exists. Furthermore, if $-\omega \leq \xi(v)\leq \omega$, $\sigma_{+}^2(v) = \sigma_{-}^2(v) = 0$. In this case, $\eta_+(v) = \eta_-(v) = 0$, which imposes sparsity and indicates that there is no significant correlation between these two imaging modalities at this specific location. On the other hand, if $\xi(v)$ were not introduced, the spatial correlation is given by:
\vspace{-3mm}
\begin{equation*}
\rho(v) = \frac{\sigma^2_+(v)-\sigma^2_-(v)}{\left[\sigma^2_+(v)+\sigma^2_-(v)+\tau_1^2(v)\right]^{1 / 2}\left[\sigma^2_+(v)+\sigma^2_-(v)+\tau_2^2(v)\right]^{1 / 2}}. 
\end{equation*}
In the instances where $\rho(v) = 0$, it is possible for $\sigma^2_+(v)=\sigma^2_-(v)$ to take very large values, suggesting that two sets of $\sigma^2_+(v)$ and $\sigma^2_-(v)$ could yield the same distribution of $Y_{1i}$ and $Y_{2i}$. This unidentifiability of $\sigma^2_+(v)$ and $\sigma^2_-(v)$ poses serious challenges for subsequent MCMC convergence. At some locations, $\sigma^2_+(v)$ and $\sigma^2_-(v)$ may escalate significantly while $\rho(v)$ remains zero. Therefore, it is crucial to introduce $\xi(v)$ to ensure the identifiability. 
}

\subsection{Equivalent model representation}
\label{sec:equiv-model}

To facilitate both theoretical investigation and posterior computation, we next derive an equivalent representations of model \eqref{eq:svcm} under the prior specifications \eqref{eq:sdgp} to \eqref{eq:varfunc}.  

We first note that, from \eqref{eq:varfunc} and \eqref{eq:spat_cor}, $\sigma_+(v)$ and $\sigma_-(v)$ can be uniquely determined by $\rho(v)$, in that, given $\tau_1^2(v)$ and $\tau_2^2(v)$, 
\vspace{-3mm}
\begin{align} \label{eq: rho_trans}
\begin{split}
& \sigma_+(v) = G_\omega\{\xi(v)\} = {s}\left\{ \rho(v);\tau_1^2(v),\tau_2^2(v) \right\}, \\
& \sigma_-(v)  = G_\omega\{-\xi(v)\} = s\left\{ -\rho(v);\tau_1^2(v),\tau_2^2(v) \right\}, 
\end{split}
\end{align}
where 
\vspace{-3mm}
\begin{align*} 
s(x; t_1,t_2)  = \left[{\frac{2 t_1 t_2}{{\left\{(t_1-t_2)^2 + 4 x^{-2} t_1 t_2\right\}}^{1/2}-(t_1+t_2)}}\right]^{1/2}I(x > 0), 
\end{align*}
for any $x \in [-1,1], t_1, t_2 >0$. We next consider a transformation of the observed images $\{Y_{1,i}(v), Y_{2,i}(v)\}$, the average $Y_{+,i}(v) = \{Y_{1,i} (v) + Y_{2,i}(v)\}/2$, and the contrast $Y_{-,i}(v) = \{Y_{1,i} (v) - Y_{2,i}(v)\}/2$. Denote 
\vspace{-3mm}
\begin{align} \label{eq: E-defi}
E_{+,i}(v) = \frac{\eta_{+,i}(v)}{\sigma_{+}(v)}, \qquad E_{-,i}(v) = \frac{\eta_{-,i}(v)}{\sigma_{-}(v)}.
\end{align} 
By \eqref{eq: rho_trans}, model \eqref{eq:svcm} is equivalent to 
\vspace{-3mm}
\begin{align}\label{eq:plus_minus}
\begin{split}
Y_{+,i}(v)  & = s\left\{ \rho(v);\tau^2_1(v),\tau^2_2(v) \right\}  E_{+,i}(v) + \varepsilon_{+,i}(v), \\
Y_{-,i}(v)   & = s\left\{ -\rho(v);\tau^2_1(v),\tau^2_2(v) \right\}  E_{-,i}(v) + \varepsilon_{-,i}(v),
\end{split}
\end{align}
where $\varepsilon_{+,i}(v)$ and $\varepsilon_{-,i}(v)$ are random noises that are independent over $i, v$, and follow a normal distribution with mean zero and variance $\{\tau_1^2(v)+\tau^2_2(v)\}/4$. The covariance between $\varepsilon_{+,i}(v)$ and $\varepsilon_{-,i}(v)$ is $\{\tau_1^2(v)-\tau^2_2(v)\}/\{\tau_1^2(v)+\tau^2_2(v)\}$.

\begin{figure}[t!]
\centering
\includegraphics[width=0.99\linewidth,height=2.7in]{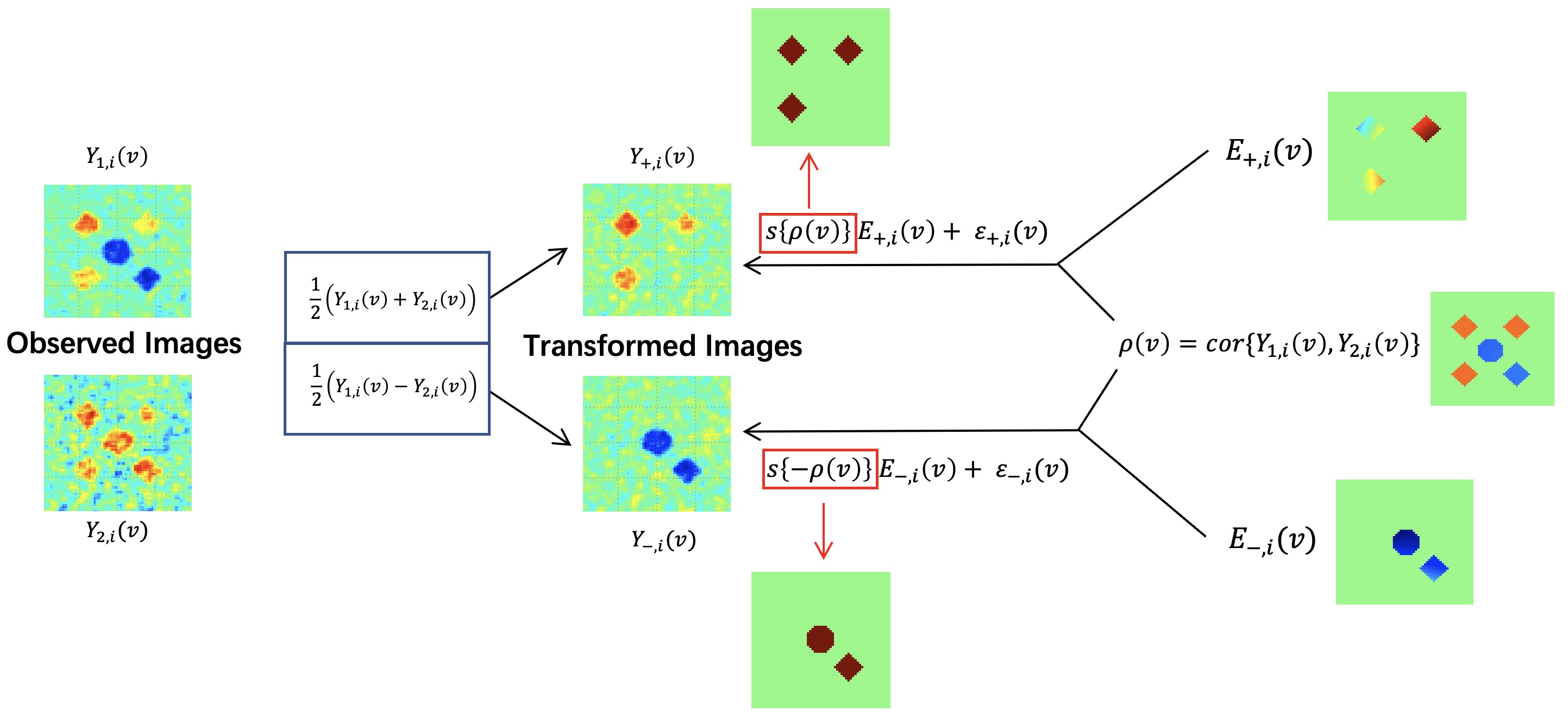}
\caption{Graphical illustration of the proposed Bayesian spatially varying correlation model. The transformed image $Y_{\pm, i}(v)$ are modeled based on \eqref{eq:plus_minus}.}
\label{fig:model}
\end{figure}

Following the prior specifications \eqref{eq:sdgp} to \eqref{eq:varfunc}, we have the equivalent prior specifications for $E_{+,i}(v)$, $E_{-,i}(v)$, and $\rho(v)$ as,  
\vspace{-3mm}
\begin{equation}\label{eq: prior_E}
\resizebox{0.88\hsize}{!}{
$E_{+,i} \sim \mathrm{GP}(0,\kappa),\quad E_{-,i}\sim \mathrm{GP}(0,\kappa), \quad \rho \mid \tau_1^2,\tau_2^2 \sim  \mathrm{TCGP}(\omega, \kappa, \tau^2_1,\tau^2_2),$}
\end{equation}
where $\kappa(\cdot, \cdot)$ is the correlation kernel as specified in \eqref{eq:covkern} and \eqref{eq:varfunc}, and in our modeling process, we use the Mat{\'e}rn kernel as specified in \eqref{eq: matern} for $\kappa(\cdot, \cdot)$.  Figure~\ref{fig:model} gives a graphical illustration of our nonparametric Bayesian spatially varying correlation model. In our subsequent theoretical and numerical analysis, we focus on the equivalent transformed model.

\change{Before the proposed correlation analysis, we normalize $\{Y_{k,i}(v)\}_{i=1}^n$ across all the subject for each voxel, thus these two imaging modalities are in the same scale and variation. This ensures the validity of our analysis. In addition, we spatially align the two imaging modalities. The general idea is to map the two images into some common representation space. There are multiple ways to achieve such an alignment, including image registration \citep{chen2021deep,gholipour2007brain}, feature-based alignment \citep{toews2013feature, calhoun2008feature,zhu2014fusing}, and machine learning-based alignment \citep{wang2021role, liu2014multimodal}. In Section \ref{sec:realdata}, as a specific example, we give more details on how we align resting-state and task fMRI images. }

\section{Theory}
\label{sec:theory}

In this section, we study the model identifiability, derive the large support property, and establish the posterior and selection consistency.

\subsection{Notations and definitions}
\label{sec:def}

We begin with some notations and definitions. For any vector $v=(v_{1}, \ldots, v_{d})\trans \in \mathbb{R}^{d}$, let $\|v\|_{p}=\left(\sum_{l=1}^{d}\left|v_{l}\right|^{p}\right)^{1 / p}$ denote the $ L_{p}$-norm, $p \geq 1$, and $\|v\|_{\infty}=\max _{l=1}^{d}\left|v_{l}\right|$ the supremum norm. For any real function $f$ on the region $\Bcal$, let $\|f\|_{p}=\left\{\int_{\mathcal{B}}|f(v)|^{p} \mathrm{~d} v\right\}^{1 / p}$ denote the $L_{p}$-norm, $p \geq 1$, and $\|f\|_{\infty}=\sup _{v \in \mathcal{B}}|f(v)|$ be the supremum norm. Suppose $\Bcal$ is a compact convex set. Recall there are $n$ subjects, and $m$ spatial locations for each image. Denote $Y_\pm = \{Y_{\pm,1}\trans, \ldots, Y_{\pm,n}\trans\}\trans$, where $Y_{\pm,i} = \{Y_{\pm,i}(v_1), \ldots, Y_{\pm,i}(v_m)\}\trans$. Furthermore, denote our parameter of interest as $\theta(\cdot)=\left\{\rho(\cdot), E_{+}\trans(\cdot), E_{-}\trans(\cdot)\right\}\trans$, where $E_{\pm}(\cdot)=\left\{E_{\pm,1}(\cdot), \ldots, E_{\pm,n}(\cdot)\right\}\trans$, and the true parameter $\theta_0(\cdot) = \left\{\rho_0(\cdot), E_{+,0}\trans(\cdot), E_{-,0}\trans(\cdot)\right\}\trans$. 

\begin{definition}
\label{def:diff_function}
Define $\mathcal{C}^{q}(\mathcal{B})$ as a set of differentiable functions {of order $q$} defined on $\mathcal{B}$, such that a function $f\in \mathcal{C}^{q}(\mathcal{B})$ has the partial derivative, 
\begin{align*}
D^{b} f(v)=\frac{\partial^{\|b\|_{1}} f}{v_{1}^{b_{1}} \ldots v_{d}^{b_{d}}}(v)=\sum_{\|a\|_{1}+\|b\|_{1} \leq q} \frac{D^{b+a} f(u)}{a !}(v-u)^{a}+R_{q}(v, u),
\end{align*}
where $b=\left(b_{1}, \ldots, b_{d}\right)\trans \in \mathbb{Z}_{+}^{d}, a \in \mathbb{Z}_{+}^{d}$, $\mathbb{Z}_{+}$ denotes the set of non-negative integers, $u \in \mathbb{R}^{d}$, and the remainder $R_q(v,u)$ satisfies the following properties: (i) Given any point $v_0$ of $\mathcal{B}$ and any constant $\epsilon >0$, there is a constant $\delta>0$, such that if {$v$ and $u$ are any two points of $\Bcal$} with $\|v-v_0\|_1<\delta$ and $\|u-v_0\|_1<\delta$, then $|R_q(v,u)|\leq \|v-u\|_1^{q-\|b\|_1}\epsilon$; (ii) If $\|D^{b}f\|_\infty \leq {C} < \infty$ {for some constant $C$}, then $|R_q(v,u)|\leq ({C} \|v-u\|_1^{q+1})/(q+1)!$. 
\end{definition}
 
\begin{definition}\label{def:rho}
Define $\Theta_\rho = \big\{ \rho(v) \in (-1,1): v\in\mathcal{B} \big\}$ as a collection of spatially varying correlation functions that satisfy the following properties: (i) There exist two disjoint non-empty open sets $\mathcal{R}_{-1}$ and $\mathcal{R}_{1}$ with $\overline{\mathcal{R}}_{1} \cap \overline{\mathcal{R}}_{-1}=\emptyset$, such that $\rho(v)$ is smooth over $\overline{\mathcal{R}}_{-1} \cup \overline{\mathcal{R}}_{1}$, i.e., $\rho(v) I\left(v \in \overline{\mathcal{R}}_{-1} \cup \overline{\mathcal{R}}_{1}\right) \in \mathcal{C}^{\alpha}\left(\overline{\mathcal{R}}_{-1} \cup \overline{\mathcal{R}}_{1}\right)$, with $\alpha=\lceil d / 2\rceil + 1$, the least integer greater than or equal to $d/2$; (ii) $\rho(v)=0$ for $v\in \mathcal{R}_0  $, $\rho(v)>0$ for $v\in \mathcal{R}_1$, and $\rho(v)<0$ for $v\in \mathcal{R}_{-1}$, where $\mathcal{R}_{0}=\mathcal{B}-\left(\mathcal{R}_{-1} \cup \mathcal{R}_{1}\right) \text { and } \mathcal{R}_{0}-\left(\partial \mathcal{R}_{1} \cup \partial \mathcal{R}_{-1}\right) \neq \emptyset$; (iii) 
$\rho(v)$ is a discontinuous function and is bounded away from zero for any $v\notin \mathcal{R}_{0}$, i.e., 
${\gamma = } \inf _{v \notin  \mathcal{R}_{0} }|\rho(v)|>0$.
\end{definition}

\begin{definition}
\label{def:E(v)}
Define $\Theta_E = \{E(v) \in \mathbb{R}^n: \|E(v)\|^2_2 = C_v\}$ for some constant $C_v < \infty$.
\end{definition}

In summary, $\Theta_\rho$ is the collection of all piecewise smooth, sparse, and jump discontinuous correlation functions $\rho(v)$ defined on $\mathcal{B}$, where $\gamma$ in Definition \ref{def:rho} represents the minimum nonzero effect size of the correlation functions that have discontinuity jumps, and $\Theta_E$ is the collection of the spatially varying functions $E(v)$ that satisfy the second moment constraints.

\subsection{Model identifiability and large support}
\label{subs: identifiability}

We first show that model \eqref{eq:plus_minus} is identifiable, then show that the prior specification in \eqref{eq: prior_E} has a large support. We begin with a regularity condition. 

\begin{assumption}
\label{cond: rho}
The true correlation function $\rho_0$ is piecewise smooth, sparse, and jump discontinuous, in that $\rho_0 \in \Theta_\rho$. In addition, the true functions $E_{+,0}$ and $E_{-,0}$ have constant second moments with respect to the location $v$, i.e., $E_{+,0}\in \Theta_{E}$ and $E_{-,0}\in \Theta_{E}$.
\end{assumption} 

\noindent 
Assumption \ref{cond: rho} essentially specifies the class of true functions that we target. Denote $\mathcal{V}(\rho) = \{v: \rho(v) \neq 0\}$, and $\mathcal{V}(\rho') = \{v: \rho'(v) \neq 0\}$. 
The next proposition shows that model \eqref{eq:plus_minus} is identifiable. Specifically, $\rho(v)$ is identifiable for all $v\in \mathcal{B}$, and $E_+(v), E_-(v)$ are identifiable for $v \in \mathcal{V}(\rho) \cup \mathcal{V}(\rho')$. The identifiability of $E_+(v)$ and $E_-(v)$ is constrained on $\mathcal{V}(\rho) \cup \mathcal{V}(\rho')$ because when $\rho(v) = 0$, $s\{\rho(v)\}=0$ in model \eqref{eq:plus_minus}.
 
\begin{proposition}
\label{prop: 2}
(Identifiability) 
Suppose Assumption \ref{cond: rho} holds. Then model \eqref{eq:plus_minus} is identifiable. That is, if the probability distributions of $\{Y_{+}, Y_{-}\}$ under $\theta = \left\{ \rho, E_+\trans, E_-\trans \right\}\trans$ and $\theta' = \left\{ \rho', {E'_+}\trans, {E'_-}\trans \right\}\trans$ are equal, then we have $\rho = \rho'$ for $v\in \mathcal{B}_m$, and $\theta = \theta'$ for $v\in \mathcal{V}(\rho) \cup \mathcal{V}(\rho') $.
\end{proposition}

To ensure the large-support property, we introduce another condition on the correlation kernel function $\kappa(\cdot, \cdot)$. The same condition was imposed in  \cite{ghosal2006posterior} as well.
\begin{assumption}
\label{cond: kernel}
The correlation kernel $\kappa(\cdot, \cdot)$ satisfies that, for any $v \in \mathcal{B}$, $\kappa(v,\cdot)$ has continuous partial derivatives up to order $2\alpha+2$, where $\alpha=\lceil d / 2\rceil + 1$.  In addition, suppose $\kappa(v, v')=\prod_{l=1}^{d} \kappa_{l} (v_{l}-v'_{l} ; \nu_{l})$, for any $v=(v_{1}, \ldots, v_{d})\trans$, and $v'=(v'_{1}, \ldots, v'_{d})\trans \in[0,1]^{d}$, where $\kappa_{l}(\cdot; \nu_{l})$ is a continuous, nowhere zero, symmetric density function on $\mathbb{R}$ with parameter $\nu_{l} \in \mathbb{R}^{+}$, for $l=1, \ldots, d$. 
\end{assumption}
 
The next theorem shows that our prior specification in \eqref{eq: prior_E} is desirable, in that it has support over a large class of sparse, piecewise smooth and jump discontinuous spatially varying correlation functions. That is, there is a positive probability that $\theta = \left\{ \rho, E_+\trans, E_-\trans \right\}\trans$ concentrates on an arbitrarily small neighborhood of any true parameter in the parameter space $\Theta = \Theta_\rho \times \Theta_E \times \Theta_E$.

\begin{theorem}\label{thm:largsup}
(Large Support)  
Suppose Assumptions \ref{cond: rho} and \ref{cond: kernel} hold. Under the prior specification in~\eqref{eq: prior_E}, for any $\epsilon>0$, $pr\big( \|\theta - \theta_0\|_{\infty} < \epsilon \big) > 0$, where $pr(\cdot)$ denotes a probability measure on the Borel set of $\Theta$. 
\end{theorem}

\subsection{Posterior consistency}
\label{subs: consistency}

Next, we establish the posterior consistency, then the selection consistency. 

\begin{assumption}
\label{cond: mn}
There exist constants $d /(2 \alpha)<\nu_{0}<1$, $C_{0} > 0, C_{1} > 0$, and $N \geq 1$, with $\alpha=\lceil d / 2\rceil + 1$, such that $C_{0} n^{d} \leq m \leq C_{1} n^{2 \alpha \nu_0}$ for all $n>N$.
\end{assumption}

\noindent
Assumption \ref{cond: mn} imposes that the number of spatial locations $m$ should be of the polynomial order of the sample size $n$. The lower bound indicates that $m$ needs to be sufficiently large to ensure that the posterior distribution of the spatially varying coefficient function concentrates around the true value. The upper bound ensures that a  sufficient amount of information is collected across subjects to identify the population level true parameters.

The next theorem shows that, under the proposed prior, the posterior distribution of $\theta$ concentrates in an arbitrarily small neighborhood of the true parameter $\theta_0$, when the number of subjects $n$ and the number of spatial locations $m$ are sufficiently large. 
 
\begin{theorem}\label{thm:consistency}
(Posterior Consistency) 
Suppose Assumptions \ref{cond: rho}, \ref{cond: kernel} and \ref{cond: mn} hold. Under model \eqref{eq:plus_minus} and the prior specification in \eqref{eq: prior_E}, for any $\epsilon > 0$, as $m \to \infty$ and $n \to \infty$, 
\vspace{-3mm}
\begin{align*}
pr\Big( \{\theta \in \Theta: \| \theta - \theta_0\|_1 < \epsilon\} \, \big | \, Y_+, Y_- \Big) \to 1 \; \textrm{ in } \; \mathbb{P}^{(m,n)}_{\theta_0}\textrm{-probability}, 
\end{align*} 
where $\mathbb{P}^{(m,n)}_{\theta_0}$ denotes the distribution of $\{Y_+,Y_-\}$ given the true parameter $\theta_{0}$, and $pr(\cdot \mid Y_+,Y_-)$ denotes the posterior probability measure on the Borel set of $\Theta$ given data $\{Y_+,Y_-\}$.
\end{theorem}

The next theorem shows that, with probability tending to one, we can identify the true activation regions that have positive correlations, negative correlations, and no correlations, respectively, when both $n$ and $m$ tend to infinity. 

\begin{theorem}\label{thm:signconsistency}
Suppose the same conditions in Theorem \ref{thm:consistency} hold. Then, as $m\to\infty$ and $n\to \infty$,
\vspace{-3mm}
\begin{align*}
pr\Big( \mathrm{sgn}\{\rho(v)\} = \mathrm{sgn}\{\rho_0(v)\}, v \in \mathcal{B} \, \big | \, Y_+, Y_- \Big) \to 1 \; \textrm{ in } \;  \mathbb{P}^{(m,n)}_{\theta_0}\textrm{-probability},
\end{align*} 
where $\mathrm{sgn}(x) = 1$ if $x>0$, $\mathrm{sgn}(x) = -1$ if $x<0$ and $\mathrm{sgn}(0) = 0$.
\end{theorem}

\section{Posterior Computation}
\label{sec:estimation}

In this section, we first adopt the Karhunen-Lo{\`e}ve expansion to simplify the model to a finite number of parameters. We next derive the full conditional distributions of the model parameters, and develop an efficient Gibbs sampling algorithm. We also propose a hybrid mini-batch MCMC to further improve the computational efficiency.

\subsection{Karhunen-Lo{\`e}ve approximation}
\label{subsec:model_repre}

Model \eqref{eq:plus_minus} involves three Gaussian processes, for $E_{+,i}(v), E_{-,i}(v)$, and $\xi(v)$, respectively, and all hinge on the infinite dimensional correlation kernel function $\kappa(\cdot, \cdot)$. We first adopt the usual strategy of Karhunen-Lo{\`e}ve expansion to simplify the model to a finite number of parameters. Specifically, consider the spectral decomposition of the kernel function, $\kappa\left(v, v^{\prime}\right)=\sum_{l=1}^{\infty} \lambda_l \psi_l(v) \psi_l\left(v^{\prime}\right)$, where $\{\lambda_{l}\}_{l=1}^{\infty}$ are the eigenvalues in descending order, and $\{\psi_{l}(v)\}_{l=1}^{\infty}$ are the corresponding orthonormal eigenfunctions. By Mercer’s Theorem \citep{mercer1909xvi}, we can represent the Gaussian processes in our model by the Karhunen-Lo{\`e}ve (KL) expansion, $E_{+,i}(v) = \sum_{l=1}^\infty e_{i,l,+}  \psi_l(v)$,  $ E_{-,i}(v) = \sum_{l=1}^\infty e_{i,l,-}  \psi_l(v)$ and  $\xi(v) = \sum_{l=1}^\infty c_l \psi_l(v)$, where $c_l, e_{i,l,\pm}$ are Karhunen-Lo{\`e}ve coefficients. We further truncate the above expansions by focusing on the leading $L$ eigenvalues and eigenfunctions, where $L$ can be determined following the usual practice of principal components analysis that retains a certain percentage of total variation. \change{That is, we compute the variance percentage as $R=\sum_{l=1}^L \lambda_l / \sum_{l=1}^{\infty} \lambda_l$, where $\lambda_l$ is the $l$th largest eigenvalue of the covariance kernel, and we approximate the denominator $\sum_{l=1}^{\infty} \lambda_l$ by a summation of truncated series $\sum_{l=1}^{L^{\prime}} \lambda_l$. Following \citet{wu2024bayesian}, we choose the smallest $L$ such that the variance percentage $R$ exceeds 60\%, while we set $L^{\prime} = 900$. In our numerical analyses, we have found such a choice achieves a good balance between computational cost and model fitting performance.}

Based on the Karhunen-Lo{\`e}ve truncation, model \eqref{eq:svcm} becomes, 
\begin{align} \label{eq:working_model}
\begin{split}
Y_{+,i}(v) & = G_\omega\left\{ \sum_{l=1}^L c_l \psi_l(v) \right\} \left\{ \sum_{l=1}^L e_{i,l,+}  \psi_l(v) \right\} + \varepsilon_{+,i}(v), \\
Y_{-,i}(v) & = G_\omega\left\{ - \sum_{l=1}^L c_l \psi_l(v)\right\} \left\{ \sum_{l=1}^L e_{i,l,-}  \psi_l(v) \right\} + \varepsilon_{-,i}(v). 
\end{split}
\end{align}
Recall that $\Bcal_m = \{v_1, \ldots, v_m\}$ denotes the set of locations where the imaging data are observed, and let $Y = \{ Y_{1,i}(v), Y_{2,i}(v), i = 1,\ldots, n, v \in \Bcal_m \}$ denote the imaging data observed at the set of voxels in $\Bcal_m$. Then all the parameters in our model include: 
\begin{align} \label{eqn:parameters}
\Tilde{\Theta} = \Big\{ \{c_l\}_{l=1}^{L}, \; \big\{ \{e_{i,l,+}\}_{l=1}^{L}, \{e_{i,l,-}\}_{l=1}^{L} \big\}_{i=1}^{n}, \; \{ \tau_1^2(v), \tau_2^2(v) \}_{v \in \Bcal_m}, \;\omega \Big\}.
\end{align}
We specify their prior distributions as, 
\vspace{-3mm}
\begin{equation} \label{eq: prior}
\resizebox{0.88\hsize}{!}{
$c_l \sim \mathrm{N}(0, \lambda_l), \;\; e_{i,l,\pm} \sim \mathrm{N}(0, \lambda_l), \;\;  
\tau^2_1(v), \tau^2_2(v) \sim  \mathrm{IG}(a_\tau, b_\tau), \;\; \omega \sim \mathrm{U}(a_\omega, b_\omega),$}
\end{equation}
That is, we impose a normal distribution for the Karhunen-Lo{\`e}ve coefficients $c_l, e_{i,l,\pm}$, where $\lambda_l$ is the eigenvalue of the kernel $\kappa(v,v')$ as specified above. We impose an inverse Gamma prior for the variance terms $\tau^2_1(v)$, $\tau^2_2(v)$, with shape $a_\tau$ and scale $b_\tau$, and we choose some small values for $a_\tau, b_\tau$, so that this prior is non-informative. We also impose a uniform prior for the thresholding parameter $\omega$, with range from $a_\omega$ to $b_\omega$, and we choose $a_\omega, b_\omega$ based on the quantiles of $|\xi(v)|_{v\in \mathcal{B}_m}$. It is also possible to consider other types of prior for $\omega$, e.g., an exponential distribution. Note that the conditional prior for $\omega$ allows it to be adaptively learnt in a fully Bayesian way in our Gibbs sampling. This is different from the gradient based MCMC methods, which require a smooth approximation of the thresholding function.

\subsection{Gibbs sampling}

We first present a general result that is useful for deriving the full conditional distributions of some of our key parameters. We note that this result is both new and general, and can be applied to deriving the Gibbs sampler for other types of models involving thresholded Gaussian processes. 

\begin{proposition}
\label{prop1}
Consider a random variable $\theta$, and two sets of functions $f_p(\theta) = a_{1p}\theta^2 + a_{2p}\theta + a_{3p}$, and $h_k(\theta) = b_{1k}\theta^2 + b_{2k}\theta + b_{3k}$, where $a_{lp},  b_{lk}$ are some coefficients, $l=1,2,3$, $p = 1, \ldots, P, k = 1, \ldots, K$. Suppose the density of $\theta$ is proportional to
\vspace{-3mm}
\begin{equation}\label{eq: prop1}
\exp\left\{{\sum_{p=1}^P f_p(\theta) I(\theta > L_p) + \sum_{k=1}^K h_k(\theta) I(\theta < U_k)}\right\},
\end{equation}
where $U_k, L_p$ are some thresholding coefficients. Then,  
\begin{enumerate}[(i)]
\item If at least one of $\{a_{1p}, \ldots, a_{1P}, b_{1k}, \cdots, b_{1K}\}$ is not equal to 0, then $\theta$ follows a mixture of truncated normal distributions.
\item  If $a_{1p} = b_{1k} = a_{2p} = b_{2k} = 0$ for all $p,k$, and at least one of $\{a_{3p}, \ldots, a_{3P},$ $ b_{3k}, \ldots, b_{3K}\}$ is not equal to 0, then $\theta$ follows a mixture of uniform distributions. 
\item If $a_{1p} = b_{1k} = 0$ for all $p,k$, and at least one of $\{a_{2p}, \cdots, a_{2P}, b_{2k}, \cdots, b_{2K}\}$ is not equal to 0, then $\theta$ follows a mixture of exponential distributions.
\end{enumerate}
\end{proposition}

We next derive the full conditional distributions of our model parameter $\tilde\Theta$ in \eqref{eqn:parameters}. Specifically, we first derive the full conditionals of $\{c_l\}_{l=1}^L$ and $\omega$, both of which are based on Proposition \ref{prop1}. We then derive the full conditionals of $\{e_{i,l,\pm}\}_{l=1,i=1}^{L,n}$ and $\{ \tau_1^2(v), \tau_2^2(v) \}_{v \in \Bcal_m}$, both of which have closed forms thanks to their conjugate priors. Let $\tilde\Theta_{\backslash \theta}$ be the set of parameters in $\tilde\Theta$ but without $\theta$. 

The full conditional of $c_l$ is a mixture of truncated normal distributions, as we show in the Supplementary Material, Section \ref{appen: c}. This is because the density of $c_l$ is of the form, 
\begin{equation*}
\resizebox{0.99\hsize}{!}{
$\begin{split}
\pi(c_l \mid Y, \tilde{\Theta}_{\backslash c_l}) \propto & \exp\left(\sum_{\substack{j=1 \\ \psi_l(v_j)>0}}^m \big[ g_{+}(c_l; v_j)I\{c_l>T_{+}(v_j)\}+ g_{-}(c_l; v_j)I\{c_l<T_{-}(v_j)\} \big]\right.\\
& + \left.\sum_{\substack{j=1 \\ \psi_l(v_j)<0}}^m\big[ g_{+}(c_l; v_j)I\{c_l<T_{+}(v_j)\}+ g_{-}(c_l; v_j)I\{c_l>T_{-}(v_j)\} \big]\right).
\end{split}$}
\end{equation*}
Given the location $v_j$, $g_{\pm}(c_l; v_j)$ are two quadratic functions of $c_l$, and $T_{\pm}(v_j)$ are two scalars, whose detailed forms are given in the Supplementary Material, Section \ref{appen: c}. If we set $f_p(\theta) = g_{+}(c_l; v_j)$, $h_k(\theta) = g_{-}(c_l; v_j)$ for those locations $v_j$ satisfying $\psi_1(v_j)>0$, and set $f_p(\theta) = g_{-}(c_l; v_j)$, $h_k(\theta) = g_{+}(c_l; v_j)$ otherwise, then the density of $c_l$ satisfies the condition of Proposition \ref{prop1}(i), and thus it follows a mixture of truncated normal distributions. 

The full conditional  of $\omega$ is a mixture of uniform distributions, as we show in the Supplementary Material, Section \ref{appen: omega}. This is because the density of $\omega$ is of the form, 
\begin{equation*}
\resizebox{0.99\hsize}{!}{
$\omega \mid Y, \tilde{\Theta}_{\backslash \omega}\sim \exp\left[\sum_{\substack{j=1 \\ a_\omega<\xi(v_j)<b_\omega}}^m C_{+}(v_j)I\{\omega < \xi(v_j)\} + \sum_{\substack{j=1 \\ a_\omega<-\xi(v_j)<b_\omega}}^m C_{-}(v_j)I\{\omega < -\xi(v_j)\}\right]$}.
\end{equation*}
Given the location $v_j$, $\xi(v_j) = \sum_{l = 1}^{L} c_l \psi_l(v_j)$, $C_{\pm}(v_j)$ are two scalars, whose detailed forms are given in the Supplementary Material, Section \ref{appen: omega}.  
If we set $h_k(\theta) = C_{+}(v_j)$ for those $v_j$ satisfying $a_\omega<\xi(v_j)<b_\omega$, and set $h_k(\theta) = C_{-}(v_j)$ for those locations satisfying $a_\omega<-\xi(v_j)<b_\omega$, then the density of $\omega$ satisfies the condition of Proposition \ref{prop1}(ii), and thus it follows a mixture of uniform distributions. We make two additional remarks. First, we specify the prior of $\omega$ as U$(a_\omega, b_\omega)$, where we choose $a_\omega, b_\omega$ to have a non-informative prior. In practice, we may adopt the empirical Bayes idea, by running the Gibbs sampling once with a non-informative prior first, then using the quantile values of the sorted $\{|\xi(v)|\}_{v \in \Bcal_m}$ to refine the range of the uniform distribution for $\omega$. \change{The refinement is fully included in the posterior sampling procedure, where $a_\omega$ and $b_\omega$ are adaptively changed based on the value of $\{|\xi(v)|\}_{v\in\mathcal{B}}$ in each iteration. This can further improve the convergence behavior of the algorithm.}  Second, if we specify the prior of $\omega$ as an exponential distribution, then we may apply Proposition \ref{prop1}(iii) to obtain the full conditional of $\omega$. 

We present the derivations of the full conditionals of $e_{i,l,\pm}$ and $\tau_k^2(v)$ and summarize the Gibbs sampling for the TCGP in Algorithm \ref{alg:proof of pro} in the Supplementary Material, Section \ref{appen: full}.

\subsection{Hybrid mini-batch MCMC}
\label{subs:mini}

The proposed Gibbs sampler is computationally efficient in general. Meanwhile, the complexity of computing the full conditional  of $c_l$ is $O(m^2)$, where $m$ is the total number of voxels. When $m$ is large, this step can be expensive. We next propose a hybrid mini-batch MCMC, with two key components, to further improve the computational efficiency. 

The first component is to develop an adaptive proposal function in Gibbs sampling. We note that the Gibbs sampler can be viewed as a special case of Metropolis–Hastings, in which the newly proposed state is always accepted with probability one, and the proposal function in Metropolis–Hastings corresponds to the full conditional distribution in the Gibbs sampler. There have been some recent progress developing scalable MCMC methods \citep{li2017mini, wu2020mini}. {However, those algorithms mainly focus on how to more efficiently evaluate the ratio of the likelihood function at each iteration, instead of focusing on the proposal function. Moreover, their aims are not to perform Bayesian inference from the exact posterior, but rather to exploit the tempered posterior with an efficient MCMC sampler to obtain a better solution from the global optimization. } 

We propose an adaptive proposal function, by subsampling voxel locations, instead of individual subjects. More specifically, let $\Bcal_{m_s} \subset \Bcal_m$ denote a random subset of all the observed locations $\Bcal_m$, $Y_{m_s}$ the corresponding imaging data observed at those voxels in $\Bcal_{m_s}$, and $m_s = |\Bcal_{m_s}|$ the cardinality of $\Bcal_{m_s}$. Recognizing that the Gibbs sampler is a special case of Metropolis–Hastings, the proposal function for the parameter $\theta \in \Tilde{\Theta} $ is $P\{\theta | \Tilde{\Theta}_{\backslash \theta}, Y\}$, which is the full conditional distribution of $\theta$. Instead of using the entire imaging data $Y$ to derive the full conditional distribution of $\theta$, we propose to use a mini-batch of data $Y_{m_s}$ to obtain the proposal function $P\{\theta | \Tilde{\Theta} _{\backslash \theta}, Y_{m_s}\}$. The acceptance ratio of $\theta$ is, 
\begin{equation*} \label{subsample}
\phi(\theta^\prime, \theta) = \min\left[ 1, \dfrac{\Pi_{v \notin \Bcal_{m_s}}P\{ Y(v) | \theta^\prime, \Tilde{\Theta}_{\backslash \theta} \} }{\Pi_{v \notin \Bcal_{m_s}} P\{ Y(v) | \theta, \Tilde{\Theta}_{\backslash \theta} \} }\right],
\end{equation*}
whose derivation is given in the Supplementary Material, Section \ref{appen: mini}. In this case, the computational complexity of sampling $c_l$ is reduced from $O(m^2)$ to $O(m_s^2)$.

The second component is to consider a hybrid version of mini-batch. This is because, when keeping using the mini-batch of voxels during the whole sampling process, the Markov chain may converge to local modes, and may also converge slowly. To overcome these issues, we propose to use the full dataset after, say, every $T_0$ iterations of using the mini-batch data. 

We summarize the hybrid mini-batch MCMC procedure in Algorithm \ref{alg: hybridsampling_c} in the Supplementary Material, Section \ref{appen: full}. In our implementation, we set $m_s = m/16$ and $T_0 = 20$, which leads to a good empirical performance. We also carry out a sensitivity analysis in the Supplementary Material, Section \ref{appen: sensitivity-analysis}, and find that the result is not sensitive to $m_s$ and $T_0$, as long as they are in a reasonable range.

\section{Simulations}
\label{sec:simulations}

We consider a $d=3$ example that mimics the Human Connectome Project data analyzed in Section \ref{sec:realdata}. More specifically, we obtain the posterior means of $c_l, e_{i,l,\pm}, \omega$ and $\tau^2_k(v), k = 1,2$ from our Human Connectome Project data analysis, then generate $Y_{+}, Y_{-}$ following model \eqref{eq:working_model}. We simulate the noise $\varepsilon_{k,i}$ from the normal distribution with mean zero and variance $\zeta_k\tau^2_k(v)$, $k= 1,2$.

To apply the proposed method, we employ the Mat{\'e}rn kernel in \eqref{eq: matern} in our data analysis. We set the prior hyperparameters $a_\tau = b_\tau = 0.001$ to obtain a non-informative prior, and choose $a_\omega$ and $b_\omega$ adaptively as the minimum and the maximum of ${|\xi(v)|}_{v\in\mathcal{B}_m}$, respectively, from each iteration.  We run the Gibbs sampler for 1000 iterations, with the first 200 iterations as the burn-in. We also run the hybrid mini-batch MCMC for 1200 iterations, with the first 400 iterations as the burn-in. We claim a voxel having a nonzero correlation by simply thresholding the posterior inclusion probability at 0.5, an approach commonly used in Bayesian analysis. We also compare with a number of alternative solutions, including the voxel-wise analysis, the region-wise analysis, and the integrated analysis method of \citet{li2018spatially} with two different thresholding values, 0.90 and 0.95, following the analysis in \citet{li2018spatially}. We evaluate the performance of each method by the sensitivity, specificity, and false discovery rate. 
   
We first set $\zeta_k = 1$ which represents the same noise level as the Human Connectome Project data and vary the sample size. We consider two cases where $n = 500$ and $n = 900$. The image resolution is $91 \times 109 \times 91$ with $m = 117,293$ voxels in the brain region. Table \ref{tab:sim3d_size} reports the results averaged over 100 data replications, and Figure ~\ref{tab:sim3d_size} visualizes the result for one data replication when $n = 900$. The red, yellow and blue regions represent the true positive, the false negative, and the false positive regions, respectively. We see that our proposed method clearly outperforms the alternative solutions. In particular, the voxel-wise analysis suffers from a low detection power, the region-wise analysis yields a high false discovery rate, and the integrated method of \citet{li2018spatially} is sensitive to the thresholding parameter. With the 90\% threshold, the integrated method enjoys a better sensitivity and specificity, but yields a larger false discovery rate, whereas with the 95\% threshold, it can well control the false discovery rate, is not as powerful. 
We also observe that the proposed method TCGP performs the best across different values of $n$. Meanwhile, it maintains a competitive performance even when $n$ is relatively small. Besides, in Figure ~\ref{tab:sim3d_size}, we only show the result for the positively correlated regions, as the result for the negative correlated regions is very similar. Finally, we briefly remark on the computational time of the two Gibbs samplers. On a laptop with 2 cores, 3.1GHz clock speed and 8GB memory, the Gibbs sampler algorithm took about 150 minutes for one data replication, while the hybrid algorithm took about 50 minutes, with the mean acceptance ratio around 0.3.  

We then consider two noise levels, or equivalently the signal strengths, with $\zeta_k = 5$ for a weak signal, and $\zeta_k = 0.5$ for a strong signal when the sample size is set as $n=904$ following the Human Connectome Project data. We also conduct 2D simulations to further illustrate the superiority of the proposed method. All the results are  reported in the Supplementary Material, Section \ref{subs: 2dsim}. 

\begin{table}[t!]
\captionsetup{font=small}
\caption{Simulation results of the 3D image example with the varying sample size $n$. Reported are the average sensitivity, specificity, and FDR, with standard error in the parenthesis, based on 100 data replications. Six methods are compared: the voxel-wise analysis, the region-wise analysis, the integrated method of \citet{li2018spatially} with two thresholding values, 0.95 and 0.90, and the proposed Bayesian method with the Gibbs sampler and the hybrid mini-batch MCMC. }
\label{tab:sim3d_size}
\vspace{-0.15in}
\begin{center}
\resizebox{\textwidth}{!}{
\begin{tabular}{clcccccc}
\toprule 
\multirow{2}{*}{Sample size}&\multirow{2}{*}{Method}&\multicolumn{3}{c}{Positive Correlation}&\multicolumn{3}{c}{Negative Correlation}\\
 & & Sensitivity & Specificity & FDR & Sensitivity & Specificity & FDR \\
\midrule
  n=500 &  Voxel-wise        &  0.023 (0.001) & 1.000 (0.000) &   0.004 (0.001)&  0.041 (0.005) &  1.000 (0.000) &   0.005 (0.002)\\
        &   Region-wise      &  0.307 (0.003) & 0.944 (0.002) &   0.330 (0.010)&  0.485 (0.003) &  0.966 (0.003) &   0.398 (0.010)\\ 
        &   Integrated(0.95) &  0.623 (0.005) & 0.990 (0.005) &   0.145 (0.005)&  0.877 (0.004) &  0.988 (0.004) &   0.190 (0.009)\\
        &   Integrated(0.90) &  0.897 (0.010) & 0.950 (0.005) &   0.305 (0.008)&  0.905 (0.004) &  0.948 (0.005) &   0.320 (0.010)\\
        &   TCGP (Gibbs)     &  0.903 (0.005) & 0.991 (0.001) &   0.074 (0.007)&  0.911 (0.002) &  0.975 (0.003) &   0.075 (0.003)\\
         &  TCGP (Hybrid)    &  0.901 (0.002) & 0.990 (0.002) &   0.076 (0.006)&  0.910 (0.004) &  0.973 (0.002) &   0.081 (0.002)\\
\midrule
  n=900 &  Voxel-wise        &  0.120 (0.005) & 1.000 (0.001) &   0.002 (0.001)&  0.182 (0.001) &  1.000 (0.000) &   0.002 (0.001)\\
        &  Region-wise       &  0.587 (0.003) & 0.901 (0.002) &   0.531 (0.010)&  0.627 (0.006) &  0.824 (0.005) &   0.532 (0.003)\\
        &  Integrated(0.95)  &  0.731 (0.005) & 0.995 (0.001) &   0.020 (0.004)&  0.892 (0.005) &  0.970 (0.000) &   0.092 (0.002)\\
        &  Integrated(0.90)  &  0.938 (0.010) & 0.974 (0.003) &   0.300 (0.005)&  0.917 (0.008) &  0.966 (0.002) &   0.311 (0.003)\\
        &  TCGP (Gibbs)      &  0.931 (0.002) & 0.994 (0.001) &   0.047 (0.003)&  0.925 (0.001) &  0.987 (0.001) &   0.058 (0.002)\\
        &  TCGP (Hybrid)     &  0.931 (0.003) & 0.994 (0.001) &   0.049 (0.003)&  0.921 (0.003) &  0.977 (0.001) &   0.081 (0.002)\\
\bottomrule 
\end{tabular}
}
\end{center}
\end{table}

\begin{figure}[t!]
\captionsetup{font=small}
\centering
\small
\begin{tabular}{rc}
\raisebox{7mm}{Voxel-wise}&\includegraphics[width=0.8\textwidth]{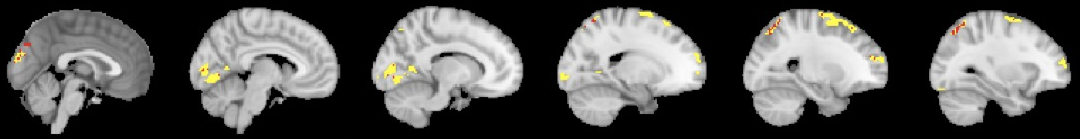}\\
\raisebox{7mm}{Region-wise}&\includegraphics[width=0.8\textwidth]{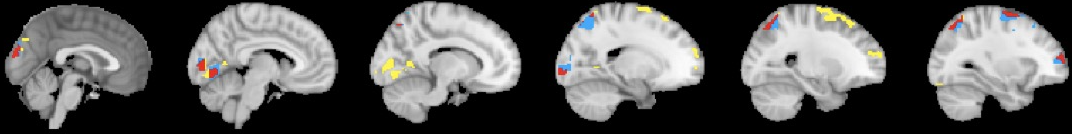}\\
\raisebox{7mm}{
\begin{tabular}{r}
Integrated\\ (0.95)
\end{tabular}}&\includegraphics[width=0.8\textwidth]{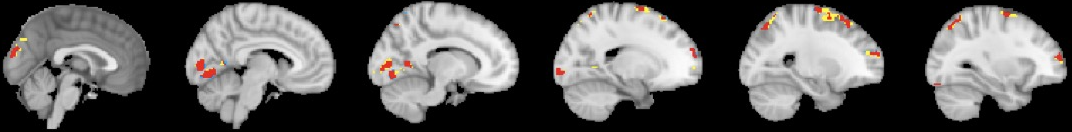}\\
\raisebox{7mm}{\begin{tabular}{r}
Integrated \\(0.90)
\end{tabular}
}&\includegraphics[width=0.8\textwidth]{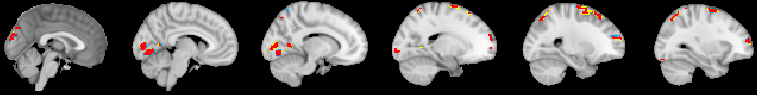}\\
\raisebox{7mm}{\begin{tabular}{r}
TCGP\\ 
(Gibbs)
\end{tabular}}&\includegraphics[width=0.8\textwidth]{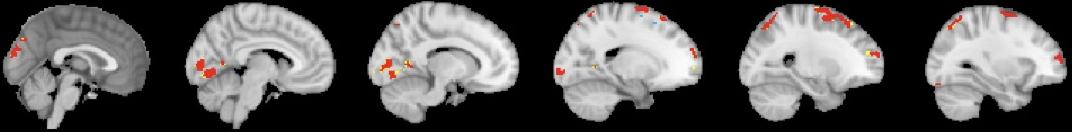}\\
\raisebox{7mm}{\begin{tabular}{r}TCGP \\
(Hybrid)\\
\end{tabular}}&\includegraphics[width=0.8\textwidth]{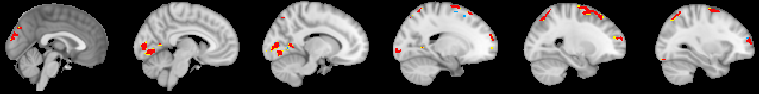}\\
\end{tabular}
\caption{Simulation results of the 3D image example for $n=900$ with $\zeta_k = 1$. The 2D slices of positively correlated regions identified by the voxel-wise analysis, the region-wise analysis, the integrated method of \citet{li2018spatially} with two thresholding values, 0.95 and 0.90, and the proposed Bayesian method with the Gibbs sampler and the hybrid mini-batch MCMC. The red, yellow and blue regions represent the true positive, the false negative, and the false positive regions, respectively.}
\label{fig:sim3d}
\end{figure}

\section{Analysis of Human Connectome Project Data}
\label{sec:realdata}

In this section, we further illustrate our method with an fMRI dataset from the Human Connectome Project. Our specific goal is to study the association between the resting-state fMRI and the memory task-related fMRI, and identify brain regions where the resting-state and task-related brain activities are strongly associated. This type of analysis is useful, as there has been increasing interest in recent years to predict task-related brain activations from resting-state fMRI \citep{tavor2016task, jones2017resting, cohen2020regression}. It also reveals numerous brain regions and offers useful insights to understand brain activities during rest and working memory tasks. 

The dataset we analyze consists of $n=904$ subjects with both resting-state and task fMRI scans. We preprocess both types of images following the usual pipelines. The preprocessing of resting-state fMRI includes correction for distortions and head motion, removal of slowest temporal drifts and structured non-neuronal artifact \citep{smith2013resting}, whereas the preprocessing of task fMRI includes gradient unwarping, motion correction, distortion correction, and grand-mean intensity normalization \citep{barch2013function}. \change{Moreover, we map both resting-state and task fMRI originally in 4D space onto some common 3D space. Specifically, for resting-state fMRI, we compute the fractional amplitude of low-frequency fluctuations (fALFF) at each voxel \citep{zou2008improved}, which quantifies the amplitude of the low frequency oscillations in fMRI signals to reflect the local brain activities during the resting state. For task fMRI, we employ statistical parametric mapping \citep{penny2011statistical}, by fitting a generalized linear model for the time series at each voxel on the stimulus design matrix convoluted by the hemodynamic function, where the intensity values at individual voxels are the estimated regression coefficients that represent the strength of brain activation in response to the task. For both types of fMRI, the resulting images reside in a 3D spatial space, whereas the temporal dimension has been collapsed. In addition, we regress out potential confounding variables of age and sex using the image-on-scalar approach \citep{zhu2014spatially}. Finally, we register and align both images to the standard MNI space \citep{mazziotta2001probabilistic}, and the resulting image dimensions for both resting-state and task fMRI are $91 \times 109 \times 91$, with $m = 117,293$ voxels in the brain mask.} 

\begin{table}[t!]
\begin{center}
\footnotesize
\caption{Results of Human Connectome Project data analysis. Reported are the activation regions containing more than 100 voxels that are declared having a nonzero correlation. AAL refers to the automatic anatomical labeling.}
\label{tab:HCP}
\scriptsize
\begin{tabular}{p{3.5cm}ccccc}
\multicolumn{6}{c}{Regions with positive correlations}\\
\toprule 
 AAL Regions & \makecell[c]{Cluster Size}& \makecell[c]{Activation Center}& \makecell[c]{Mean}& \makecell[c]{Std.}& \makecell[c]{PIP}\\
\midrule
Precentral-L &   385& (45.4, 6.75, 42.95)& 0.39& 0.07& 0.79\\
Frontal-Sup-R &   141& (-25.6, 60.5, 19.2)& 0.30& 0.05& 0.64\\ 
Frontal-Sup-R &   329& (-26.4, 8.0, 65.2)& 0.35& 0.08& 0.62\\ 
Frontal-Mid-L &   643& (33.7, 32.9, 42.2)& 0.35& 0.04& 0.59\\ 
Frontal-Inf-Tri-R &   218& (-51.9, 28.0, 22.6)& 0.21& 0.06& 0.64\\ 
Calcarine-R &   200& (-13.8, -87.4, 3.64)& 0.37& 0.05& 0.69\\ 
Cuneus-L &   120& (-0.4, -87.4, 22.6)& 0.33& 0.04& 0.65\\ 
Lingual-R &   144& (-10.4, -75.3, -4.5)& 0.35& 0.05& 0.62\\ 
Parietal-Sup-L &   187& (20.0, -67.4, 53.7)& 0.35& 0.05& 0.53\\ 
Parietal-Sup-L &   108& (26.0, -52.8, 62.4)& 0.40& 0.06& 0.58\\ 
Parietal-Sup-R &   165& (-29.2, -20.9, 68.3)& 0.30& 0.08& 0.76\\ 
Parietal-Inf-L &   253& (47.2, -46.1, 49.7)& 0.40& 0.05& 0.59\\ 
Angular-R &   209& (-46.9, -60.2, 44.7)& 0.43& 0.03& 0.70\\ 
Temporal-Sup-L &   331& (54.3, -31.8, 18.0)& 0.40& 0.05& 0.82\\ 
Temporal-Mid-L &   104& (63.1, -25.7, 1.38)& 0.41& 0.07& 0.56\\
\bottomrule 
\end{tabular}

\smallskip 

\begin{tabular}{p{3.5cm}ccccc}
\multicolumn{6}{c}{Regions with negative correlations}\\
\toprule 
AAL Regions& \makecell[c]{Cluster Size}& \makecell[c]{Activation Center}& \makecell[c]{Mean}& \makecell[c]{Std.}& \makecell[c]{PIP}\\
 \midrule
Precentral-L &   115& (28.6, -23.1, 65.4)& -0.44& 0.03& 0.90\\
Precentral-R &   183& (-54.4, 8.0, 36.0)& -0.40& 0.08& 0.59\\
Frontal-Mid-L &   191& (28.2, 52.2, 12.7)& -0.39& 0.06& 0.78\\
Rolandic-Oper-L &   186& (-45.6, -14.5, 15.9)& -0.36& 0.14& 0.58\\
Supp-Motor-Area-L &   120& (1.1, -7.9, 66.1)& -0.36& 0.05& 0.71\\
Supp-Motor-Area-R &   143& (-6.9, -13.3, 69.5)& -0.38& 0.07& 0.85\\
Calcarine-R &   183& (-15.4, -68.8, 10.5)& -0.32& 0.06& 0.65\\
Lingual-L &   292& (10.5, -75.0, -5.5)& -0.28& 0.04& 0.79\\
Lingual-R &   286& (-21.2, -86.3, -9.0)& -0.38& 0.05& 0.80\\ 
Occipital-Sup-L &   111& (16.0, -89.8, 25.0)& -0.40& 0.06& 0.80\\ 
Occipital-Sup-R &   147& (-25.2, -89.9, 26.2)& -0.37& 0.04& 0.77\\ 
Occipital-Inf-R &   122& (-38.8, -81.7, -3.2)& -0.44& 0.05& 0.56\\ 
SupraMarginal-L & 191& (58.8, -25.7, 30.8)& -0.37& 0.04& 0.83\\ 
SupraMarginal-R & 121& (-58.2, -36.9, 28.3)& -0.38& 0.03& 0.98\\ 
Paracentral-Lobule-R & 147& (-5.7,-30.5, 70.4)& -0.33& 0.04& 0.63\\ 
Temporal-Mid-L & 109& (58.6, -36.3, 7.7)& -0.43& 0.05& 0.55\\
\bottomrule 
\end{tabular}
\end{center}
\vspace{-0.15in}
\end{table}  

\begin{figure}[th!]
\centering
\subfigure[Positive correlation map]{
\label{fig: pos}
\includegraphics[width=0.7\columnwidth]{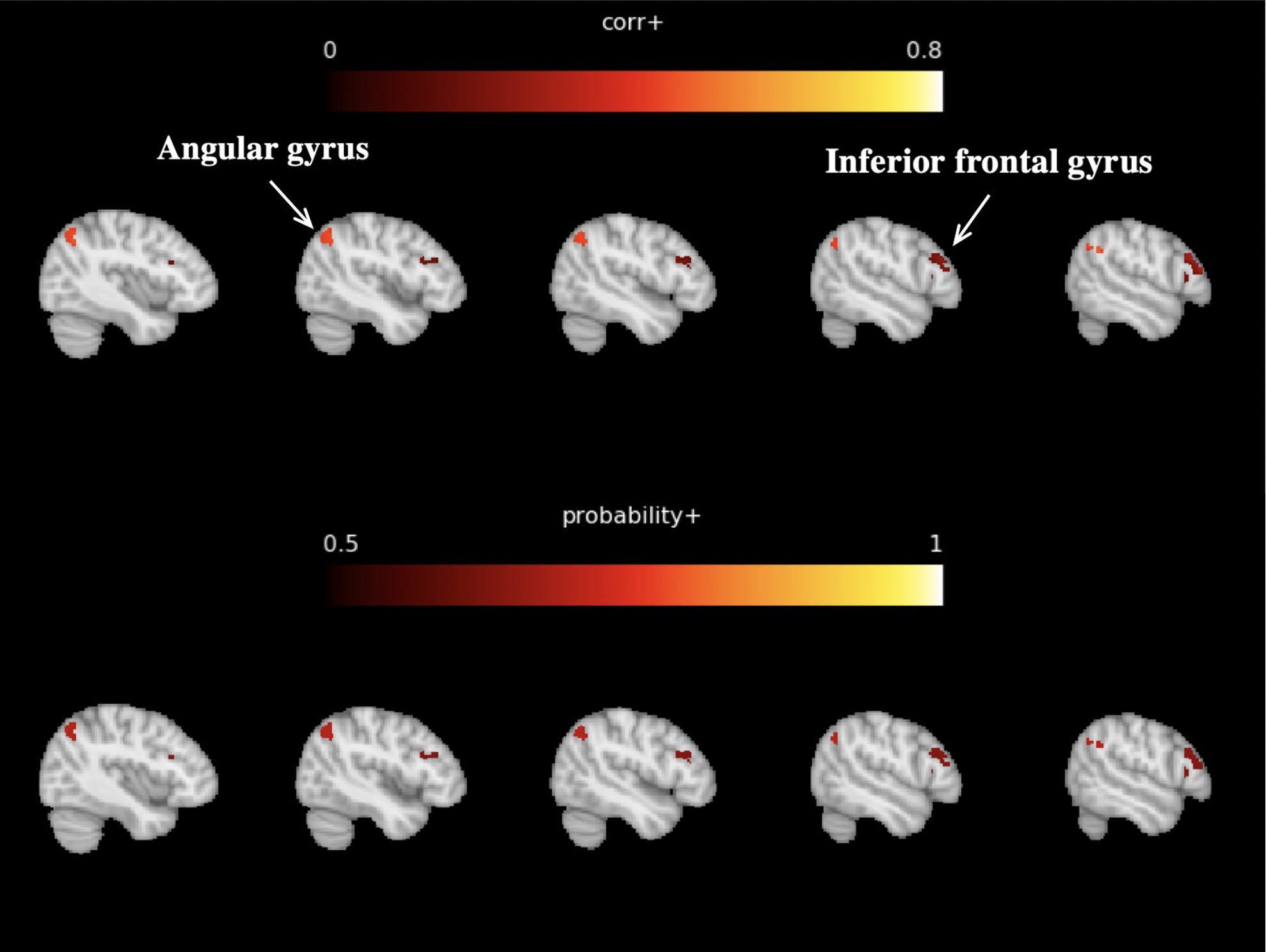}
}
\subfigure[Negative correlation map]{
\label{fig: neg}
\includegraphics[width=0.7\columnwidth]{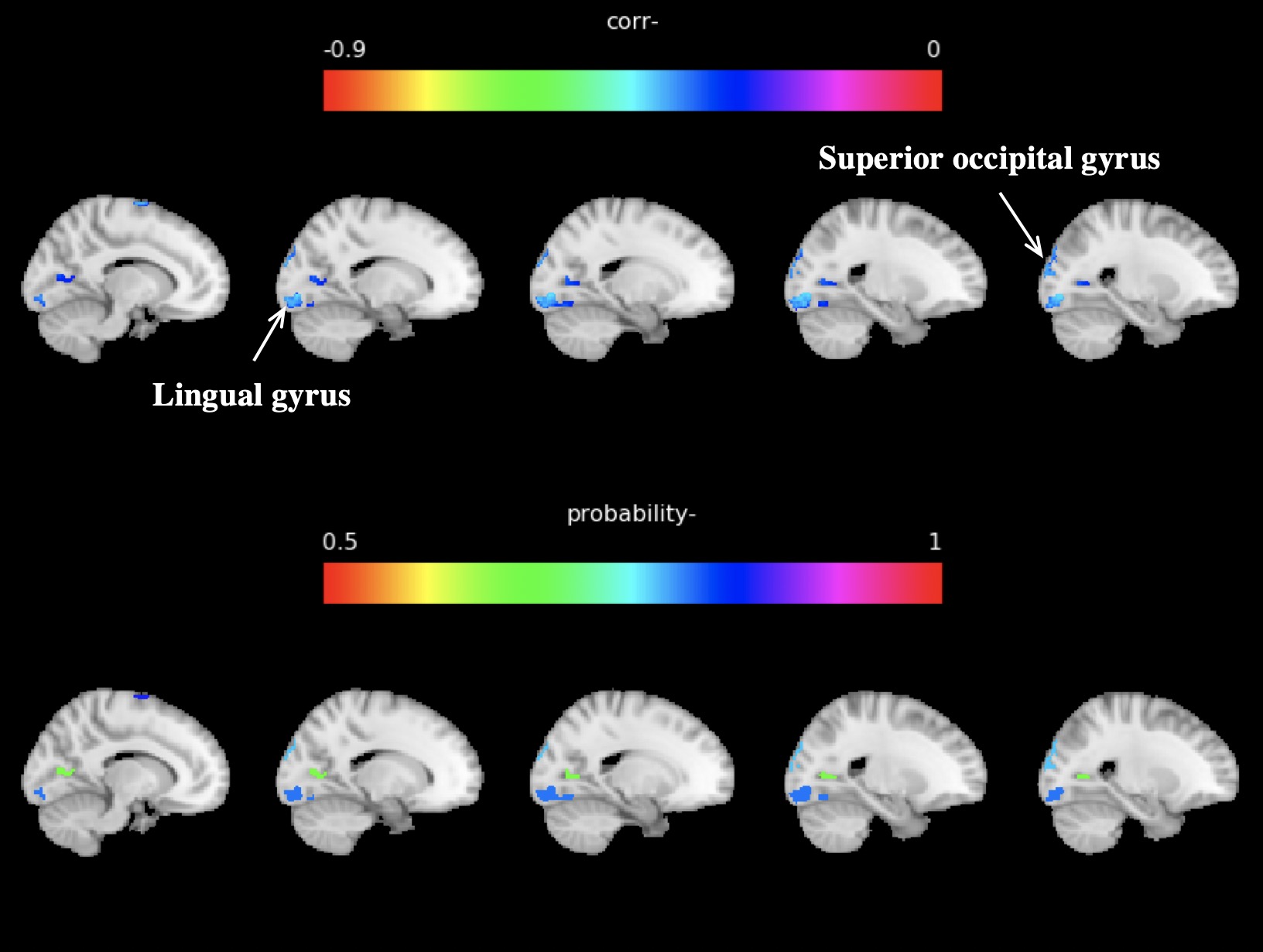}
}
\caption{Results of Human Connectome Project  data analysis. The sagittal slices of activation regions with significant correlations. Panel (a) shows the slices of positive correlation map and the associated inclusion probability map. Panel (b) shows the slices of negative correlation map and the associated posterior inclusion probability map.}
\label{fig:HCPslice}
\end{figure}

We apply the proposed methods to this data. We employ the Mat{\'e}rn kernel in \eqref{eq: matern} in our data analysis. \change{We choose the number of leading eigenvalues $L$ as the smallest value such that the variance percentage exceeds $60\%$ and obtain $L = 540$.  We set the prior hyperparameters $a_\tau= 0.001$, $b_\tau = 0.001$, and choose $a_\omega$ and $b_\omega$ as the $75\%$ quantile and $100\%$ quantile of $\{|\xi(v)|\}_{v\in \mathcal{B}}$, respectively. The choice of $a_\omega$ is based on the belief that at most $25\%$ voxels have non-zero correlations.} We perform sensitivity analysis  on the small changes of prior specification for the threshold parameters in the Supplementary Material, Section \ref{appen: aomega}. We run the Gibbs sampler for 1000 iterations, with the first 200 iterations as the burn-in. We also run the hybrid mini-batch MCMC for 1200 iterations, with the first 400 iterations as the burn-in. We again claim a voxel having a nonzero correlation by simply thresholding the posterior inclusion probability at 0.5. 

Table \ref{tab:HCP} summarizes the estimated activation regions with strong positive or negative correlations. Here we only report those regions containing more than 100 voxels that are declared having a nonzero correlation.  We also map those estimated regions to the automatic anatomical labeling (AAL) brain atlas, and report where each estimated activation region is located, the cluster size, the activation center coordinates, the mean and the standard deviation of the correlation in a specific cluster, and the posterior inclusion probability. We make the following observations. We identify a region in angular gyrus that has the highest positive mean correlation. This finding agrees with the literature, as intensive research has shown that angular gyrus is involved in cognitive processes related to language, number processing, spatial cognition, memory retrieval, and attention \citep{farrer2008angular, seghier2013angular}. We also identify a region with strong positive correlations in middle temporal gyrus and superior parietal gyrus. The former region is  connected with numerous cognitive processes including recognition of known faces, audio-visual emotional recognition, and accessing word meaning while reading \citep{acheson2013stimulating}, and the latter is critically involved in information manipulation in working memory \citep{koenigs2009superior}. In addition, we identify two regions in lingual gyrus with strong negative correlation, while lingual gyrus is believed to play an important role in visual memory and word processing \citep{leshikar2012task}. Figure~\ref{fig:HCPslice} shows the identified activation regions with significant correlations.

\section*{Supplementary Materials}

In the supplement materials, we first present the proofs of all the theoretical results in the paper, along with a number of useful lemmas. We next derive the full conditional distributions of the model parameters, and present some additional numerical results. 

\par

\par


\bibhang=1.7pc
\bibsep=2pt
\fontsize{9}{14pt plus.8pt minus .6pt}\selectfont
\renewcommand\bibname{\large \bf References}
\expandafter\ifx\csname
natexlab\endcsname\relax\def\natexlab#1{#1}\fi
\expandafter\ifx\csname url\endcsname\relax
  \def\url#1{\texttt{#1}}\fi
\expandafter\ifx\csname urlprefix\endcsname\relax\def\urlprefix{URL}\fi





\bibliographystyle{chicago}      
\bibliography{reference}   

\clearpage


\appendix




\newcommand{\Ebb}{\mathbb{E}}

\renewcommand{\baselinestretch}{2}

\markright{ \hbox{\footnotesize\rm Statistica Sinica
}\hfill\\[-13pt]
\hbox{\footnotesize\rm
}\hfill }

\markboth{\hfill{\footnotesize\rm FIRSTNAME1 LASTNAME1 AND FIRSTNAME2 LASTNAME2} \hfill}
{\hfill {\footnotesize\rm FILL IN A SHORT RUNNING TITLE} \hfill}

\renewcommand{\thefootnote}{}
$\ $\par







\section{Proofs}

\subsection{Proof of Proposition 1}
\label{appen: proof_prop2}

Given $\tau_1^2(v)$ and $\tau_2^2(v)$, if $\pi({Y}_{+,i}(v), {Y}_{-,i}(v) \mid \theta)=\pi\left({Y}_{+,i}(v), {Y}_{-,i}(v) \mid  \theta^{\prime}\right)$, for any $i = 1,\ldots,n$, $v \in \mathcal{B}_m$, and since $\left\{{Y}_{+,i}(v), {Y}_{-,i}(v)\right\}$ follows a bivariate normal distribution, we have that $\mu_{+,i}(v) = \mu'_{+,i}(v)$, and $\mu_{-,i}(v) = \mu'_{-,i}(v)$, i.e., $s\{\rho(v)\} E_{+,i}(v)= s\{\rho'(v)\} E'_{+,i}(v)$, and $s\{-\rho(v)\} E_{-,i}(v)= s\{-\rho'(v)\} E'_{-,i}(v)$, for any $i = 1,\ldots,n$, $v \in \mathcal{B}_m$. 

Furthermore, we have that, 
\vspace{-3mm}
\begin{equation*}
\resizebox{0.9\hsize}{!}{
$
    \begin{split}
        0 &= \sum_{i=1}^n\left[s\{\rho(v)\} E_{+,i}(v) -s\{\rho'(v)\}E'_{+,i}(v) \right]^2 \\
        & = \sum_{i=1}^n\left[s\{\rho(v)\}^2 E_{+,i}(v)^2 -2s\{\rho(v)\} s\{\rho'(v)\}E_{+,i}(v) E'_{+,i}(v) + s\{\rho'(v)\}^2 E'_{+,i}(v)^2\right] \\
        & =\left[s\{\rho(v)\} -  s\{\rho'(v)\}\right]^2\sum_{i=1}^n E^2_{+,i}(v) + s\{\rho'(v)\}s\{\rho(v)\}\sum_{i=1}^n \{E_{+,i}(v) - E'_{+,i}(v)\}^2 \\
        & \quad + s\{\rho'(v)\}\left[s\{\rho(v)\}- s\{\rho'(v)\}\right] \sum_{i=1}^n \left\{E_{+,i}(v)^2 - E'_{+,i}(v)^2\right\}\\
    \end{split}$}
\end{equation*}
By Definition (4), we have $\sum_{i=1}^n E_{+,i}(v)^2 = \sum_{i=1}^n E'_{+,i}(v)^2$.

When $v\in \mathcal{V}(\rho)\cup \mathcal{V}(\rho')$, we have $s\{\rho(v)\} \geq 0$, $s\{\rho'(v)\} \geq 0$, and at least one of $s\{\rho(v)\}$ and $s\{\rho'(v)\}$ is not equal to 0. Therefore, $s\{\rho(v)\}  =  s\{\rho'(v)\}$, and $E_{+,i}(v) = E'_{+,i}(v)$, for any $i = 1,\ldots,n, \; v \in \mathcal{B}_m$. On the other hand, if $v\notin \mathcal{V}(\rho)\cup \mathcal{V}(\rho')$, then $s\{\rho(v)\}  =  s\{\rho'(v)\} = 0$. Similarly, we have $E_{-,i}(v) = E'_{-,i}(v)= 0$, for any $i = 1,\ldots,n$, $v \in \mathcal{B}_m$. 

Since $s(\cdot)$ is a monotonic function, we have $\rho(v) = \rho'(v)$ for all $v \in \mathcal{B}_m$. This completes the proof of Proposition 1. 
\eop

\subsection{Proof of Theorem 1}
\label{appen:largesup}

By Lemma \ref{lemma1}, we have $\rho(v) = T_\omega\{\xi(v)\} = H[R_\omega\{\xi(v)\}]$, where $H(t) = t^2/(t^2+1)$ when $\xi(v) > \omega$, $H(t) = -t^2/(t^2+1)$ when $\xi(v) < -\omega$, and $H(t) = 0$ otherwise, and $R_\omega(x) = G_\omega(x) - G_\omega(-x)$ is the hard thresholded function. Therefore, we have that, 
\vspace{-3mm}
\begin{equation*}
\begin{split}
        pr\left(\|\rho-\rho_0\|_{\infty}<\varepsilon\right) &= pr\left(\|H[R_\omega\{\xi(v)\}] - H[R_\omega\{\xi_0(v)\}]\|<\epsilon \right)\\
        & \geq pr\left(\|R_\omega\{\xi(v)\} - R_\omega\{\xi_0(v)\}\|<\epsilon\right), 
\end{split}
\end{equation*}
by the  Lipschitz continuity of $H(\cdot)$. Given the assumptions for $\rho_0({v})$, we have that $\xi({v})$ is bounded away from 0 for $v\notin \mathcal{R}_0$. Henceforth, 
\vspace{-3mm}
\begin{equation}\label{eq:lemma11}
\begin{split}
        & \quad pr(\|R_\omega(\xi({v} )) - R_\omega(\xi_0({v} ))\|<\epsilon)\\ 
        & \geq pr\left(\sup_{{v}  \notin \mathcal{R}_0}\left|{\xi}({v})-\xi_{0}({v})\right|<\epsilon, \inf_{v \notin \mathcal{R}_0}|\xi({v})|>\omega, \sup_{v \in \mathcal{R}_0}|\xi({v})| \leq \omega\right).
\end{split}
\end{equation}
Without loss of generality, we only consider $0<\epsilon<\omega - \omega_0$, where $\omega_0 =\inf _{{v} \notin  \mathcal{R}_{0}}|\rho({v})|$. Note that for all ${v} \notin \mathcal{R}_0$, $|\xi({v}) - \xi_0({v})| <\epsilon$ and $|\xi_0({v})|\geq \omega_0$, which implies that $|\xi({v})| \geq \omega_0 - \epsilon > \omega $. Then \eqref{eq:lemma11} is equivalent to 
\begin{equation*}
    pr(\|\rho({v}) - \rho_0({v})\|<\epsilon) \geq pr\left(\sup_{{v}  \notin \mathcal{R}_0} |\xi({v}) - \xi_0({v})| < \epsilon, \sup_{{v}  \in \mathcal{R}_0} |\xi({v})| \leq \omega\right).
\end{equation*}

Let $\psi_l({v})$ and $\lambda_l$ be the normalized eigenfunctions and eigenvalues of the kernel function $\kappa(\cdot,\cdot)$. The KL expansions of $\xi({v})$ and $\xi_0({v})$ 
are $\xi({v}) = \sum_{l=1}^\infty c_l\psi_l({v})$, $\xi_0({v}) = \sum_{l=1}^\infty c_{l0}\psi_l({v})$. 

For ${v} \notin \mathcal{R}_0$, we have that, 
\vspace{-5mm}
\begin{equation*}
\resizebox{0.99\hsize}{!}{
$\sup_{{v} \notin \mathcal{R}_0}|\xi({v}) - \xi_0({v})| \leq \sup_{{v} \notin \mathcal{R}_0} |\xi_L({v}) - \xi_L^0({v})| + \sup_{{v} \notin \mathcal{R}_0} |\xi({v}) - \xi_L({v})| + \sup_{{v} \notin \mathcal{R}_0} |\xi_L^0({v}) - \xi_0({v})|.$}
\end{equation*}
Since the RKHS of $\kappa(\cdot, \cdot)$ is the space of the continuous functions on $\mathcal{R}$, $\xi({v})$ is uniformly continuous on $\mathcal{B}\backslash \mathcal{R}_0$ with probability 1. Then by Theorem 3.1.2 of \cite{adler2009random}, $\lim_{L\rightarrow \infty}\sup_{{v} \notin \mathcal{R}_0}|\xi({v}) - \xi_L({v})| = 0$ with probability 1. By the uniform convergence of the series $\sum_{l=1}^L c_{l0}\psi_l({v})$ to $\xi_0({v})$ on $\mathcal{B}\backslash \mathcal{R}_0$, as $L\rightarrow \infty$, we have $\lim_{L\rightarrow \infty}\sup_{v\notin \mathcal{R}_0}|\xi_0({v}) - \xi_L^0({v})| = 0$. Then we can find a finite integer $L^\prime$, such that, for all $L > L^\prime$, $\sup_{v\notin \mathcal{R}_0} |\xi({v}) - \xi_L({v})| < {\epsilon}/{3}$ with probability 1, and $\sup_{v\notin \mathcal{R}_0} |\xi_0({v}) - \xi_L^0({v})| < {\epsilon}/{3}$. Since $\psi_l({v}), l = 1,\ldots,L$, are all continuous functions in $\mathcal{R}$, we have $\max_{1 \leq l \leq L}\left\|\psi_{l}({v})\right\|_{\infty}<M_{\psi, L}$, for some constant $M_{\psi, L}$. When $|c_l - c_{l0}| < \epsilon/(3LM_{\psi, L})$ for all $l = 1, \ldots, L$, we have $\sup_{{v}\notin \mathcal{R}_0}|\xi_L({v}) - \xi_L^0({v})| \leq {\epsilon}/{3}$. Therefore, $|c_l - c_{l0}| < \epsilon/(3LM_{\psi, L})$, $l = 1, \ldots, L$, guarantees that $\sup_{{v}\notin \mathcal{R}_0}|\xi({v}) - \xi_0({v})| \leq \epsilon$ with probability one.

For ${v} \in \mathcal{R}_0$, we have that, 
\vspace{-5mm}
\begin{equation*}
    \sup_{{v}\in \mathcal{R}_0}|\xi({v})| \leq \sup_{{v}\in \mathcal{R}_0}|\xi({v}) - \xi_L({v})| + \sup_{{v}\in \mathcal{R}_0} |\xi_L({v})|.
\end{equation*}
Similarly, we can find $L$ and $M_{\psi, L}$, such that $|c_l| \leq \omega/(2LM_{\psi,L})$, $l = 1, \ldots, L$, guarantees that $\sup_{{v}\in \mathcal{R}_0}|\xi({v})| \leq \omega$ with probability 1. 

Then we have that, 
\vspace{-5mm}
\begin{equation*}
\resizebox{0.95\hsize}{!}{
$\begin{split}
    pr\left(\|\rho-\rho_0\|_{\infty}<\varepsilon\right) &\geq pr\left(\{|c_l - c_{l0}| < \dfrac{\epsilon}{3LM_{\psi,L}}: L = 1, 2, \ldots,L\ when\  {v}\notin \mathcal{R}_0\}\right.\\
    &\cup \left.\{|c_l| \leq \dfrac{\omega}{2LM_{\psi,L}}: L = 1, 2, \ldots,L\ when\ {v}\in \mathcal{R}_0\} \right).
\end{split}$}
\end{equation*}
This completes the proof of Theorem 1. 
\eop

\subsection{Proof of Theorem 2}
\label{appen:consistency}
  
Based on Theorem 1, Lemma \ref{lemma: pos_prior} shows the positivity of prior neighborhoods. We then construct sieves for $\theta(v)$ as follows:
\vspace{-3mm}
\begin{equation}
\label{eq: sieves}
\resizebox{0.88\hsize}{!}{
$\begin{split}
        \Theta_{n} =& \bigg \{\rho \in \Theta_\rho, {E}_{+},{E}_{-} \in \Theta_E : \\
        & \|\rho\|_{\infty} \leq H\left(m^{1/(2d)}\right),  \sup _{v \in \mathcal{R}_{1} \cup \mathcal{R}_{-1}}\left|D^{\tau} \rho(v)\right| \leq m^{1/(2d)},\ 1 \leq\|\tau\|_{1} \leq \alpha\\
        & \|E_{+,i}\|_{\infty} \leq m^{1/(2d)},\sup _{v \in \mathcal{R}_{1} \cup \mathcal{R}_{-1}}\left|D^{\tau} E_{+,i}(v)\right| \leq m^{1/(2d)},  \\
        &\|E_{-,i}\|_{\infty} \leq m^{1/(2d)}, \sup _{v \in \mathcal{R}_{1} \cup \mathcal{R}_{-1}}\left|D^{\tau} E_{-,i}(v)\right| \leq m^{1/(2d)},\ \mbox{for}\ i = 1,\ldots, n \bigg \},
\end{split}$}
\end{equation}
where $\alpha$ and $m$ are defined in Assumption 3. 

We can then find an upper bound for the tail probability, and construct the uniform consistent tests in Lemmas \ref{test1}, \ref{test2}, \ref{test3} and \ref{test4}. These lemmas verify the three key conditions in Theorem A1 of \cite{choudhuri2004bayesian}, which leads to the posterior consistency. That is, by Lemmas \ref{test1}, \ref{test2}, \ref{test3} and \ref{test4}, as $n \rightarrow \infty$, $m \rightarrow \infty$, we have that, 
\vspace{-3mm}
\begin{equation*}
\begin{aligned} 
& \Ebb_{\theta_{0}}\left(\Psi_{n}\right) \rightarrow 0, \\ 
& \sup _{\theta \in \mathcal{U}_{\epsilon}^{C} \cap \Theta_{n}} \Ebb_{\theta}\left(1-\Psi_{n}\right) \leq C_{0} \exp \left(-C_{1} n\right), \\ 
& pr\left(\Theta_{n}^{C}\right) \leq K \exp \left(-b m^{1 / d}\right) \leq K \exp \left(-C_{3} n\right).
\end{aligned}
\end{equation*}
where $\mathcal{U}_{\epsilon}=\left\{\theta \in \Theta:\left\|\theta-\theta_{0}\right\|_{1}<\epsilon\right\}$ for any $\epsilon>0$, and $\Psi_n$ is the test statistic defined in \eqref{eq: testing_stat}. This completes the proof of Theorem 2. 
\eop

\subsection{Proof of Theorem 3}
\label{appen: seleconsistency}

Let $\mathcal{R}_{0} = \left\{v: \rho_{0}(v)=0\right\}$, $\mathcal{R}_{1}=\left\{v: \rho_{0}(v)>0\right\}$, and $\mathcal{R}_{-1}=\left\{v: \rho_{0}(v)<0\right\}$. For any $\mathcal{A} \subset \mathcal{B}$ and any integer $k\geq 1$, define 
\begin{equation*}
\mathcal{F}_{k}(\mathcal{A})=\left\{\rho \in \Theta_\rho: \int_{\mathcal{A}}\left|\rho(v)-\rho_{0}(v)\right| \mathrm{d} v<\frac{1}{k}\right\}.
\end{equation*}
Then $\mathcal{F}_{k+1}(\mathcal{A}) \subseteq \mathcal{F}_{k}(\mathcal{A})$ for all $k$, and $\mathcal{F}_{k}(\mathcal{B}) \subseteq \mathcal{F}_{k}(\mathcal{A})$. Consider
\[
\mathcal{F}_{k}\left(\mathcal{R}_{0}\right)=\left\{\rho \in \Theta_\rho: \int_{\mathcal{R}_{0}}|\rho(v)| \mathrm{d} v<\frac{1}{k}\right\}.
\]
Define $\mathcal{U}_{\epsilon}^\rho=\left\{\rho \in \Theta_\rho:\left\|\rho-\rho_{0}\right\|_{1}<\epsilon\right\}$. By Theorem 2 and the fact that $\mathcal{U}_{1 / k}^\rho=\mathcal{F}_{k}(\mathcal{B})$, we have
\[
pr\left\{\mathcal{F}_{k}\left(\mathcal{R}_{0}\right) \mid Y_+, Y_-\right\} \geq pr\left(\mathcal{U}_{1 / k}^\rho \mid Y_+, Y_-\right) \rightarrow 1, \ \mbox{as}\ n \rightarrow \infty.
\]
In addition,
\[
\left\{\rho(v)=0, \text { for all } v \in \mathcal{R}_{0}\right\}=\left\{\int_{\mathcal{R}_{0}}|\rho(v)| \mathrm{d} v=0\right\}=\bigcap_{k=1}^{\infty} \mathcal{F}_{k}\left(\mathcal{R}_{0}\right).
\]
By the monotonic continuity of the probability measure, we have,
\[
pr\big\{ \rho(v)=0, \text { for all } v \in \mathcal{R}_{0} \mid Y_+, Y_- \big\} = \lim _{k \rightarrow \infty} pr\big\{\mathcal{F}_{k}\left(\mathcal{R}_{0}\right) \mid Y_+, Y_- \big\} = 1,\ \mbox{as}\ n \rightarrow \infty.
\]

For any $v_0 \in \mathcal{R}_{1}$ and any integer $k \geq 1$, there exists $\delta_0 > 0$, such that $\left|\rho\left(v_1\right)-\rho\left(v_0\right)\right|<1/2k$, for any $v_1 \in \mathcal{B}\left(v_0, \delta_0\right)=\left\{v:\left\|v_1-v_0\right\|_1<\delta_0\right\}$. As $\mathcal{R}_1$ is an open set, there exists $\delta_1>0$, such that $\mathcal{B}\left(v_0, \delta_1\right) \subseteq \mathcal{R}_1$. Let $\delta=\min \left\{\delta_1, \delta_0\right\}>0$, we have that, 
\vspace{-3mm}
\begin{equation*}
\resizebox{0.98\hsize}{!}{
$\begin{split}
& \left\{\rho\left(v_0\right)>-\frac{1}{k}, \text { for all } v_0 \in \mathcal{R}_1\right\} \\
& \supseteq\left\{\rho\left(v_0\right)>\rho\left(v_1\right)-\frac{1}{2 k} \text { and } \rho\left(v_1\right)>-\frac{1}{2 k}, \text { for some } v_1 \in \mathcal{B}\left(v_0, \delta\right), \text { for all } v_0 \in \mathcal{R}_1\right\} \\
& \supseteq\left\{\int_{\mathcal{B}\left(v_0, \delta\right)} \rho(v)dv>-\frac{1}{2 k}, \text { for all } v_0 \in \mathcal{R}_1\right\} \\
& \supseteq\left\{\int_{\mathcal{B}\left(v_0, \delta\right)} \rho(v)dv>\int_{\mathcal{B}\left(v_0, \delta\right)} \rho_0(v) d v-\frac{1}{2 k}, \text { for all } v_0 \in \mathcal{R}_1\right\} \\
& \supseteq \mathcal{F}_{2 k}\left[\mathcal{B}\left(v_0, \delta\right)\right] \supseteq \mathcal{U}_{1 / 2 k}^\rho .
\end{split}$}
\end{equation*}
Therefore, 
$$pr\left\{\rho\left(v_0\right)>-{1}/{k}, \text { for all } v_0 \in \mathcal{R}_1 \mid Y_+, Y_-\right\} \geq pr\left(\mathcal{U}_{1 / 2 k}^\rho \mid Y_+, Y_-\right) \rightarrow 1, $$ as $n \rightarrow \infty$. By the monotonic continuity of the probability measure, we have that, 
\begin{equation*}
\resizebox{0.99\hsize}{!}{
$pr\left\{\rho(v)>0, \text { for all } v \in \mathcal{R}_1 \mid Y_+, Y_-\right\} 
= \lim _{k \rightarrow \infty} pr\left\{\rho\left(v_0\right)>-\frac{1}{k}, \text { for all } v_0 \in \mathcal{R}_1 \mid Y_+, Y_-\right\} \rightarrow 1,$}
\end{equation*}
as $n \rightarrow \infty$. Similarly, we can obtain that $pr\left\{\rho(v)<0 \text {, for all } v \in \mathcal{R}_{-1} \mid Y_+, Y_-\right\} \rightarrow 1, n \rightarrow \infty$. This completes the proof of Theorem 3.
\eop

\subsection{Proof of Proposition 2}
\label{appen:prop1}

We prove this proposition by sorting all the thresholding values, and derive the unnormalized density on each interval, respectively. We then obtain the full conditional density function of $\theta$ by normalizing the function on each interval as the density function.

We sort $\left(L_{1},\ldots, L_{P}, U_{1},\ldots, U_{K} \right)$ in ascending order, which leads to $P+K+1$ intervals, and denoted them as $I_1, I_2, \ldots, I_{P+K+1}$. For each interval $I_i$, $i=1,\ldots,P+K + 1$, the full conditional distribution of $\theta$ is proportional to $\exp(-D_i\theta^2 - E_i\theta - F_i)$. We initialize $D_i = E_i = F_i=0$, then loop through $p = 1, \ldots, P$ and $k = 1, \ldots, K$ to update $D_i$, $E_i$ and $F_i$. More specifically, if $I_i \subset [L_p, +\infty)$, we update $D_i = D_i+a_{1p}$, $E_i =E_i+a_{2p}$, and $F_i=F_i+a_{3p}$. If $I_i \subset (-\infty, U_k]$, we update $D_i = D_i+b_{1k}$, $E_i =E_i+ b_{2k}$, and $F_i =F_i+ b_{3k}$. We consider three specific cases.

\begin{itemize}
\item If at least one of $\{a_{1p}, \ldots, a_{1P}, b_{1k}, \ldots, b_{1K}\}$ is not equal to 0, then $D_i \neq 0$, for any $i=1,\ldots,P+K + 1$. Therefore, when $\theta \in I_i$, the full conditional distribution of $\theta$ is $\mathrm{N}\{-E_i/(2D_i), - 1/(2 D_i)\}$. Incorporating the normalizing constant $M_i$ for each interval, which is independent of $\theta$, the full conditional distribution of $\theta$ is the mixture of truncated normal distributions, $\sum_{i = 1}^{P+K + 1} M_i\cdot \mathrm{Truncated Normal}_{I_i}\{-E_i/(2D_i), - 1/(2 D_i)\}$. 

\item If $a_{1p} = b_{1k} = 0$, for any $p = 1, \ldots, P$ and $k = 1, \ldots, K$,  and at least one of $\{a_{2p}, \ldots, a_{2P}, b_{2k}, \ldots, b_{2K}\}$ is not equal to 0, then $D_i = 0$ and $E_i \neq 0$, for any $i=1,\ldots,P+K + 1$. Therefore, when $\theta \in I_i$, the full conditional distribution of $\theta$ is the exponential distribution $\mbox{Exp}(E_i)$. Incorporating the normalizing constant $M_i$, the full conditional distribution of $\theta$ is $\sum_{i = 1}^{P+K + 1} M_i\cdot \mbox{Exponential}_{I_i}(E_i)$.

\item  If $a_{1p} = b_{1k} = a_{2p} = b_{2k} = 0$, for any $p = 1, \ldots, P$ and $k = 1, \ldots, K$, and at least one of $\{a_{3p}, \ldots, a_{3P}, b_{3k}, \ldots, b_{3K}\}$ is not equal to 0, then $D_i=E_i = 0$, and at least one of $F_i \neq 0$, for any $i=1,\ldots,P+K + 1$. Therefore, when $\theta \in I_i$, the full conditional distribution of $\theta$ is proportional to the uniform distribution on $I_i = \left[u_{1i}, u_{2i}\right]$. Incorporating the normalizing constant $M_i$, the full conditional distribution of $\theta$ is $\sum_{i = 1}^{P+K + 1} M_i\cdot \mathrm{U}(u_{1i}, u_{2i})$.
\end{itemize}
This completes the proof of Proposition 2. 
\eop

\section{Additional Lemmas}
\label{appen: lemma}

\begin{lemma}
\label{lemma1}
Rewrite $\rho(v) = T_\omega\{\xi(v); \tau_1^2(v),\tau_2^2(v)\}$ in Equation \eqref{eq:spat_cor}. Then $T_\omega(\cdot)$ is a piecewise Lipschitz continuous function for any $\omega$. 
\end{lemma}

\noindent\textit{Proof}:
From Equation \eqref{eq:spat_cor}, 
it is straightforward to verify that $\rho(v)$ can be written as 
\begin{footnotesize}
\begin{align*}\label{eqAppen:GP_prior}
\rho(v) & = \mathrm{Corr}\{Y_{1,i}(v),Y_{2,i}(v) \}\nonumber \\
& = \frac{G^2_\omega\{\xi(v)\}-G^2_\omega\{-\xi(v)\}}{\left[{G^2_\omega\{\xi(v)\}+G^2_\omega\{-\xi(v)\}+\tau^2_1(v)}\right]^{1/2}\left[{G^2_\omega\{\xi(v)\}+G^2_\omega\{-\xi(v)\}+\tau^2_2(v)}\right]^{1/2}} \notag \\
& = \frac{\textrm{sgn}\{\xi(v)\}R_\omega^2\{\xi(v)\}}{\left[{R_\omega^2\{\xi(v)\}+\tau^2_1(v)}\right]^{1/2}\left[{R_\omega^2\{\xi(v)\}+\tau^2_2(v)}\right]^{1/2}}, 
\end{align*}
\end{footnotesize}
where $R_\omega(x) = G_\omega(x) - G_\omega(-x)$. Without loss of generality, suppose $\tau_1^2(v)$ and $\tau_2^2(v)$ are both equal to one. Then $T_\omega(x) = H\{R_\omega(x)\}$, where $H(t) = t^2/(t^2+1)$ when $\xi(v) > \omega$, $H(t) = -t^2/(t^2+1)$ when $\xi(v) < -\omega$, and $H(t) = 0$ otherwise. Since $H(t)$ is continuous and $|H^\prime(t)| \leq 1/(2\omega)$, $H(t)$ is Lipschitz continuous. As $R_\omega(x)$ is the hard thresholding function, which is piecewise Lipschitz continuous function, $T_\omega(x)=H\{R_\omega(x)\}$ is also a piecewise Lipschitz continuous function. This completes the proof of Lemma \ref{lemma1}. 
\eop

\begin{lemma}
\label{lemma: transform}
Given $\rho(v) = T_\omega\{\xi(v);\tau_1^2(v),\tau_2^2(v)\}$ in (6), there exist a piecewise Lipschitz continuous function $s(\cdot)$, such that $G_\omega\{\xi(v)\} = s\{\rho(v);\tau_1^2(v), \tau_2^2(v)\}$.
\end{lemma}

\noindent\textit{Proof}:
It is straightforward to show that $G_\omega\{\xi(v)\}=s\left\{\rho(v) ; \tau_1^2(v), \tau_2^2(v)\right\}$, and $G_\omega\{-\xi(v)\}=s\left\{-\rho(v) ; \tau_1^2(v), \tau_2^2(v)\right\}$, where $s(x ; t_1, t_2)$ is as given in (7). Therefore, $s(\cdot)$ is a piecewise Lipschitz continuous function. This completes the proof of Lemma \ref{lemma: transform}. 
\eop

\begin{lemma}
\label{lemma: pos_prior}
Let $\Pi_{n,i}(\cdot; \theta)$ denote the density function of $Z_{n,i} = ({Y}_{+,i}, {Y}_{-,i})$. Define $\Lambda_{n, i}(\cdot ; \theta_{0}, \theta)$ $= \log \pi_{n, i}(\cdot ; \theta)-\log \pi_{n, i}(\cdot ; \theta_{0})$, \\$K_{n, i}(\theta_{0}, \theta) = \Ebb_{\theta_{0}}\left\{\Lambda_{n, i}\left(Z_{n, i} ; \theta_{0}, \theta\right)\right\}$, and $V_{n, i}\left(\theta_{0}, \theta\right) = \operatorname{var}_{\theta_{0}}$ $\{\Lambda_{n, i} (Z_{n, i}; \theta_{0}, \theta) \}$. There exists a set $O$ with $\Pi(O)>0$, such that, for any $\epsilon > 0$, 
\begin{footnotesize}
\[
\liminf _{n \rightarrow \infty} \Pi\left[\left\{\theta \in O, n^{-1} \sum_{i=1}^{n} K_{n, i}\left(\theta_{0}, \theta\right)<\epsilon\right\}\right]>0 \ \mbox{and}\  n^{-2} \sum_{i=1}^{n} V_{n, i}\left(\theta_{0}, \theta\right) \rightarrow 0 \ \mbox{for}\  \theta \in O.
\]
\end{footnotesize}
\end{lemma}

\noindent\textit{Proof}:
The density function is of the form, 
\vspace{-3mm}
\[
\Pi_{n,i}(Z_{n,i}; \theta) =\sum_{v\in \mathcal{B}_m} \dfrac{1}{2\pi u^2(v) \{1-r^2(v)\}^{1/2}}\cdot \exp\left[-\dfrac{ W_i(v)}{2\{1-r^2(v)\}u^2(v)}\right],
\]
where $W_i(v) =\{Y_{+,i}(v) - \mu_{+,i}(v)\}^2 + \{Y_{-,i}(v) - \mu_{-,i}(v)\}^2 +2r(v)\{Y_{+,i}(v)\mu_{-,i}(v) + Y_{-,i}(v)\mu_{+,i}(v)\}$, $r(v) = \{\tau^2_1(v) - \tau^2_2(v)\}/\{\tau^2_1(v) + \tau^2_2(v)\}$, and $u^2(v) =  \{\tau^2_1(v) + \tau^2_2(v)\}/4$. Therefore, we have, 
\vspace{-3mm}
\begin{footnotesize}
\begin{equation*}
\begin{split}
    \Lambda_{n, i}\left(Z_{n, i} ; \theta_{0}, \theta\right) & = \log\Pi(Z_{n,i}; \theta) - \log\Pi(Z_{n,i}; \theta_0)\\
        & = \sum_{v\in \mathcal{B}_m}\left[-\dfrac{1}{2\{1-r^2(v)\}u^2(v)}\right] \Bigl[\mu_{+,i}^2(v) - \mu_{+,i,0}^2(v) + \mu_{-,i}^2(v) - \mu_{+,i,0}^2(v) \\
        & \quad+ 2Y_{+,i}(v)\{\mu_{+,i,0}(v) - \mu_{+,i}(v)\} + 2Y_{-,i}(v)\{\mu_{-,i,0}(v) - \mu_{-,i}(v)\}(v)\\
        & \quad+ 2rY_{+,i}(v)\{\mu_{-,i}(v) - \mu_{-,i,0}(v)\}+ 2rY_{-,i}(v)\{\mu_{+,i}(v) - \mu_{+,i,0}(v)\}\Bigr],
\end{split}
\end{equation*}
\vspace{-3mm}
\begin{equation*}
\begin{split}
K_{n, i}\left(\theta_{0}, \theta\right) & = \Ebb_{\theta_{0}}\left\{\Lambda_{n, i}\left(Z_{n, i} ; \theta_{0}, \theta\right)\right\}\\
        & = \sum_{v\in \mathcal{B}_m} \Bigl(-\dfrac{1}{2\{1-r^2(v)\}u^2(v)} \Bigl[\{\mu_{+,i}(v) - \mu_{+,i,0}(v)\}^2+ \{\mu_{-,i}(v) - \mu_{-,i,0}(v)\}^2 \\
        & \quad + 2r(v)\mu_{+,i,0}(v)\mu_{-,i}(v) + 2r(v)\mu_{-,i,0}(v)\mu_{+,i}(v) \\
        & \quad - 2r(v)\mu_{+,i,0}(v)\mu_{-,i,0}(v) - 2r(v)\mu_{-,i,0}(v)\mu_{+,i,0}(v)\Bigr]\Bigr).
\end{split}
\end{equation*}
\end{footnotesize}
Given any $\zeta>0$, let $O(\zeta)=\{\theta: \|\theta-\theta_{0}\|_\infty< \zeta\}$, with
\vspace{-3mm}
\begin{footnotesize}
\begin{equation*}
\left\|{\theta}-{\theta}_0\right\|_{\infty}  =\max_{v\in \mathcal{V}(\rho) \cup \mathcal{V}(\rho_0)} \left\{\|\rho-\rho_0\|_\infty, \max _{1 \leq i \leq n} \|E_{-,i}-E_{-,i,0}\|_\infty, \max _{1 \leq i \leq n} \|E_{+,i}-E_{+,i,0}\|_\infty\right\}, 
\end{equation*}
\end{footnotesize}
and $\mathcal{V}(\rho) = \{v: \rho(v) \neq 0\}$, $\mathcal{V}(\rho_0) = \{v: \rho_0(v) \neq 0\}$, then, for any $v\in O(\zeta)$,
\vspace{-3mm}
\begin{equation*}
\begin{split}
& \left|\mu_{i,+}(v) - \mu_{i,+,0}(v)\right| \leq \left|s\{\rho(v)\}E_{+,i}(v) - s\{\rho_0(v)\}E_{i,+,0}(v)\right|\\
\leq & \left|E_{+,i}(v)\left(s\{\rho(v)\} - s\{\rho_0(v)\}\right)\right| + \left|s\{\rho_0(v)\}\left(E_{+,i}(v)- E_{i,+,0}(v)\right)\right| \leq K_1\zeta,
\end{split}
\end{equation*}
where the last inequality is due to the compactness and convexity of $\mathcal{B}_m$, and 
\vspace{-3mm}
\[
K_1 = \max\limits_{v\in \mathcal{V}(\rho) \cup \mathcal{V}(\rho_0)}\left\{E_{+,i}(v),s\{\rho_0(v)\}\right\}.
\]
Similarly, we have $\left|\mu_{i,-}(v) - \mu_{i,-,0}(v)\right| \leq K_2\zeta$, for any $v$, where 
\vspace{-3mm}
\[
K_2 = \max\limits_{v\in \mathcal{V}(\rho) \cup \mathcal{V}(\rho_0)}\left\{E_{-,i}(v),s\{-\rho_0(v)\}\right\}.
\]

Therefore, we have that, 
\begin{equation*}
\footnotesize
\begin{split}
\left|\sum_{i=1}^n K_{n,i}(\theta, \theta_0)\right| & \leq \sum_{{v\in \mathcal{V}(\rho) \cup \mathcal{V}(\rho_0)}} \dfrac{1}{2\left\{1-r^2(v)\right\}u^2(v)} \bigg (\sum_{i=1}^n|\mu_{i,+}(v) - \mu_{i,+,0}(v)|^2 \\
        & \quad + \sum_{i=1}^n\left|\mu_{i,-}(v) - \mu_{i,-,0}(v)\right|^2 \\
        & \quad + 2r(v)M\sum_{i=1}^n\left|\mu_{i,-}(v) - \mu_{i,-,0}(v)\right| + 2r(v)M\sum_{i=1}^n\left|\mu_{i,+}(v) - \mu_{i,+,0}(v)\right|\bigg)\\
        & \leq \sum_{v\in \mathcal{V}(\rho) \cup \mathcal{V}(\rho_0)}\dfrac{1}{2\{1-r^2(v)\}u^2(v)}\left(nK_1^2\zeta^2+ nK_2^2\zeta^2 + 2|r(v)|Mn(K_1+K_2)\zeta\right)\\
        & \leq An\zeta^2 + Bn\zeta,
    \end{split}
\end{equation*}
where 
\vspace{-5mm}
\begin{eqnarray*}
M & = & \max\limits_{{v\in \mathcal{V}(\rho) \cup \mathcal{V}_0(\rho_0)}, \forall i}\{\mu_{+,i,0}(v), \mu_{-,i,0}(v)\}, \\
A & = & (K_1^2 + K_2^2)\sum\limits_{v\in \mathcal{V}(\rho) \cup \mathcal{V}(\rho_0)} \dfrac{1}{2\{1-r^2(v)\}u^2(v)}, \\
B & = & M(K_1 + K_2)\sum\limits_{v\in \mathcal{V}(\rho) \cup \mathcal{V}(\rho_0)} \dfrac{|r(v)|}{2\{1-r^2(v)\}u^2(v)}.
\end{eqnarray*}
Henceforth, for any $\epsilon>0$, we obtain that,  
\begin{equation*}
\liminf _{n \rightarrow \infty} \Pi\left[\left\{\theta \in O, n^{-1} \sum_{i=1}^{n} K_{n, i}\left(\theta_{0}, \theta\right)<\epsilon\right\}\right]>0.
\end{equation*}

Similarly, we have that, 
\vspace{-3mm}
\begin{equation*}
\footnotesize
\begin{split}
V_{n, i}\left(\theta_{0}, \theta\right) & = \sum\limits_{v\in \mathcal{V}(\rho) \cup \mathcal{V}(\rho_0)} \dfrac{1}{\{1-r^2(v)\}u^2(v)}\left[\{\mu_{+,i}(v) - \mu_{+,i,0}(v)\}^2 + \{\mu_{-,i}(v) - \mu_{-,i,0}(v)\}^2 \right.\\
& \left.\quad +\{r^3(v)-3r(v)\}\{\mu_{+,i}(v) - \mu_{+,i,0}(v)\}\{\mu_{-,i}(v) - \mu_{-,i,0}(v)\}\right], \\
|V_{n, i}\left(\theta_{0}, \theta\right)|  & \leq \sum\limits_{v\in \mathcal{V}(\rho) \cup \mathcal{V}(\rho_0)} \dfrac{1}{\{1-r^2(v)\}u^2(v)}\left(K_1^2\zeta^2+K_2^2\zeta^2 + |r^3(v)-3r(v)| K_1K_2\zeta^2\right)\leq C\zeta^2,
\end{split}
\end{equation*}
where 
\vspace{-3mm}
\[
C = (K_1^2+K_2^2)\sum\limits_{v\in \mathcal{V}(\rho) \cup \mathcal{V}(\rho_0)}\dfrac{1}{\{1-r^2(v)\}u^2(v)} + K_1K_2\sum\limits_{v\in \mathcal{V}(\rho) \cup \mathcal{V}(\rho_0)}\dfrac{|r^3(v)-3r(v)|}{\{1-r^2(v)\}u^2(v)}.
\]
Henceforth, we obtain that, 
\vspace{-3mm}
\begin{equation*}
\left|\sum_{i=1}^n V_{n, i}\left(\theta_{0}, \theta\right)\right| \leq nC\zeta^2\ \mbox{and}\ \frac{1}{n^{2}} \sum_{i=1}^{n} V_{i, n}\left(\theta_{0}, \theta\right) \rightarrow 0, \mbox{as}\  n \rightarrow \infty.
\end{equation*}
This completes the proof of Lemma \ref{lemma: pos_prior}. 
\eop

\vspace{0.2in}
Given the sieves we construct in Equation \eqref{eq: sieves}, we next derive an upper bound for the tail probability, and construct the uniform consistent tests in Lemmas \ref{test1}, \ref{test2}, \ref{test3} and \ref{test4}.

\begin{lemma}
\label{test1}
Suppose $\rho \sim \mathrm{TCGP}(\omega_0, \kappa)$ with $\omega_0 > 0$, the kernel function $\kappa$ satisfies Assumption 2, and $E_{+,i}, E_{-,i} \sim \mathcal{GP}(0,I)$, for $i = 1, \ldots, n$. Then there exist constants $K$ and $b$, such that $pr\left(\Theta_{n}^{C}\right) \leq K \exp (-C_3 n)$.
\end{lemma}

\noindent\textit{Proof}:
Following the same notation as that in the proof of Lemma \ref{lemma1}, we have $\rho(v)=T_\omega\{\xi(v)\} = H[R_\omega\{\xi(v)\}]$. Let $\mathcal{R}_{1}=\{v: \rho(v)>0\}$, and $\mathcal{R}_{-1}=\{v: \rho(v)<0\}$. We have $R_\omega\{\xi(v)\} = \xi(v) > \omega$ when $v\in \mathcal{R}_1$, and $R_\omega\{\xi(v)\} = \xi(v) < -\omega $ when $v\in \mathcal{R}_{-1}$. Then
\begin{footnotesize}
\begin{align}\label{eq: lemma4}
pr\left(\Theta_{n}^{C}\right) \leq \; & pr\left\{\sup _{v \in \mathcal{R}_{1} \cup \mathcal{R}_{-1}}\left|H(\xi(v) )\right|>H\left(m^{{1}/{2d}}\right)\right\} \\
& + \sum_{\tau: 1 \leq\|\tau\|_{1} \leq \alpha} pr\left\{\sup _{v \in \mathcal{R}_{1} \cup \mathcal{R}_{-1}}\left|D^{\tau} H(\xi(v))\right|>m^{1 /2 d}\right\} \nonumber\\
    & + \sum_{i=1}^n pr\left\{\sup _{v \in \mathcal{R}_{1} \cup \mathcal{R}_{-1}}\left|E_{+,i}\right|> m^{{1}/{2d}}\right\} + \sum_{i=1}^n pr\left\{\sup _{v \in \mathcal{R}_{1} \cup \mathcal{R}_{-1}}\left|E_{-,i}\right|> m^{{1}/{2d}}\right\} \nonumber \\
    & + \sum_{i=1}^n \sum_{\tau: 1 \leq\|\tau\|_{1} \leq \alpha} pr\left\{\sup _{v \in \mathcal{R}_{1} \cup \mathcal{R}_{-1}}\left|D^{\tau} E_{+,i}\right|>m^{{1}/{2d}}\right\} \nonumber \\
    & + \sum_{i=1}^n \sum_{\tau: 1 \leq\|\tau\|_{1} \leq \alpha} pr\left\{\sup _{v \in \mathcal{R}_{1} \cup \mathcal{R}_{-1}}\left|D^{\tau} E_{-,i}\right|>m^{{1}/{2d}}\right\}. 
\end{align}
\end{footnotesize}

Since $H(t)$ is a monotonic function,
\begin{equation*}
\footnotesize
    \begin{split}
     pr\left\{\sup _{v \in \mathcal{R}_{1} \cup  \mathcal{R}_{-1}}\left|H(\xi(v) )\right|>H(m^{{1}/{2d}})\right\}
    & \leq pr\left\{\sup _{v \in \mathcal{R}_{1} \cup  \mathcal{R}_{-1}}\left|\xi(v) \right|>m^{{1}/{2d}}\right\} \\
    & \leq K_{1} \exp \left(-b_{1} m^{1 / d}\right) + K_{-1} \exp \left(-b_{-1} m^{1 / d}\right),
     \end{split}
\end{equation*}
where the existence of $K_1, K_{-1}, b_1, b_{-1}$ in the second inequality is ensured by Theorem 5 of \citet{ghosal2006posterior}. 

We next consider the second term in \eqref{eq: lemma4}. Since $|H^\prime(t)| \leq 1$ and $|H''(x)| \leq 2$, we have, 
\begin{equation*}
\footnotesize
    \begin{split}
    & \sum_{\tau: 1 \leq\|\tau\|_{1} \leq \alpha} pr\left\{\sup _{v \in \mathcal{R}_{1} \cup \mathcal{R}_{-1}}\left|D^{\tau} H(\xi(v) - \omega)\right|>m^{{1}/{2d}}\right\}\\
    &\leq pr\left\{\sup _{v \in \mathcal{R}_{1} \cup \mathcal{R}_{-1}}\left|D^{\tau} \xi(v)\right|>{m^{{1}/{2d}}}\right\} + pr\left\{\sup _{v \in \mathcal{R}_{1} \cup \mathcal{R}_{-1}}\left|2\cdot D^{\tau} \xi(v)\right|>{m^{{1}/{2d}}}\right\}\\
    &\leq \sum_{\tau: 0<\|\tau\|_1 \leq \alpha} K_{\tau} \exp \left(-b_{\tau} m^{1 / d}\right).
    \end{split}
\end{equation*}

Denote the sum of the last four terms in \eqref{eq: lemma4} as $S_E$. By Theorem 5 of \citet{ghosal2006posterior} again, there exist $K_{E_+}$, $b_{E_+}$,$K_{E_-}$, $b_{E_-}$, $K_{E_\tau}$ and $b_{E_\tau}$, such that
\begin{footnotesize} 
\[
S_E \leq K_{E_+} \exp(-b_{E_+}m^{1/d}) + K_{E_-} \exp(-b_{E_-} m^{1/d}) + \sum_{\tau: 0<\|\tau\|_1 \leq \alpha} K_{E_\tau} \exp \left(-b_{E_\tau} m^{1 / d}\right).
\]
\end{footnotesize}

Taking $K = K_{-1}+K_{1}+K_{E_+}+K_{E_-}+\sum_{\tau: 0<\|\tau\| \leq \alpha} K_{\tau}+\sum_{\tau: 0<\|\tau\| \leq \alpha} K_{E_\tau}$, and $b = \min \left\{b_{-1}, b_{1},b_{E_+}, b_{E_-}, \min _{1 \leq|\tau| \leq \alpha} b_{\tau},  \min _{1 \leq|\tau| \leq \alpha} b_{E_\tau}\right\}$, we have, 
\[
pr\left(\Theta_{n}^{C}\right)\leq  K \exp \left(-b {m}^{1/d}\right) \leq K \exp \left(-C_3 n\right). 
\] 
This completes the proof of Lemma \ref{test1}. 
\eop

\begin{lemma}
\label{test2}
Suppose Assumption 1 holds. The hypothesis testing problem, 
\begin{equation*}\label{eq: rho}
\begin{split}
& H_0: \rho(v) = \rho_0(v), \;\; E_{\pm,i}(v) = E_{\pm,i,0}(v), \quad i = 1, \ldots, n, \ v \in \mathcal{V}(\rho_1)\cup \mathcal{V}(\rho_0), \\
& H_1: \rho(v) = \rho_1(v), \;\; E_{\pm,i}(v) = E_{\pm,i,1}(v), 
\end{split}
\end{equation*}
is equivalent to the hypothesis testing problem,
\begin{equation*}\label{eq: mu}
\begin{split}
& H^*_0: \mu_{\pm,i}(v) = \mu_{\pm,i,0}(v), \quad i = 1, \ldots, n, \ v \in \mathcal{V}(\rho_1)\cup \mathcal{V}(\rho_0), \\
& H^*_1: \mu_{\pm,i}(v) = \mu_{\pm,i,1}(v),
\end{split}
\end{equation*}
\end{lemma}
where $\mathcal{V}(\rho_1) =\{v: \rho_1(v)\neq 0\}$ and $\mathcal{V}(\rho_0) =\{v: \rho_0(v)\neq 0\}$.

\noindent\textit{Proof}: For any $k\in\{0,1\}$, it is straightforward to see that if $H_k$ holds, then $H_k^*$ also holds. We show that, if $H^*_k$ holds, then $H_k$ also holds. For any $v\in \mathcal{B}_m$, 
\begin{equation*}
\footnotesize
    \begin{split}
        0 &= \sum_{i=1}^n\left[s\{\rho(v)\} E_{+,i}(v) -s\{\rho_k(v)\}E_{+,i,k}(v) \right]^2 \\
        & = \sum_{i=1}^n\left[s\{\rho(v)\}^2 E_{+,i}(v)^2 -2s\{\rho(v)\} s\{\rho_k(v)\}E_{+,i,1}(v) E_{+,i,k}(v) + s\{\rho_k(v)\}^2 E_{+,i,0}(v)^2\right] \\
        & =\left[s\{\rho(v)\} -  s\{\rho_0(v)\}\right]^2\sum_{i=1}^n E^2_{+,i,k}(v) + s\{\rho_k(v)\}s\{\rho(v)\}\sum_{i=1}^n \left\{ E_{+,i}(v) - E_{+,i,k}(v) \right\}^2 \\
        & \quad + s\{\rho_0(v)\}\left[s\{\rho(v)\}- s\{\rho_0(v)\}\right] \sum_{i=1}^n \left\{ E_{+,i}(v)^2 - E_{+,i,k}(v)^2 \right\}, 
    \end{split}
\end{equation*}

By Definition 4, we have $\sum_{i=1}^nE_{+,i}(v)^2 = \sum_{i=1}^nE_{+,i,0}(v)^2 = \sum_{i=1}^nE_{+,i,1}(v)^2$. When $v\in \mathcal{V}(\rho_1)\cup \mathcal{V}(\rho_0)$, $s\{\rho_0(v)\} \geq 0$, $s\{\rho_1(v)\} \geq 0$, and at least one of $s\{\rho_0(v)\}$ and $s\{\rho_1(v)\}$ is not equal to 0,
\begin{equation*}
s\{\rho(v)\}  -  s\{\rho_k(v)\} = 0, \quad  E_{+,i}(v) - E_{+,i,k}(v) = 0, \quad i = 1, \ldots, n.
\end{equation*}
Similarly, we have that $E_{-,i}(v) - E_{-,i,k}(v)= 0 $ for any $v\in \mathcal{V}(\rho_1)\cup \mathcal{V}(\rho_0)$, $i = 1,\ldots, n$. Since $s(\cdot)$ is a monotonic function, $\rho(v) = \rho_k(v)$ for any $v \in \mathcal{B}_m$, which ccompletes the proof of Lemma \ref{test2}.
\eop

\begin{lemma}
\label{test3}
For the hypothesis testing problem,
 \begin{equation*}
        \begin{split}
         & H_0: \mu_{\pm,i}(v_j) = \mu_{\pm,i,0} (v_j), \quad i =1, \ldots, n,\ v_j \in \mathcal{V}(\rho_1)\cup \mathcal{V}(\rho_0),\ j = 1,\ldots, m,\\
          & H_1: \mu_{\pm,i}(v_j) = \mu_{\pm,i,1}(v_j), 
        \end{split}
\end{equation*}
construct the testing statistic, $\Psi_{n} = \Psi_{+ n}+\Psi_{- n}-\Psi_{+ n} \Psi_{- n}$, where 
\begin{align*}
\Psi_{\pm n} &= \max_{i = 1,\ldots, n} \left\{ I\left(\sum_{j=1}^m\delta_{\pm,i}(v_j)(Y_{\pm,i}(v_j) - \mu_{\pm,i,0}(v_j))> 2\left(\frac{m}{C_0}\right)^{\frac{\nu}{d} + \frac{1}{2d}}\right) \right \},
\end{align*}
$\delta_{\pm,i}(v_j)=2I\{\mu_{\pm,i,1}(v_j) \geq \mu_{\pm,i,0}(v_j)\} - 1$, $\nu_0 / 2<\nu<1 / 2$, and $\nu_0, d$, $C_0$ are as defined in  Assumption 3. Write $\mu = \{\mu_{i,\pm}(v_j)\}$, and $\mu_{k} = \{\mu_{i,\pm, k}(v_j)\}$ for $k = 0, 1$. Then, for any $\epsilon_0>0$, there exist constants $C_0$, $C_1$ and $i_* \in\{1,\ldots, n\}$, such that, for any $\mu_{1}$ and $\mu_{0}$ satisfying that $\sum_{j=1}^m |\mu_{+,i_*,1}(v_j)-\mu_{+,i_*,0}(v_j)| >m\epsilon_0$, or $\sum_{j=1}^m|\mu_{-,i_*,1}(v_j)-\mu_{-,i_*,0}(v_j)| >m \epsilon_0$, and $ \mu$ satisfying that $\|\mu -\mu_1 \|_\infty< \epsilon_0/4 $, we have $\Ebb_{\mu_{0}}(\Psi_{n})<C_{0} \exp(-2 n^{2\nu})$ and $\Ebb_{\mu}(\Psi_{n})<C_{0} \exp(-C_1 n)$.
\end{lemma}

\noindent\textit{Proof}: 
To bound the type I error, we have $\Ebb_{\mu_0}(\Psi_n) \leq \Ebb_{\mu_0}(\Psi_{+n}) + \Ebb_{\mu_0}(\Psi_{-n})$. By Assumption 3, we have $\left({m}/{C_0}\right)^{{\nu}/{d}} \geq n^{{\nu}}$. By the definition of $\Psi_{+ n}$, we have that, 
\begin{equation*}
\footnotesize
\begin{split}
    \Ebb_{\mu_0}(\Psi_{+ n}) &\leq {pr}\left(\sum_{j = 1}^m\delta_{+,i_*}(v_j)\{Y_{+,i_*}(v_j) - \mu_{+,i_*,0}(v_j)\}> 2\left(\frac{m}{C_0}\right)^{\frac{\nu}{d} + \frac{1}{2d}}\right)\\
    &= {pr}\left(\sqrt{\frac{C_0}{{m^d}}}\sum_{j=1}^m\delta_{+,i_*}(v_j)\{Y_{+,i_*}(v_j) - \mu_{+,i_*,0}(v_j)\} > 2\left(\frac{m}{C_0}\right)^{\frac{\nu}{d}}\right)\\
    & = 1-\Phi\left(2\left(\frac{m}{C_0}\right)^{\frac{\nu}{d}}\right) \leq 1-\Phi\left(2n^{{\nu}}\right) \leq \dfrac{\phi(2n^\nu)}{2n^\nu} = \dfrac{1}{2\sqrt{2\pi}}\dfrac{\exp(-2n^{2\nu})}{n^{\nu}}.
\end{split}
\end{equation*}

Similarly, we have that $\Ebb_{\mu_0}(\Psi_{-n}) \leq \dfrac{1}{2\sqrt{2\pi}}\dfrac{\exp(-2n^{2\nu})}{n^{\nu}}$. Therefore, 
\vspace{-0.01in}
\[
\Ebb_{\mu_0}(\Psi_n) \leq \dfrac{1}{\sqrt{2\pi}}\dfrac{\exp(-2n^{2\nu})}{n^{\nu}}.
\]

To bound the type II error, we have that,
\begin{equation*}
    \Ebb_{\mu}\left[1-\Psi_{n}\right] \leq \min \left\{\Ebb_{\mu}\left(1-\Psi_{+ n}\right), \Ebb_{\mu}\left(1-\Psi_{- n}\right)\right\}.
\end{equation*}
As such, we only need to show that at least one of the type II error probabilities for $\Psi_{+n}$ and $\Psi_{-n}$ is exponentially small. Suppose $\sum_{j=1}^m|\mu_{+,i_*,0}(v_j)- \mu_{+,i_*,1}(v_j)|>m\epsilon_0$. Since $\sum_{j=1}^m|\mu_{+,i_*}(v_j) - \mu_{+,i_*,1}(v_j)| < m\epsilon_0/4$, we have,
\begin{equation*}
\footnotesize
\begin{split}
        &\Ebb_{\mu}(1-\Psi_{+n}) \\
        &\leq {pr}\left(\sum_{j = 1}^m\delta_{+,i_*}(v_j)\{Y_{+, i_*}(v_j) - \mu_{i_*,+,0}(v_j)\}> 2\left(\frac{m}{C_0}\right)^{\frac{\nu}{d} + \frac{1}{2d}}\right)\\
        & = {pr}\left(\sqrt{\frac{C_0}{m^d}}\sum_{j = 1}^m\delta_{+,i_*}(v_j)\{Y_{+,i}(v_j) - \mu_{+,i,0}(v_j)\} \leq 2\left(\frac{m}{C_0}\right)^{\frac{\nu}{d}}\right)\\
        & = {pr}\left(\sqrt{\frac{C_0}{m^d}}\sum_{j = 1}^m\delta_{+,i_*}(v_j)\{Y_{+,i}(v_j) - \mu_{+,i}(v_j)\} + \sqrt{\frac{C_0}{m^d}}\sum_{j = 1}^m\delta_{+,i_*}(v_j)\{\mu_{+,i}(v_j) - \mu_{+,i,1}(v_j)\}\right.\\
         & \quad + \left.\sqrt{\frac{C_0}{m^d}}\sum_{j = 1}^m\delta_{+,i_*}(v_j)\{\mu_{+,i,1}(v_j) - \mu_{+,i,0}(v_j)\} < 2(m/C_0)^{\nu/d}\right)\\
         & \leq {pr}\left(\sqrt{\frac{C_0}{m^d}}\sum_{j = 1}^m\delta_{+,i_*}(v_j)\{Y_{+,i}(v_j) - \mu_{+,i}(v_j)\} \leq \frac{C_0\epsilon_{0} m^{1 / 2d}}{4}-C_0\epsilon_{0} m^{1 / 2d}+2 (m/C_0)^{\nu/d}\right).
\end{split}
\end{equation*}
Since $\nu < 1/2$, there exists $N > N_0$, such that, for all $n \geq N$, $(m/C_0)^{\nu/d}<C_0 m^{1 / 2d}\epsilon_0/4$. By Assumption 3, this further implies that, 
\begin{equation*}
\footnotesize
\begin{split}
\Ebb_{\mu}(1-\Psi_{+n}) & \leq {pr}\left(\sqrt{\frac{C_0}{m^d}}\sum_{j = 1}^m\delta_{+,i_*}(v_j)\{Y_{+,i_*}(v_j) - \mu_{+,i_*}(v_j)\} \leq -\frac{C_0\epsilon_{0} m^{1 / 2d}}{4}\right)\\
& \leq \Phi\left(-\frac{C_0\epsilon_{0} m^{1 / 2d}}{4}\right) \leq \Phi\left(-\frac{\epsilon_{0} n^{1 / 2}}{4}\right) 
\leq \frac{4 }{\epsilon_{0}(2 \pi n)^{1 / 2}} \exp \left(-\frac{n \epsilon_{0}^{2}}{32}\right).
\end{split}
\end{equation*}

Taking $C_{0}=\max \left\{2^{-1}(2 \pi)^{-1 / 2}, 4 \epsilon_{0}^{-1}(2 \pi)^{-1 / 2}\right\}$ and $C_{1}=\epsilon_{0}^{2}/32$ completes the proof of Lemma \ref{test3}.
\eop

\begin{lemma}
\label{bound}
Suppose Assumption 1, 2 and 3 hold. For any $\epsilon > 0$, there exist $N$, $i$ and $\epsilon_0> 0$, such that, for all $n \geq N$ and all $\theta \in \Theta_n$ that $\left\|\theta-\theta_0\right\|_1>\varepsilon$, we have $\sum_{j=1}^{m}\left|\mu_{\pm,i}\left(v_{j}\right)-\mu_{\pm,i,0}\left(v_{j}\right)\right|>\epsilon_0 m$.
\end{lemma}

\noindent\textit{Proof}: 
We first note that, 
\begin{equation}\label{eq:theta_1norm}
\footnotesize
    \begin{split}
        \left\|\theta-\theta_0\right\|_1 &= \sum\limits_{v\in \mathcal{V}(\rho) \cup \mathcal{V}(\rho_0)}|\rho(v) - \rho_0(v)| + \max\limits_{i =1,\ldots,n}\sum\limits_{v\in \mathcal{V}(\rho) \cup \mathcal{V}(\rho_0)}|E_{+,i}(v) - E_{+,i,0}(v)| \\
        & +\max\limits_{i =1,\ldots,n}\sum\limits_{v\in \mathcal{V}(\rho) \cup \mathcal{V}(\rho_0)}|E_{-,i}(v) - E_{-,i,0}(v)|
    \end{split}
\end{equation}
Since $\left\|\theta-\theta_0\right\|_1 > \epsilon$, at lease one of the three terms in \eqref{eq:theta_1norm} is greater than $\epsilon/3$. Without loss of generality, suppose $\max\limits_{i =1,\ldots,n}\left\{\sum\limits_{v\in \mathcal{V}(\rho) \cup \mathcal{V}(\rho_0)}|E_{+,i}(v) - E_{+,i,0}(v)|\right\} > \epsilon/3$. Then there exist $i$, such that $\sum\limits_{v\in \mathcal{V}(\rho) \cup \mathcal{V}(\rho_0)}|E_{+,i}(v) - E_{+,i,0}(v)| > \epsilon/3$. Therefore, 
\begin{equation}
\footnotesize
\begin{split}
& \sum_{j=1}^{m}\left|\mu_{\pm,i}\left(v_{j}\right)-\mu_{\pm,i,0}\left(v_{j}\right)\right| = \sum_{j=1}^{m}\left|s\{\rho(v_j)\}E_{+,i}(v)- s\{\rho_0(v_j)\}E_{+,i,0}(v)\right|\\
= \; & \sum_{j=1}^{m}\left|s\{\rho(v_j)\}\left\{E_{+,i}(v) - E_{+,i,0}(v)\right\}+ E_{+,i,0}(v)\left[s\{\rho(v_j)\} - s\{\rho_0(v_j)\}\right]\right|\\
> \; & \sum_{j=1}^{m} \left|s\{\rho(v_j)\}\right| \left|E_{+, i}(v_j) - E_{+, i,0}(v_j)\right| - \sum_{j=1}^{m} \left|E_{+, i,0}(v_j)\right| \left|s(\rho(v_j)) - s(\rho_0(v_j))\right|
\end{split}
\end{equation}
By Definition 3, there exists $C_\rho > 0$, such that $|s\{\rho(v_j)\}| > C_\rho$ when $v_j \in \mathcal{V}(\rho) \cup \mathcal{V}(\rho_0)$. By the compactness of $\mathcal{V}(\rho) \cup \mathcal{V}(\rho_0)$, there exists $C$, such that $\max\limits_{j=1,\ldots, m} \left|E_{+, i,0}(v_j)\right| \left|s(\rho(v_j)) - s(\rho_0(v_j))\right|$ $< C$. Therefore,  
\begin{equation*}
\sum_{j=1}^{m}\left|\mu_{+,i}\left(v_{j}\right)-\mu_{+,i,0}\left(v_{j}\right)\right| > C_\rho m\epsilon/3 - mC
\end{equation*}
Taking $\epsilon_0 = C_\rho \epsilon/3 - C$ completes the proof of Lemma \ref{bound}.
\eop

\begin{lemma}
\label{test4}
For any $\epsilon^\star>0$ and $\nu_0<\nu<\frac{1}{2}$, there exist $N, C_0, C_1$ and $C_2$, such that, for all $n > N$ and $\theta \in \Theta_n$, if $\|\theta - \theta_0\|_1 > \epsilon^\star$, a test function $\Psi_n$ can be constructed satisfying that $\Ebb_{\theta_{0}}\left(\Psi_{n}\right) \leq C_{0} \exp \left(-C_{2} n^{2 \nu}\right)$ and $\Ebb_{\theta}\left(1-\Psi_{n}\right) \leq C_{0} \exp \left(-C_{1} n\right)$, where $\nu_0$ is as defined in Assumption 3.
\end{lemma}

\noindent\textit{Proof}:
Let $N_t$ be the $t$ covering number of $\Theta_n$ in the supremum norm. Let $\theta^{1}, \ldots, \theta^{N_{t}} \in \Theta_{n}$ satisfy that, for each $\theta\in \Theta_n$, there exist at least one $l$ such that $\left\|\theta-\theta^{l}\right\|_{\infty}<t$. For any $\theta\in \Theta_n$, define
\begin{equation}
\label{eq: testing_stat}
\Psi_{n}=\max _{1 \leq l \leq N_{t}} \Psi_{n}\left(\theta_{0}, \theta^{l}\right),
\end{equation}
where $\Psi_{n}\left(\theta_{0}, \theta^{l}\right)$ is the test statistic constructed in Lemma~\ref{test3} for the hypothesis testing problem $H_0: \theta = \theta_0$ versus $H_1: \theta = \theta^l$. 
If $\left\|\theta-\theta_{0}\right\|_{1}>\epsilon^\star$, then for $\theta^{l}$ satisfying that $\left\|\theta-\theta^{l}\right\|_{1}<t \leq \epsilon^\star / 2$, we have $\left\|\theta^{l}-\theta_{0}\right\|_{1}>\epsilon^\star / 2$. By Lemma \ref{bound}, there exist $N_{0}^{*}, i$ and $\epsilon>0$, such that $\sum_{j=1}^{m}\left|\mu_{+,i}^{l}\left(v_{j}\right)-\mu_{+,i,0}\left(v_{j}\right)\right|>\epsilon m$. By Lemma \ref{test3}, we can choose $\epsilon_0$, such that 
\begin{equation*}
\footnotesize
\begin{split}
\Ebb_{\theta_{0}}\left\{\Psi_{n}\left(\theta_{0}, \theta^{l}\right)\right\} \leq C_{0} \exp \left(-2 n^{2 \nu}\right), \;\; \textrm{ and } \;\; \Ebb_{\theta}\left\{1-\Psi_{n}\left(\theta_{0}, \theta^{l}\right)\right\} \leq C_{0} \exp \left(-C_{1} n\right).
\end{split}
\end{equation*}

Furthermore, we have, 
\begin{equation*}
\begin{split}
\Ebb_{\theta_{0}}\left(\Psi_{n}\right) & \leq \sum_{l=1}^{N_{t}} \Psi_{n}\left(\theta_{0}, \theta^{l}\right) \leq C_{0} N_{t} \exp \left(-2 n^{2 \nu}\right)=C_{0} \exp \left(\log N_{t}-2 n^{2 \nu}\right) \\ 
& \leq C_{0} \exp \left\{C n^{1 /(2 \alpha)} t^{-d / \alpha}-2 n^{2 \nu}\right\} 
\leq C_{0} \exp \left(C n^{\nu_{0}} t^{-d / \alpha}-2 n^{2 \nu}\right) \\
& = C_{0} \exp \left\{-\left(2-C n^{\nu_{0}-2 \nu} t^{-d / \alpha}\right) n^{2 \nu}\right\}.
\end{split}
\end{equation*}
When $C t^{-d / \alpha}<2$, $\Ebb_{\theta_{0}}\left(\Psi_{n}\right) \leq C_{0} \exp \left\{-\left(2-C t^{-d / \alpha}\right) n^{2 \nu}\right\}$. When $C t^{-d / \alpha} \geq 2$, since $\nu_{0}-2 \nu<0$, there exists $N_1^\star$, such that, for all $n>N_1^*$, $C n^{\nu_{0}-2 \nu} t^{-d / \alpha}<1$. Then 
$\Ebb_{\theta_{0}}\left(\Psi_{n}\right) \leq C_{0} \exp \left\{-n^{2 \nu}\right\}$.
In addition, 
\begin{footnotesize}
    \begin{equation*}
        \Ebb_{\theta}\left(1-\Psi_{n}\right)=\Ebb_{\theta}\left[\min _{1 \leq l \leq N_{t}}\left\{1-\Psi_{n}\left(\theta_{0}, \theta^{l}\right)\right\}\right] \leq \Ebb_{\theta}\left[\left\{1-\Psi_{n}\left(\theta_{0}, \theta^{l}\right)\right\}\right] \leq C_{0} \exp \left(-C_{1} n\right)
    \end{equation*}
\end{footnotesize}

Taking $C_{2}=\left(2-C t^{-d / \alpha}\right) I\left(C t^{-d / \alpha}<2\right)+I\left(C t^{-d / \alpha} \geq 2\right)>0$, and $N=\max \left\{N_{1}^{*}, N_{0}^{*}\right\}$ completes the proof of Lemma \ref{test4}.
\eop

\section{Derivations of Posterior Computation}
\label{appen: full}

\subsection{Full conditional distribution}

We first summarize in Algorithm \ref{alg:proof of pro} the general procedure of deriving the full conditional distribution of $\theta$ using Proposition 2. The main steps are to first rewrite the density of $\theta$ in the form of (15), where $\{L_p\}_{p=1}^P$, $\{U_k\}_{k=1}^K$, $\{f_p(\theta)\}_{p=1}^P$, $\{h_k(\theta)\}_{k=1}^K$ are the input to Algorithm \ref{alg:proof of pro}. We then sort $\left(L_{1},\ldots, L_{P}, U_{1},\ldots, U_{K} \right)$ in ascending order, which leads to $P+K + 1$ intervals. We next loop through all the intervals, and update the coefficient of $H_i(\theta)$. Finally, after obtaining the unnormalized conditional density function of $\theta$ on each interval, we  derive the full conditional density of $\theta$ by incorporating the corresponding normalizing constants.

\begin{algorithm}[t!]
\caption{Full conditional distribution of $\theta$}
\label{alg:proof of pro}
\textbf{Input}: $\{L_p\}_{p=1}^P$, $\{U_k\}_{k=1}^K$, $\{f_p(\theta)\}_{p=1}^P$, $\{h_k(\theta)\}_{k=1}^K$. \\
\textbf{Output}: the full conditional distribution of $\theta$.\\
    {Sort $\left(L_{1},\ldots, L_{P}, U_{1},\ldots, U_{K} \right)$ in ascending order, which leads to $P+K + 1$ intervals, denoted as $I_1, I_2, \ldots, I_{P+K+1}$.}\\
   \For{interval $I_i$, $i=1,\ldots,P+K + 1$}{
        {Initialize $D_i= E_i=F_i=0$}\\
        \For{$p = 1,\ldots, P$, $k = 1,\ldots, K$}{
            \uIf{$I_i \subset [L_p, +\infty)$}{
                {$D_i = D_i+a_{1p}$, $E_i =E_i+a_{2p}$, $F_i=F_i+a_{3p}$.}
            }
            \uIf{$I_i \subset (-\infty, U_k]$}{
                {$D_i = D_i+b_{1k}$, $E_i =E_i+ b_{2k}$, $F_i =F_i+ b_{3k}$.}
            }
        }
        {Write $H_i(\theta) = D_i \theta^2 + E_i \theta + F_i$.}\label{state:h}
   }
    \uIf{there exists $i$, such that $D_i \neq 0$}{
       {the full conditional distribution of $\theta$ is a mixture of truncated normal distributions.}
    }
    \uIf{$D_i= 0$ for all $i$, and there exists $i$, such that $E_i \neq 0$}{
       {the full conditional distribution of $\theta$ is a mixture of truncated exponential distributions.}
    }
    \uIf{$D_i=E_i= 0$ for all $i$, and there exists $i$, such that $F_i \neq 0$}
       {the full conditional distribution of $\theta$ is a mixture of uniform distributions.}
\end{algorithm}

\subsection{Full conditional distribution of $c_l$}
\label{appen: c}

Without loss of generality, we only consider $c_1$ in the following discussion. By model (9) and the Karhunen-Lo{\`e}ve expansion, we have  \\$ \mu_{\pm,i}(v) = G_\omega\left\{\pm\xi(v)\right\} E_{\pm,i}(v)$, $ \xi(v)  = \sum_{l = 1}^{L}c_l \psi_l(v)$, \\and $E_{\pm,i}(v) = \sum_{l=1}^{L} e_{i,l,\pm} \psi_l(v)$. Given $Y_+, Y_-$, $\tilde{\Theta}_{\backslash c_1}$, the full conditional density of $c_1$ is,  
\begin{equation}\label{eq: full conditional of c}
\pi(c_1\mid Y_{+},Y_{-}, \tilde{\Theta}_{\backslash c_1})\propto \exp\left(-\sum_{v\in \mathcal{B}_m}\dfrac{\sum_{i=1}^n W_i(v)}{K(v)}\right)\cdot \exp\left(-\dfrac{c_1^2}{2\lambda_l}\right),
\end{equation}
where $W_i(v) =\{ Y_{+,i}(v) - \mu_{+,i}(v )\}^2 + \{ Y_{-,i}(v) - \mu_{-,i}(v) \}^2 +\\ 2r(v) \{ Y_{+,i}(v)\mu_{-,i}(v) + Y_{-,i}(v)\mu_{+,i}(v) \}$, and $K(v) =2\{1-r^2(v)\}u^2(v)$, \\with $r(v) = \{\tau^2_1(v) - \tau^2_2(v)\}/\{\tau^2_1(v) + \tau^2_2(v)\}$ and $u^2(v) =  \{\tau^2_1(v) + \tau^2_2(v)\}/4$. Write  $T_{\pm}(v) = \{\pm\lambda_1 - \sum_{l = 2}^{L}c_l\psi_l(v)\}/\{\psi_1(v)\}$. According to the sign of $\psi_1(v)$, we have two different representations of $\sum_{i=1}^n W_i(v)$. 

When $\psi_1(v) > 0$,  
\begin{equation*} \label{eq: Wi_first situation}
\begin{split}
\sum_{i = 1}^n W_i(v) &= \{A_+(v)c_1^{2} + B_+(v)c_1 + C_+(v)\}I\{c_1 > T_{+}(v)\}\\
&\qquad + \{A_-(v)c_1^{2} + B_-(v)c_1 + C_-(v)\}I\{c_1 < T_{-}(v)\}.
\end{split}
\end{equation*}

When $\psi_1(v) < 0$,  
\begin{equation*} \label{eq: Wi_second situation}
\begin{split}
\sum_{i = 1}^n W_i(v) &= \{A_+(v)c_1^{2} + B_+(v)c_1 + C_+(v)\}I\{c_1 < T_{+}(v)\}\\
&\qquad + \{A_-(v)c_1^{2} + B_-(v)c_1 + C_-(v)\}I\{c_1 > T_{-}(v)\}. 
\end{split}
\end{equation*}
where $A_{\pm}(v), B_{\pm}(v), C_{\pm}(v)$ are all functions of $\Tilde{\Theta}_{\backslash c_1}$, and are of the form, 
\begin{equation*}
\footnotesize
\begin{split}
A_\pm(v) & = \left\{\sum_{i=1}^n E_{\pm,,i}(v)^2\right\} \cdot \psi_1^2(v),\\
B_\pm(v) & = 2\psi_1(v)\left[\left\{\sum_{l = 2}^{L}c_l\psi_1(v)\right\}\left\{\sum_{i=1}^n E_{\pm,i}(v)^2\right\}\right.  \\
 &\left.\mp \sum_{i=1}^n \left\{Y_{\pm,i}(v)\cdot E_{\pm,i}(v)\right\} \mp r(v)\sum_{i=1}^n \left\{Y_{\mp,i}(v) \cdot E_{\pm,i}(v)\right\}\right],\\
C_\pm(v) & = \left\{\sum_{l = 2}^{L}c_l\psi_1(v)\right\}^2\left\{\sum_{i=1}^n E_{\pm,i}(v)^2 \mp \dfrac{2\cdot\sum_{i=1}^n Y_{\pm,i}(v)E_{\pm,i}(v)}{\sum_{l = 2}^{L}c_l\psi_1(v)} \pm \dfrac{2r(v)\sum_{i=1}^n Y_{\mp,i}(v)E_{\pm,i}(v)}{\sum_{l = 2}^{L}c_l\psi_1(v)}\right\}.
 \end{split}
\end{equation*}

\begin{algorithm}[t!]
\caption{Full conditional distribution of $c_l$}
\label{alg: procedure of c}
\textbf{Input}: $P = K = m$, where $m$ is the number of spatial locations, \\
$L_p=\left\{\begin{array}{ll} T_{+}(v_j) & \text{if } \psi_l(v_j) > 0 \\ T_{-}(v_j) & \text{if } \psi_l(v_j) < 0 \end{array}\right.$, $U_k=\left\{\begin{array}{ll} T_{-}(v_j) & \text{if } \psi_l(v_j) > 0 \\ T_{+}(v_j) & \text{if } \psi_l(v_j) < 0 \end{array}\right.$, \\
$f_p(\theta)=\left\{\begin{array}{ll} g_{+}(c_l; v_j) & \text{if } \psi_l(v_j) > 0 \\ g_{-}(c_l; v_j) & \text{if } \psi_l(v_j) < 0 \end{array}\right.$, $h_k(\theta)=\left\{\begin{array}{ll} g_{-}(c_l; v_j) & \text{if } \psi_l(v_j) > 0 \\ g_{+}(c_l; v_j) & \text{if } \psi_l(v_j) < 0 \end{array}\right.$. \\
\textbf{Output}: the full conditional distribution of $c_l$.\\
{Follow the procedure in Algorithm \ref{alg:proof of pro}.}
\end{algorithm}
Therefore, given $Y_{+}$, $Y_{-}$, $\tilde{\Theta}_{\backslash c_1}$ and the eigenfunctions $\{\psi_1(v_j)\}_{j=1}^m$ evaluated on $\mathcal{B}_m$,  
\begin{equation*}
\footnotesize
\begin{split}
\pi(c_1\mid Y_{+}, Y_{-}, \tilde{\Theta}_{\backslash c_1}) \propto & \exp\left(\sum_{\substack{j=1 \\ \psi_1(v_j)>0}}^m \left[g_{+}(c_1; v_j)I\{c_1>T_{+}(v_j)\}+ g_{-}(c_1; v_j)I\{c_1<T_{-}(v_j)\}\right]\right.\\
& + \left.\sum_{\substack{j=1 \\ \psi_1(v_j)<0}}^m\left[g_{+}(c_1; v_j)I\{c_1<T_{+}(v_j)\}+ g_{-}(c_1; v_j)I\{c_1>T_{-}(v_j)\}\right]\right),
\end{split}
\end{equation*}
where 
\vspace{-0.1in}
\begin{equation*}
\begin{split}
g_{\pm}(c_1; v_j) & = \left\{-\dfrac{A_\pm(v_j)}{K(v_j)} - \dfrac{1}{2\lambda_1^2}\right\}c_1^2 + \dfrac{B_\pm(v_j)}{K(v_j)}c_1 + \dfrac{C_\pm(v_j)}{K(v_j)}. 
\end{split}
\end{equation*}

By Proposition 1, the full conditional distribution of $c_1$ is a mixture of truncated normal distributions. We summarize the procedure of obtaining this distribution in Algorithm \ref{alg: procedure of c}.

\subsection{Full conditional distribution of $\omega$}
\label{appen: omega}

Recall that the prior of $\omega$ is the uniform distribution on $[a_\omega,b_\omega]$. Then we have, 
\begin{equation}\label{eq: full conditional of lambda}
\begin{split}
\pi(\omega\mid Y_{+},Y_{-}, \tilde{\Theta}_{\backslash \omega}) \propto \exp\left\{-\sum_{v\in \mathcal{B}_m}\dfrac{\sum_{i=1}^n W_i(v)}{K(v)}\right\}\cdot \dfrac{1}{b_\omega-a_\omega} I(a_\omega \leq \omega \leq b_\omega),
\end{split}
\end{equation}
where $W_i(v)$ is defined as in \eqref{eq: full conditional of c}. Then, 
\vspace{-3mm}
\begin{equation*}
\sum_{i = 1}^n W_i(v) = Q_+(v)I\{\omega < \xi(v)\}+Q_-(v)I\{\omega < -\xi(v)\},
\end{equation*}
where
\vspace{-3mm}
\begin{equation*}
\begin{split}
    Q_\pm(v) = & \xi(v)^2\left\{\sum_{i=1}^n E_{\pm,i}(v)^2\right\} \mp 2 \xi(v)\left\{\sum_{i=1}^n Y_{\pm,i}(v)E_{\pm,i}(v)\right\} \\
    &\pm 2r(v)\xi(v)\left\{\sum_{i=1}^n Y_{\mp,i}(v)E_{\pm,i}(v)\right\}.
    \end{split}
\end{equation*}

\begin{algorithm}
\SetAlgoLined
\caption{Full conditional distribution of $\omega$}
\label{alg: procedure of lambda}
\textbf{Input}: $P=0$, $K = 2m$, \\
 $U_k=\left\{\begin{array}{ll} \xi(v_j), & \text{if } a_\omega<\xi(v_j)<b_\omega\\  -\xi(v_j) & \text{if } a_\omega<-\xi(v_j)<b_\omega \end{array}\right.$, $h_k(\theta)=\left\{\begin{array}{ll} C_{+}(v_j), & \text{if } U_k=\xi(v_j)\\ C_{-}(v_j) & \text{if } U_k=-\xi(v_j) \end{array}\right.$. \\
\textbf{Output}: the full conditional distribution of $\omega$
{Follow the procedure in Algorithm \ref{alg:proof of pro}}
\end{algorithm}

Therefore, given $Y_+, Y_-$, $\tilde{\Theta}_{\backslash \omega}$ and the eigenfunctions $\psi_l(v_j), j = 1,\ldots, m, l = 1, \ldots, L$, evaluated on $\mathcal{B}_m$, we have, 
\begin{equation*}
    \begin{split}
    &\pi(\omega\mid Y_+, Y_-, \tilde{\Theta}_{\backslash \omega})\\
    &\propto \exp\left[\sum_{\substack{j=1 \\ a_\omega<\xi(v_j)<b_\omega}}^m C_{+}(v_j)I\{\omega < \xi(v_j)\} + \sum_{\substack{j=1 \\ a_\omega<-\xi(v_j)<b_\omega}}^m C_{-}(v_j)I\{\omega < -\xi(v_j)\}\right],
    \end{split}
\end{equation*}
where $C_{\pm}(v_j) = -\dfrac{Q_\pm(v_j)}{K(v_j)} - \log(b_\omega-a_\omega)$, and we only consider those $\xi(v_j)$ and $-\xi(v_j)$ that are between $a_\omega$ and $b_\omega$.

By Proposition 1, the full conditional distribution of $\omega$ is a mixture of uniform distributions. We summarize the procedure of obtaining this distribution in Algorithm \ref{alg: procedure of lambda}.

\subsection{Full conditional distribution of $e_{i,l \pm}$}
\label{appen: e}
Since $e_{i,l,+}$ only exist in $\mu_{+,i}(v)$, we can rewrite $\mu_{+,i}(v)$ as $\mu_{+,i}(v) = a_{+,i}(v)+b_{+,i}(v)$, where $a_{+,i}(v) = G_\omega\left\{ \sum_{l=1}^{L} c_l \psi_l(v)\right\} e_{i,l,+}\psi_l(v) = C_{l,+}(v) \cdot e_{i,l,+}$, and $b_{+,i}(v) = G_\omega\left\{ \sum_{l=1}^{L} c_l\psi_l(v)\right\}$ $\sum_{l'\neq l}e_{i,l',+}\psi_{l'}(v)$.  Note that $b_{+,i}(v)$ does not depend on $e_{i,l,+}$. Henceforth, we have that,
\vspace{-3mm}
\begin{footnotesize}
\begin{align*}  
\lefteqn{\{Y_{+,i}(v) - \mu_{+,i}(v)\}^2  =}\\
&\quad Y_{+,i}^2(v) + a_{+,i}^2(v) + b_{+,i}^2(v) + 2a_{+,i}(v)b_{+,i}(v)-2Y_{+,i}(v)a_{+,i}(v)-2Y_{+,i}(v)b_{+,i}(v), \\
\lefteqn{\{Y_{+,i}(v) - \mu_{+,i}(v)\}\{Y_{-,i}(v) - \mu_{-,i}(v) \} =}\\
&\quad Y_{+,i}(v)\{Y_{-,i}(v) - \mu_{-,i}(v)\} -  a_{+,i}(v)\{Y_{-,i}(v) - \mu_{-,i}(v)\} - b_{+,i}(v)\{Y_{-,i}(v) - \mu_{-,i}(v)\}.   
\end{align*}
\end{footnotesize}
Ignoring the terms $\{Y_{-,i}(v) - \mu_{-,i}(v)\}^2$ that do not contain $e_{i,l,+}$, we have, 
\begin{equation*}
\footnotesize
\begin{split}
  \lefteqn{\pi(e_{i,l +} \mid  Y_{+},Y_{-},\tilde{\Theta}_{\backslash e_{i,l +}})}\\
& \propto \prod_{v\in \mathcal{B}_m}\exp\left({-\dfrac{a_{+,i}^2(v) + 2a_{+,i}(v)[b_{+,i}(v)-Y_{+,i}(v) - r(v)\{Y_{-,i}(v)-\mu_{-,i}(v)\}]}{2\{1-r^2(v)\}u^2(v)}}\right)\\
& \quad \cdot \exp\left(-\dfrac{e_{i,l,+}^2}{2\lambda_l}\right)\\ 
&\propto \exp\left[-\dfrac{1}{2}\dfrac{\{e_{i,l,+} - M_{i,l,+}\}^2}{V^2_{i,l,+}}\right].\label{eq: e+}
\end{split}
\end{equation*}
where the mean and the variance are
\vspace{-3mm}
\begin{eqnarray*}
M_{i,l,\pm} & = & \sum\limits_{v\in \mathcal{B}_m}\left[\{\lambda_l m_{i,l,\pm}(v)\}/\{{\lambda_l}+{\sigma^2_{i,l,\pm}(v)}\}\right], \\
V^2_{i,l,\pm} & = & \sum\limits_{v\in \mathcal{B}_m}\left[{\lambda_l\sigma^2_{i,l,\pm}(v)}/\{\lambda_l + \sigma^2_{i,l,\pm}(v)\}\right], 
\end{eqnarray*}
with $m_{i,l,\pm}(v) = -\left[{\left\{Y_{\pm,i}(v) - b_{\pm,i}(v)\right\} - r(v) \cdot \{Y_{\pm,i}(v) - \mu_{\pm,i}(v)\}}\right]/{C_{l, \pm}(v)}$, and $\sigma_{i,l,\pm}^2(v)$ $= {\{1-r^2(v)\}u^2(v)}/{C^2_{l,\pm}(v)}$. Therefore, $e_{i,l \pm}$ follows a normal distribution, i.e., 
\[
e_{i,l \pm} \mid  Y_{+},Y_{-},\tilde{\Theta}_{\backslash e_{i,l \pm}} \sim \mathrm{N}(M_{i,l,\pm}, V_{i,l,\pm}^2).
\]

\subsection{Full conditional distribution of $\tau_1^2(v)$ and $\tau_2^2(v)$}
\label{appen: tau}
For a given $v_0\in \mathcal{B}_m$, we have, 
\begin{equation*}\label{eq: full condition tau1}
    \footnotesize
    \begin{split}
        & \quad
        \pi\left\{\tau_1^2(v_0)\mid Y_{+},Y_{-}, \tilde{\Theta}_{\backslash \tau_1^2(v_0)}\right\} \\
        & \propto \prod_{i=1}^n \dfrac{1}{ \sqrt{\tau_1^2}} \cdot \exp\left[-\frac{1}{2}\left(\frac{1}{\tau_1^2} +\frac{1}{\tau_2^2}\right)\left\{\tilde Y_{+,i}(v_0)^2 + \tilde Y_{-,i}(v_0)^2-2\frac{\tau_1^2 - \tau_2^2}{\tau_1^2 + \tau_2^2}\tilde Y_{+,i}(v_0)\tilde Y_{-,i}(v_0)\right\}\right] \\
        & \quad \cdot \Gamma_{\tau_1^2}^{-1}(a_\tau, b_\tau)\\
        & \propto \left\{\dfrac{1}{\tau_1^2(v_0)}\right\}^{\frac{n}{2}} \exp\left[-\frac{1}{2\tau_1^2(v_0)}\sum_{i=1}^n\left\{Y_{+,i}(v_0) - \mu_{+,i}(v_0) + Y_{-,i}(v_0) - \mu_{-,i}(v_0)\right\}^2\right]\\
    \end{split}
\end{equation*}
where $\tilde Y_{\pm,i}(v_0) = Y_{\pm,i}(v_0) - \mu_{\pm,i}(v_0)$. Therefore, we have, 
\vspace{-0.01in}
\begin{equation*}
\label{eq: tau1}
    \tau_1^2(v_0)\mid Y_{+},Y_{-}, \tilde{\Theta}_{\backslash \tau_1^2(v_0)} \sim \mathrm{IG}\left(a_\tau + \frac{n}{2}, \dfrac{\sum_{i=1}^n \{\tilde Y_{+,i}(v_0)+ \tilde Y_{-,i}(v_0)\}^2}{2} + nb_\tau\right).
\end{equation*}
Similarly, we have, 
\begin{equation*}
\label{eq: tau2}
    \tau_2^2(v_0)\mid Y_{+},Y_{-}, \tilde{\Theta}_{\backslash \tau_2^2(v_0)} 
    \sim \mathrm{IG}\left(a_\tau + \frac{n}{2}, \dfrac{\sum_{i=1}^n \left\{\tilde Y_{+,i}(v_0) - \tilde Y_{-,i}(v_0)\right\}^2}{2} + nb_\tau\right).
\end{equation*}

\subsection{Derivation of hybrid mini-batch MCMC}
\label{appen: mini}

We derive the acceptance ratio in the hybrid mini-batch MCMC. Let $Y = \{Y_{1i}(v), Y_{2i}(v),$ $i = 1,\ldots, n, v\in \mathcal{B}_m\}$, $Y_{m_s} = \{Y_{1i}(v), Y_{2i}(v), i = 1, \ldots, n, v \in \mathcal{B}_{m_s},\}$, and $\Tilde{\Theta} = \{\theta, \Tilde{\Theta}_{\backslash \theta}\}$, where $m_s<m$, and henceforth $\mathcal{B}_{m_s}\subset \mathcal{B}_{m}$. In the Gibbs sampler, we use the full conditional distribution $P(\theta | Y, \Tilde{\Theta}_{\backslash \theta})$ as the proposal function, with the acceptance ratio equal to 1. In the hybrid mini-batch MCMC, we use $P(\theta | Y_{m_s}, \Tilde{\Theta}_{\backslash \theta})$ as the proposal function, and the acceptance ratio becomes,
\begin{equation*}
\begin{split}
        \phi(\theta', \theta) &= \min\left\{1,\dfrac{P(Y|\theta', \Tilde{\Theta}_{\backslash\theta})}{P(Y|\theta, \Tilde{\Theta}_{\backslash\theta})}\dfrac{P(\theta|Y_{m_s}, \Tilde{\Theta}_{\backslash\theta})}{P(\theta'|Y_{m_s}, \Tilde{\Theta}_{\backslash\theta})}\right\}\\
        &= \min\left\{1,\dfrac{\prod_{v\in \mathcal{B}_m} P(Y(v)|\theta',\Tilde{\Theta}_{\backslash \theta})}{\prod_{v\in \mathcal{B}_m}  P(Y(v)|\theta,\Tilde{\Theta}_{\backslash \theta})}\cdot \dfrac{\prod_{v\in \mathcal{B}_{m_s}}  P(Y(v)|\theta,\Tilde{\Theta}_{\backslash \theta})p(\theta)}{\prod_{v\in \mathcal{B}_{m_s}}  P(Y(v)|\theta',\Tilde{\Theta}_{\backslash \theta})p(\theta')}\right\}\\
        & = \min\left\{1,\dfrac{\prod_{v\notin \mathcal{B}_{m_s}}  P(Y(v)|\theta',\Tilde{\Theta}_{\backslash \theta})}{\prod_{v\notin \mathcal{B}_{m_s}}  P(Y(v)|\theta,\Tilde{\Theta}_{\backslash \theta})}\right\}.
\end{split}
\end{equation*}

\subsection{Posterior computation algorithms}
We summarize the Gibbs sampling for the TCGP in Algorithm~\ref{alg: sampling}, and the hybrid mini-batch MCMC procedure in Algorithm \ref{alg: hybridsampling_c}
\begin{algorithm}
\SetAlgoLined
\caption{Gibbs sampling for TCGP}
\label{alg: sampling}
\textbf{Input}: the observed imaging data $Y = \{\{Y_{1,i}(v), Y_{2,i}(v)\}_{i = 1}^n, v\in \mathcal{B}_m\}$,\\
 \quad \quad \, \, the kernel function $\kappa(\cdot, \cdot)$,  \\
 \quad \quad \, \, the Karhunen-Lo{\`e}ve truncation number $L$, \\
 \quad \quad \, \, the prior hyperparameters $a_\tau, b_\tau, a_\omega, b_\omega$.\\
\textbf{Output}: the posterior samples of $\Tilde{\Theta} = \{ \{c_l\}_{l=1}^{L}, \{e_{i,l,\pm}\}_{l=1,i=1}^{L,n}, \{ \tau_1^2(v), \tau_2^2(v) \}_{v \in \mathcal{B}_m}, \omega \}$.\\
\textbf{Initialize} $\Tilde{\Theta}$: sample $\Tilde{\Theta}$ from the prior distribution.\\
\For{$t = 1,\cdots, T$}{
{parallel sample ${\tau^2_k(v)}$ from the inverse Gamma distribution, $v\in \Bcal_m, k = 1,2$.}}
\For{$l = 1, \ldots, L$}{
   {sample $c_l$ from the mixture of truncated normal distributions.}\\
   {sample $\omega$ from the mixture of uniform distributions.}\\
   {sample $e_{i,l,\pm}$ from the normal distribution, $i = 1, \ldots, n$. }}
\end{algorithm}

\begin{algorithm}
\setstretch{1.35}
\SetAlgoLined
\caption{Hybrid mini-batch MCMC for TCGP.}
\label{alg: hybridsampling_c}
\textbf{Input}: the observed imaging data $Y = \{\{Y_{1,i}(v), Y_{2,i}(v)\}_{i = 1}^n, v\in \mathcal{B}_m\}$,\\
  \quad \quad \,  \, the kernel function $\kappa(\cdot, \cdot)$, \\
  \quad \quad \,  \, the Karhunen-Lo{\`e}ve truncation number $L$, \\
  \quad \quad \,  \, the prior hyperparameters $a_\tau, b_\tau, a_\omega, b_\omega$.\\
\textbf{Output}: the posterior samples of $\Tilde{\Theta} = \{ \{c_l\}_{l=1}^{L}, \{e_{i,l,\pm}\}_{l=1,i=1}^{L,n}, \{ \tau_1^2(v), \tau_2^2(v) \}_{v \in \mathcal{B}_m}, \omega \}$.\\
{\textbf{Initialize} $\Tilde{\Theta}$: sample $\Tilde{\Theta}$ from the prior distribution.}\\
\For{$t = 1,\cdots, T$}{
    {parallel sample ${\tau^2_k(v)}$ from the inverse Gamma distribution, for all $v\in \Bcal_m, k = 1,2$.}\\
    {random sample $m_s$ locations from $\mathcal{B}_m$ and form $\mathcal{B}_{m_s}$ and $Y_{m_s}$.}}
    \For{$l = 1,\cdots, L$}{
            \uIf{$t\mod{T_0} = 0$}{
                {sample $c_l$ from the mixture of truncated normal distributions based on $Y$.}   \\
                {sample $\omega$ from the mixture of uniform distributions based on $Y$.} }  
            \uElse
            {sample $c_l^{(t)}$ from the mixture of truncated normal distributions based on $Y_{m_s}$.  \\
            {accept $c_l^{(t)}$ with probability $\phi(c_l^{(t)},c_l^{(t-1)})$.}\\
            {sample $\omega^{(t)}$ from the mixture of uniform distributions based on $Y_{m_s}$.}  \\  
            {accept $\omega^{(t)}$ with probability $\phi(\omega^{(t)},\omega^{(t-1)})$.}}
            {{parallel} sample $e_{i,l,\pm}$ from the normal distribution, $i = 1, \ldots, n$. }}
\end{algorithm}

\section{Additional numerical results}

\subsection{2D image simulation}
\label{subs: 2dsim}
We simulate the data from model \eqref{eq:svcm}, with the sample size $n = 50$, and the image resolution $m = 64 \times 64$. We simulate the mean $\mu_{k,i}$ from \eqref{eq:sdgp} and \eqref{eq:covkern}, $k= 1,2$, with $\kappa(v, v') = \exp{-0.1(v^2+v'^2)-10(v-v')^2}$, $\sigma_+^2(v) = {\zeta_+} \sum_{j=1}^3 I(\|v - u_{+,j}\|_1 < 0.1)$, where $u_{+,1} = (0.3, 0.7)$, $u_{+,2} = (0.7, 0.7)$, $u_{+,3} = (0.3, 0.3)$, and $\sigma_-^2(v) = {\zeta_-} \{I(\|v - u_{-,1}\|_1 < 0.1) + I(\|v - u_{-,2}\|_2 < 0.1) \} $, where $u_{-,1} = (0.5, 0.5)$, $u_{-,2} = (0.7, 0.3)$. Here $(\zeta_+, \zeta_-)$ controls the signal strength, and we consider two settings, with  $(\zeta_+, \zeta_-) = (0.15, 0.25)$ for a weak signal, and $(\zeta_+, \zeta_-) = (0.75, 0.85)$ for a strong signal. We simulate the noise $\varepsilon_{k,i}$ from the normal distribution with mean zero and variance $\tau^2_k(v)$, and simulate $\log(\tau_k^2(v))$ from a Gaussian process with mean zero and correlation kernel $\kappa(v, v')$, $k= 1,2$. 

\begin{table}[t!]
\caption{Results of 2D image simulations. Reported are the average sensitivity, specificity, and FDR, with standard error in the parenthesis, based on 100 data replications. Six methods are compared: the voxel-wise analysis, the region-wise analysis, the integrated method of \citet{li2018spatially} with two thresholding values, 0.95 and 0.90, and the proposed Bayesian method Thresholded Correlation Gaussian Process (TCGP) with the Gibbs sampler and the hybrid mini-batch MCMC.}
\label{tab:sim2d}
\vspace{-0.15in}
\begin{center}
\resizebox{\textwidth}{!}{
\begin{tabular}{clcccccc}
\toprule 
\multirow{2}{*}{Signal}&\multirow{2}{*}{Method}&\multicolumn{3}{c}{Positive Correlation}&\multicolumn{3}{c}{Negative Correlation}\\
 & & Sensitivity & Specificity & FDR & Sensitivity & Specificity & FDR \\
\midrule
Weak&  Voxel-wise &     0.000 (0.000) &   1.000 (0.000) &   0.020 (0.010) &     0.000 (0.001) &  1.000 (0.001) & 0.010 (0.001)\\
        &   Region-wise&    0.238 (0.001) &   0.953 (0.002) &   0.447 (0.002) &     0.473 (0.002) &  0.956 (0.003) & 0.629 (0.004)\\ 
        &   Integrated (0.95) &    0.612 (0.001) &   0.994 (0.000) &   0.134 (0.010) &     0.844 (0.003) &  0.993 (0.000) & 0.131 (0.003)\\
        &   Integrated (0.90) &    0.821 (0.001) &   0.971 (0.000) &   0.341 (0.010) &     0.963 (0.003) &  0.966 (0.000) & 0.398 (0.006)\\
        &   TCGP (Gibbs)     &    0.855 (0.003) &   0.996 (0.001) &   0.057 (0.008) &     0.997 (0.002) &  0.993 (0.001) & 0.108 (0.005)\\
        &   TCGP (Hybrid) &    0.851 (0.006) &   0.993 (0.001) &   0.092 (0.010) &     0.993 (0.002) &  0.992 (0.001) & 0.126 (0.005)\\
\midrule 
Strong&  Voxel-wise &     0.062 (0.002) &   1.000 (0.000) &   0.000 (0.014) &     0.091 (0.002) &  1.000 (0.000) & 0.000 (0.006)\\
        &  Region-wise&     0.741 (0.002) &   0.852 (0.003)  &   0.747 (0.004) &     0.479 (0.002) &  0.950 (0.002) & 0.645 (0.003)\\
        &  Integrated (0.95) &     0.773 (0.001) &   0.998 (0.000) &   0.036 (0.002) &     0.933 (0.002) &  0.996 (0.000) & 0.067 (0.001)\\
        &  Integrated (0.90) &     0.996 (0.020) &   0.959 (0.000) &   0.378 (0.017) &     0.999 (0.020) &  0.953 (0.000) & 0.468 (0.001)\\
        &  TCGP (Gibbs)     &     0.976 (0.002) &   0.999 (0.000) &   0.015 (0.004) &     1.000 (0.001) &  0.999 (0.000) & 0.018 (0.001)\\
        &  TCGP (Hybrid) &     0.960 (0.003) &   0.997 (0.001) &   0.049 (0.005) &     0.990 (0.001) &  0.999 (0.000) & 0.023 (0.002)\\
\bottomrule 
\end{tabular}
}
\end{center}
\end{table}
   
\begin{figure}[b!]
\begin{center}
\includegraphics[width=\textwidth, height=2.0in]{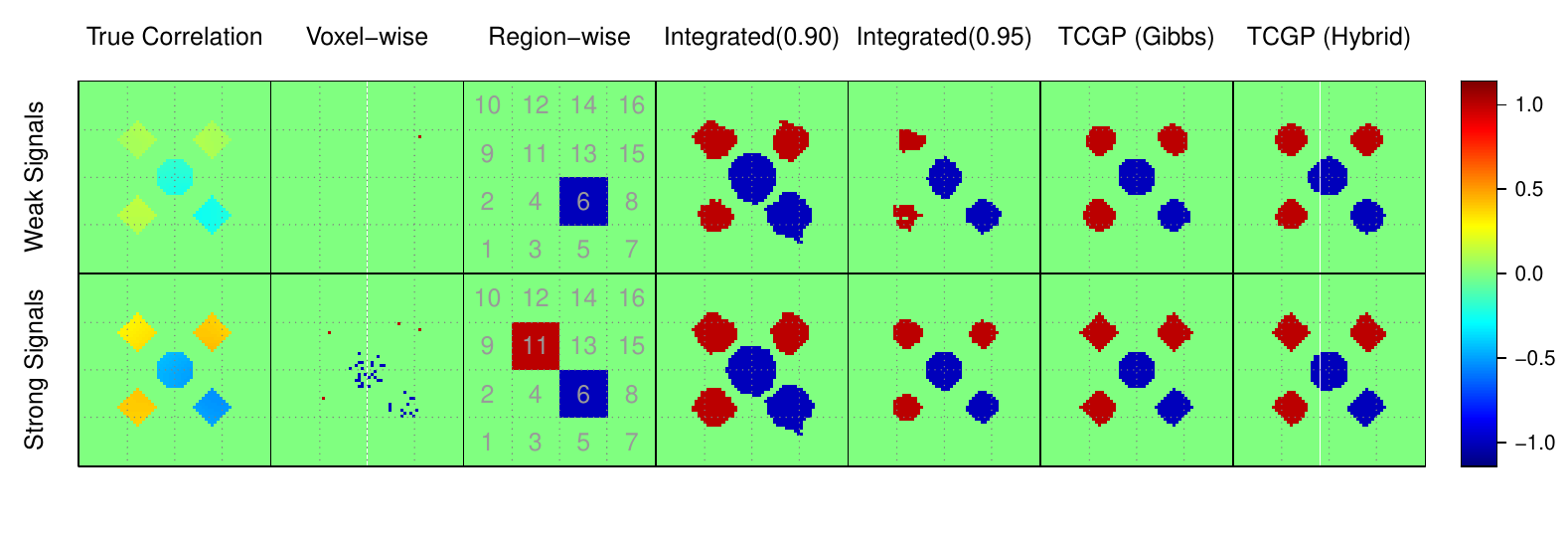}
\caption{Results of 2D image simulations. The first row is for a weak signal and the second row a strong signal. The panels from left to right show the true correlation map, the significantly positively (red) and negatively (blue) correlated regions selected by different methods. TCGP represents the proposed Thresholded Correlation Gaussian Process.}
\label{fig:sim2d}
\end{center}
\end{figure}
 
\begin{figure}[t!]
\centering
\includegraphics[width=0.8\textwidth, height=2.0in]{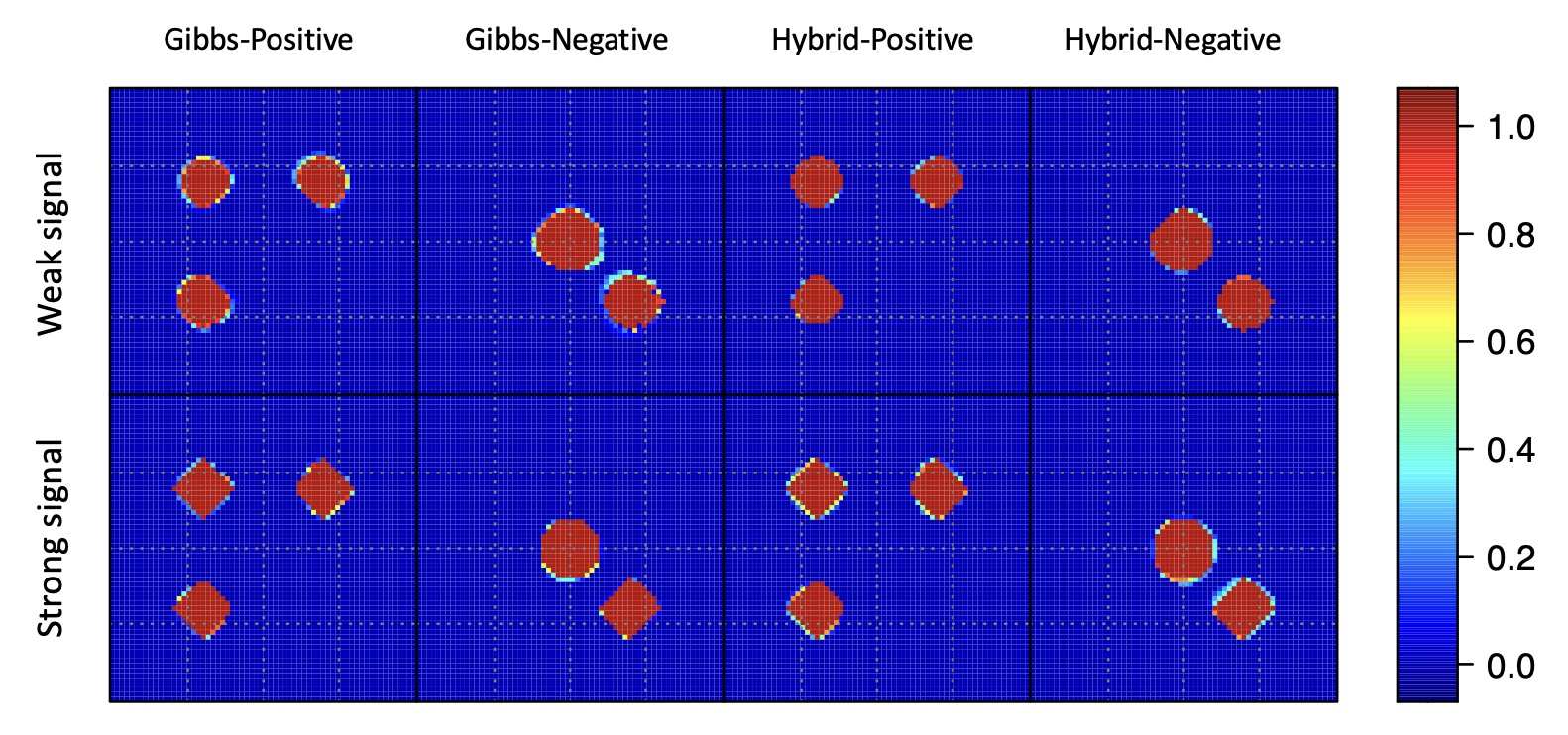}
\caption{Results of 2D image simulations. The posterior inclusion probability map of the positive and negative spatially-varying correlations using the Gibbs sampler and the hybrid mini-batch MCMC. }
\label{fig:sim2d-probmap}
\end{figure}

Table \ref{tab:sim2d} reports the results averaged over 100 data replications, and Figure \ref{fig:sim2d} visualizes the result for one data replication. We see that our proposed method clearly outperforms the alternative solutions. We observe essentially the same patterns as in the 3D example. In addition, the proposed Bayesian method is also capable of statistical inference, in that we can simulate the entire posterior distribution, compute the posterior inclusion probability, and quantify the uncertainty for the spatially varying correlation. Figure~\ref{fig:sim2d-probmap} shows the probability map of the identified positively and negatively correlated regions, which are close to the truth. 

We then vary the sample size $n = \{30, 50, 100\}$ while fixing the image resolution $m = 64 \times 64$, or vary $m = \{32 \times 32, 64 \times 64, 100 \times 100\}$ while fixing $n = 50$. Table \ref{sim1: table_diff_nm} reports the results averaged over 100 data replications. We see that our proposed method performs the best across different values of $n$ and $m$. Meanwhile, it maintains a competitive performance even when $n$ is relatively small or when $m$ is relatively large. 

\begin{table}[t!]
\caption{The 2D simulation example with the varying sample size $n$ and the varying image resolution $m$. Reported are the average sensitivity, specificity, and FDR, with standard error in the parenthesis, based on 100 data replications. Six methods are compared: the voxel-wise analysis, the region-wise analysis, the integrated method of \citet{li2018spatially} with two thresholding values, 0.95 and 0.90, and the proposed Bayesian method Thresholded Correlation Gaussian Process (TCGP) with the Gibbs sampler and the hybrid mini-batch MCMC.}
\label{sim1: table_diff_nm}
\vspace{-0.15in}
\begin{center}
\resizebox{\textwidth}{!}{
\begin{tabular}{clcccccc}
\toprule 
&\multirow{2}{*}{Method}&\multicolumn{3}{c}{Positive Correlation}&\multicolumn{3}{c}{Negative Correlation}\\
 & & Sensitivity & Specificity & FDR & Sensitivity & Specificity & FDR \\
\midrule
  $n=30$  &   Voxel-wise   &0.080(0.002) &  1.000(0.000) &   0.004(0.003)&    0.102(0.002) &  1.000(0.000) &   0.001(0.002)\\
        &   Region-wise  &0.148(0.005) &  0.971(0.002) &   0.326(0.003)&    0.473(0.006) &  0.957(0.003) &   0.624(0.003)\\ 
        &   Integrated(0.95)   &0.518(0.005) &  0.992(0.003) &   0.199(0.008)&    0.781(0.003) &  0.993(0.004) &   0.146(0.009)\\
        &   Integrated(0.90)   &0.855(0.007) &  0.960(0.005) &   0.378(0.010)&    0.871(0.004) &  0.937(0.004) &   0.392(0.011)\\
        &   TCGP (Gibbs)   &0.910(0.004) &  0.991(0.003) &   0.109(0.005)&    0.990(0.002) &  0.993(0.001) &   0.065(0.007)\\
        &   TCGP (Hybrid)   &0.890(0.005) &  0.990(0.003) &   0.111(0.008)&  0.983(0.004) &  0.990(0.002) &   0.110(0.008)\\
    \cmidrule{1-8} 
  $n=50$  &   Voxel-wise  &0.098(0.002) &  1.000(0.000) &   0.002(0.001)&   0.150(0.002) &       1.000(0.000) &   0.003(0.001)\\
        &   Region-wise &0.438(0.004) &  0.953(0.005) &   0.547(0.010)&   0.573(0.003) &       0.956(0.001) &   0.629(0.010)\\ 
        &   Integrated(0.95)  &0.659(0.003) &  0.995(0.002) &   0.130(0.008)&  0.899(0.005) &  0.997(0.001) &   0.110(0.009)\\
        &   Integrated(0.90)  &0.959(0.009) &   0.970(0.005) &   0.308(0.009)&  0.969(0.003) &   0.969(0.003) &   0.355(0.010)\\
        &   TCGP (Gibbs)   &0.941(0.004) &  0.995(0.002) &   0.081(0.005)&   0.996(0.002) &       0.992(0.001) &   0.063(0.005)\\
        &   TCGP (Hybrid) &0.931(0.005) &  0.993(0.003) &   0.092(0.005)&   0.993(0.002) &       0.992(0.002) &   0.086(0.006)\\
     \cmidrule{1-8} 
   $n=100$&   Voxel-wise&   0.102(0.004) & 1.000(0.001) &   0.002(0.003) &  0.198(0.001) &       1.000(0.000) &   0.003(0.001)\\
        &   Region-wise&    0.617(0.010) & 0.881(0.003) &  0.744(0.004)&   0.476(0.005) &       0.955(0.002) &   0.631(0.010)\\ 
        &   Integrated(0.95) &   0.714(0.005) & 0.998(0.003) &   0.099(0.005)&   0.898(0.004) &       0.997(0.002) &   0.099(0.008)\\
        &   Integrated(0.90) &   0.980(0.010) & 0.969(0.010) &   0.300(0.010)&   0.975(0.003) &       0.971(0.003) &   0.298(0.011)\\
        &   TCGP (Gibbs)     &   0.953(0.002) & 0.997(0.002) &   0.041(0.002)&   0.999(0.001) &       0.997(0.001) &   0.033(0.001)\\
        &   TCGP (Hybrid)   &   0.945(0.003) & 0.997(0.002) &   0.069(0.003)&   0.993(0.003) &       0.996(0.001) &   0.085(0.002)\\
\midrule \midrule 
  $m = 32\times32$ &  Voxel-wise &    0.017(0.001) &   1.000(0.000) &   0.005(0.001)&   0.040(0.002) &  1.000(0.000) & 0.004(0.002)\\
        &  Region-wise&           0.297(0.005) &   0.945(0.005) &   0.531(0.010)&   0.472(0.003) &  0.957(0.002) & 0.617(0.010)\\
        &  Integrated(0.95) &     0.620(0.005) &   0.989(0.004) &   0.138(0.005)&   0.852(0.004) &  0.989(0.001) & 0.198(0.009)\\
        &  Integrated(0.90) &     0.933(0.010) &   0.971(0.006) &   0.287(0.008)&   0.944(0.004) &  0.957(0.005) & 0.300(0.011)\\
        &   TCGP (Gibbs)   &        0.931(0.003) &   0.993(0.002) &   0.083(0.003)&   0.991(0.004) &  0.992(0.003) & 0.065(0.004)\\
        &  TCGP (Hybrid) &         0.922(0.005) &   0.992(0.002) &   0.082(0.005)&   0.991(0.005) &  0.991(0.002) & 0.089(0.005)\\
    \cmidrule{1-8} 
  $m = 64\times 64$ & Voxel-wise  &   0.098(0.002) &  1.000(0.000) &   0.002(0.001)&   0.150(0.002) &   1.000(0.000) &   0.003(0.001)\\
        &   Region-wise &         0.438(0.004) &  0.953(0.005) &   0.547(0.010)&   0.573(0.003) &   0.956(0.001) &   0.629(0.010)\\ 
        &   Integrated(0.95)  &   0.659(0.003) &  0.995(0.002) &   0.130(0.008)&   0.899(0.005) &   0.997(0.001) &   0.110(0.009)\\
        &   Integrated(0.90)  &   0.959(0.009) &  0.970(0.005) &   0.308(0.009)&   0.969(0.003) &   0.969(0.003) &   0.355(0.010)\\
        &   TCGP (Gibbs)   &        0.941(0.004) &  0.995(0.002) &   0.081(0.005)&   0.996(0.002) &   0.992(0.001) &   0.063(0.004)\\
        &   TCGP (Hybrid) &        0.931(0.005) &  0.993(0.003) &   0.092(0.005)&   0.993(0.002) &   0.992(0.002) &   0.086(0.006)\\
    \cmidrule{1-8}  
 $m = 100\times100$&   Voxel-wise &   0.005(0.001) &   1.000(0.000) &   0.004(0.002)&   0.011(0.001) &  1.000(0.000) &   0.000(0.001)\\
        &   Region-wise&          0.627(0.002) &   0.861(0.003) &   0.763(0.005)&   0.462(0.002) &  0.948(0.002) &   0.663(0.008)\\ 
        &   Integrated(0.95) &    0.843(0.002) &   0.998(0.003) &   0.039(0.005)&   0.952(0.004) &  0.997(0.002) &   0.052(0.007)\\
        &   Integrated(0.90)  &   0.960(0.003) &   0.965(0.005) &   0.300(0.012)&   0.977(0.002) &  0.965(0.004) &   0.298(0.011)\\
        &   TCGP (Gibbs) &          0.971(0.001) &   0.999(0.000) &   0.029(0.002)&   0.997(0.001) &  0.998(0.002) &   0.031(0.001)\\
        &   TCGP (Hybrid) &        0.964(0.001) &   0.999(0.000) &   0.033(0.001)&   0.995(0.001) &  0.997(0.002) &   0.033(0.002)\\
\bottomrule 
\end{tabular}
}
\end{center}
\end{table}


\subsection{Additional 3D simulations}
We conduct additional simulations for the 3D image example. We fix the sample size $n = 904$ follow the Human Connectome Project Data and vary the signal to noise ratio with $\zeta_k = 5$ for weak signal and $\zeta_k = 0.5$ for strong signal. Table \ref{tab:sim3d_snr} reports the results averaged over 100 data replications. We see that the proposed method outperforms other methods with different signal to noise ratio.
\begin{table}[t!]
\captionsetup{font=small}
\caption{Simulation results of the 3D image example with the varying signal to noise ratio. Reported are the average sensitivity, specificity, and FDR, with standard error in the parenthesis, based on 100 data replications. Six methods are compared: the voxel-wise analysis, the region-wise analysis, the integrated method of \citet{li2018spatially} with two thresholding values, 0.95 and 0.90, and the proposed Bayesian method with the Gibbs sampler and the hybrid mini-batch MCMC. }
\label{tab:sim3d_snr}
\vspace{-0.15in}
\begin{center}
\resizebox{\textwidth}{!}{
\begin{tabular}{clcccccc}
\toprule 
\multirow{2}{*}{Signal}&\multirow{2}{*}{Method}&\multicolumn{3}{c}{Positive Correlation}&\multicolumn{3}{c}{Negative Correlation}\\
 & & Sensitivity & Specificity & FDR & Sensitivity & Specificity & FDR \\
\midrule
  Weak&  Voxel-wise          &  0.082 (0.003) & 0.999 (0.005) &   0.001 (0.006)&  0.101 (0.005) &  0.998 (0.000) &   0.002 (0.003)\\
        &   Region-wise      &  0.366 (0.001) & 0.865 (0.002) &   0.573 (0.011)&  0.472 (0.002) &  0.892 (0.003) &   0.453 (0.004)\\ 
        &   Integrated(0.95) &  0.487 (0.002) & 0.981 (0.001) &   0.160 (0.010)&  0.582 (0.001) &  0.952 (0.005) &   0.101 (0.003)\\
        &   Integrated(0.90) &  0.873 (0.008) & 0.934 (0.001) &   0.230 (0.009)&  0.831 (0.004) &  0.946 (0.005) &   0.270 (0.004)\\
        &   TCGP (Gibbs)     &  0.890 (0.005) & 0.987 (0.001) &   0.070 (0.007)&  0.890 (0.002) &  0.975 (0.003) &   0.075 (0.001)\\
         &  TCGP (Hybrid)    &  0.884 (0.002) & 0.978 (0.001) &   0.078 (0.006)&  0.871 (0.004) &  0.965 (0.003) &   0.089 (0.002)\\
\midrule
  Strong&  Voxel-wise        &  0.220 (0.005) & 0.999 (0.002) &   0.001 (0.001)&  0.237 (0.004) &  0.999 (0.000) &   0.002 (0.001)\\
        &  Region-wise       &  0.641 (0.003) & 0.765 (0.001) &   0.587 (0.010)&  0.627 (0.006) &  0.824 (0.005) &   0.532 (0.003)\\
        &  Integrated(0.95)  &  0.550 (0.005) & 0.992 (0.000) &   0.066 (0.005)&  0.882 (0.005) &  0.970 (0.000) &   0.101 (0.002)\\
        &  Integrated(0.90)  &  0.934 (0.010) & 0.974 (0.003) &   0.244 (0.007)&  0.933 (0.010) &  0.955 (0.001) &   0.233 (0.003)\\
        &  TCGP (Gibbs)      &  0.951 (0.002) & 0.997 (0.001) &   0.052 (0.003)&  0.951 (0.002) &  0.991 (0.001) &   0.041 (0.002)\\
        &  TCGP (Hybrid)     &  0.949 (0.003) & 0.995 (0.001) &   0.058 (0.004)&  0.950 (0.003) &  0.989 (0.001) &   0.050 (0.001)\\
\bottomrule 
\end{tabular}
}
\end{center}
\end{table}

\subsection{Sensitivity analysis}
\label{appen: sensitivity-analysis}
 
In our hybrid mini-batch MCMC, we sample a subset of $m_s$ voxels and use the full dataset after every $T_0$ iterations of using the mini-batch data. We next carry out a sensitivity analysis to study the effect of $m_s$ and $T_0$. Table \ref{sim1: table_diff_msT0} reports the results averaged over 100 data replications. We see that the results are relatively stable for different values of $m_s$ and $T_0$.

\begin{table}[t!]
\caption{The sensitivity analysis of the batch size $m_s$ and the number of iterations $T_0$ for the hybrid mini-batch MCMC. Reported are the average sensitivity, specificity, and False Discorvery Rate (FDR), with standard error in the parenthesis, based on 100 data replications.}
\label{sim1: table_diff_msT0}
\vspace{-0.15in}
\begin{center}
\resizebox{\textwidth}{!}{
\begin{tabular}{ccccccccc}
\toprule 
\multirow{2}{*}{$m_s$}&\multirow{2}{*}{$T_0$}&\multicolumn{3}{c}{Positive Correlation}&\multicolumn{3}{c}{Negative Correlation}\\
 & & Sensitivity & Specificity & FDR & Sensitivity & Specificity & FDR \\
\midrule
$m/32$ &  20 &   0.950(0.003) &   1.000(0.001) &   0.015(0.003)&      0.991(0.002) &  0.989(0.003) & 0.050(0.005)\\
  \midrule
$m/16$ &  20 &   0.953(0.003) &   0.996(0.001) &   0.061(0.002)&      0.991(0.003) &  0.997(0.001) & 0.049(0.005)\\
  \midrule
$m/4$ &   20  &   0.955(0.002) &   0.997(0.001) &   0.058(0.002)&      0.990(0.001) &  0.997(0.001) & 0.047(0.003)\\
\midrule \midrule 
  $m/16$ &  50 &  0.948(0.003) &   0.998(0.001) &   0.045(0.002)&      0.990(0.002) &  0.990(0.003) & 0.062(0.003)\\
  \midrule
  $m/16$ &  20 &  0.953(0.003) &   0.996(0.001) &   0.061(0.002)&      0.991(0.003) &  0.997(0.001) & 0.049(0.005)\\
  \midrule
  $m/16$ &   10 & 0.953(0.001) &   0.995(0.001) &   0.059(0.003)&      0.993(0.002) &  0.998(0.001) & 0.041(0.004)\\
\bottomrule 
\end{tabular}}
\end{center}
\end{table}

\subsection{Prior specification for the HCP data analysis}
\label{appen: aomega}

In our HCP data analysis, we set the prior for $\omega$ as U$(a_\omega, b_\omega)$, and we choose $a_\omega$ and $b_\omega$ as the $75\%$ quantile and $100\%$ quantile of $\{|\xi(v)|\}_{v\in \mathcal{B}}$, respectively. The choice of $a_\omega$ is based on the belief that at most $25\%$ voxels have non-zero correlations. Here we vary $a_\omega = \{0.73, 0.75, 0.77\}$, and investigate the corresponding performance of our proposed method. Table \ref{hcp: aomega} reports the results, which we see that are relatively stable across different choices of $a_\omega$. 

\begin{table}[t!]
\caption{Prior specification for the Human Connectome Project data under different choices of $a_\omega$. Reported are the activation regions containing more than 100 voxels that are declared having a nonzero correlation.}
\label{hcp: aomega}
\vspace{-0.15in}
\begin{center}
\begin{tabular}{p{1cm}cccc}
\toprule
\multicolumn{5}{c}{Lingual-R}\\
\toprule 
$a_\omega$ & \makecell[c]{cluster size}& \makecell[c]{Activation center}& \makecell[c]{overlap rate}& \makecell[c]{mean correlation}\\
\midrule
 0.73 &   151&   (-10.0, -74.5, -4.0)& 0.931& 0.35\\
\midrule
0.75 &   144&   (-10.4, -75.3, -4.5)& 1.000& 0.35\\
\midrule 
0.77 &   140&   (-10.6, -75.8, -5.4)& 0.905& 0.38\\
\bottomrule
\multicolumn{5}{c}{Angular-R}\\
\toprule 
$a_\omega$ & \makecell[c]{cluster size}& \makecell[c]{cluster center}& \makecell[c]{overlap rate}& \makecell[c]{mean correlation}\\
\midrule
0.73 &   215&   (-45.9, -60.1, 45.5)& 0.910& 0.41\\
\midrule
0.75 &   209&   (-46.9, -60.2, 44.7)& 1.000& 0.43\\
\midrule 
0.77 &   200&   (-46.0, -59.9, 43.9)& 0.911& 0.43\\
\bottomrule
\multicolumn{5}{c}{Temporal-Mid-L}\\
\toprule 
$a_\omega$ & \makecell[c]{cluster size}& \makecell[c]{cluster center}& \makecell[c]{overlap rate}& \makecell[c]{mean correlation}\\
\midrule
0.73 &   110&   (62.1, -24.9, 1.3)& 0.940& 0.42\\
\midrule
0.75 &   104&   (63.1, -25.7, 1.4)& 1.000& 0.41\\
\midrule 
0.77 &   99&   (62.7, -25.5, 1.3)& 0.921& 0.43\\
\bottomrule
\multicolumn{5}{c}{Precentral-L}\\
\toprule 
$a_\omega$ & \makecell[c]{cluster size}& \makecell[c]{cluster center}& \makecell[c]{overlap rate}& \makecell[c]{mean correlation}\\
\midrule
0.73 &   130&   (29.1, -23.0, 64.5)& 0.930& -0.41\\
\midrule
0.75 &   115&    (28.6, -23.1, 65.4)& 1.000& -0.44\\
\midrule 
0.77 &   107&   (28.8, -23.1, 65.8)& 0.931& -0.42\\
\bottomrule
\multicolumn{5}{c}{Occipital-Inf-R}\\
\toprule 
$a_\omega$ & \makecell[c]{cluster size}& \makecell[c]{cluster center}& \makecell[c]{overlap rate}& \makecell[c]{mean correlation}\\
\midrule
0.73 &   130&   (-38.1, -81.0, -3.9)& 0.910& -0.45\\
\midrule
0.75 &   122&    (-38.8, -81.7, -3.2)& 1.000& -0.44\\
\midrule 
0.77 &   107&   (-38.5, -80.0, -4.0)& 0.901& -0.43\\
\bottomrule 
\end{tabular}
\end{center}
\end{table}



\end{document}